\documentclass[12pt,preprint]{aastex}
\def\etal{{\it et al.\ }}
\def\eg{{\it e.g.,}\,}
\def\ie{{\it i.e.,}}
\def\Halpha{H$\alpha$}
\def\Hbeta{H$\beta$}
\def\Hgamma{H$\gamma$}
\def\Hdelta{H$\delta$}
\def\Lya{Ly$\alpha$}
\def\Lyb{Ly$\beta$}

\newcounter{qub}
\shorttitle{Wide HI profile galaxies} \shortauthors{Brosch et al.}

\begin{document}

\def\etal{{\it et al.\, }}
\def\eg{{\it e.g.,}\,}
\def\ie{{\it i.e.,}\,}
\def\Halpha{H$\alpha$}
\def\Hbeta{H$\beta$}
\def\Hgamma{H$\gamma$}
\def\Hdelta{H$\delta$}
\def\Lya{Ly$\alpha$}
\def\Lyb{Ly$\beta$}

%-------------------------------------%
%     ENCAPSULATED POSTSCRIPT         %
%-------------------------------------%
\def\putplot#1#2#3#4#5#6#7{\begin{centering} \leavevmode
\vbox to#2{\rule{0pt}{#2}} \includegraphics{#1}
% e.g.,
% \putplot{psfile}{vspace}{angle}{hscale}{vscale}{hoffset}{voffset}
% with vspace in any TeX units, angle in degrees, scale in percent,
% and offset in PostScript points (72/in)

\end{centering}}
%-------------------------------------%
%    END ENCAPSULATED POSTSCRIPT      %
%-------------------------------------%

% to fill page with print
% \oddsidemargin=0cm
 %\evensidemargin=0cm
 %\topmargin=0cm
%\textwidth=16cm
%\textheight=22cm
%-----------------------------------------------------
%\begin{document}

\title{Galaxies with wide HI profiles}

\author{Noah Brosch, Oded Spector and Adi Zitrin}%\altaffilmark{1}}
\affil{The Wise Observatory and the Raymond and  Beverly Sackler School of Physics and
Astronomy, the Faculty of Exact Sciences, \\ Tel Aviv University, Tel Aviv 69978, Israel}
\email{noah@wise.tau.ac.il}

%\date{1 feb. 2009}
% Updated to re-create file from August 2009 that went wrong...
% Corrections to that entered 25 Nov 2009
%C:\Users\noah\Documents\Projects\ALFALFA_wide_profiles\WideALFALFAprofiles\WideProfiles.tex
% Additions following H-alpha data from Oded and model runs from Adi 27 March 2010
% Editing changes April 6, 2010
% New changes following remarks from Oded on the draft; still misses H\alpha for 2 objects and redoing the SFH by Adi (May 27 2010)
% Further changes after Oded finished reducing the last 2 galaxies and Adi ran new models
% Editing changes 12 Aug 2010
% last corrections from Oded put in 29 August 2010; paper submitted to MNRAS
% corrections from Martha Haynes 31 August 2010: acknowledgements format, AGC numbers instead of HI numbers in Table 1. Following tables retained as previously, by truncating RA+Dec
% 01 Feb 2011: changed short format names to AGC numbers throughout

%\maketitle

%%% ----------------------------------------------------------------------

\begin{abstract}

We investigate the nature of objects in a complete sample of 28 galaxies selected from the first
sky area fully covered by ALFALFA, being well-detected and  having  HI profiles wider than 550 km s$^{-1}$. The selection does not use brightness, morphology, or any other property derived from optical or
other spectral bands. %Our goal is to detect possible differences between these
%galaxies with wide HI profiles and regular galaxies. %The original selection was performed y, providing a clean HI-selected sample.

We investigate the degree of isolation, the morphology, and other properties
gathered or derived from open data bases and show that some objects have wide HI profiles probably because they are disturbed or are interacting,
or might be confused in the ALFALFA beam.
We identify a sub-sample of 14 galaxies lacking immediate
interacting neighbours and showing regular,
symmetric, two-horned HI profiles that we propose as candidate high-mass disk systems (CHMDs).

We measure the net-H$\alpha$ emission from the CHMDs and
combine this with public multispectral data to model the global star
formation (SF) properties of each galaxy. The H$\alpha$ observations show SFRs not higher
than a few M$_{\odot}$ yr$^{-1}$. Simple SF models indicate that the CHMDs
formed most of their stars almost a Hubble time ago, but most also underwent an SF
event in the last $\sim10^{6-7}$ yrs; the young stars now produce 10 to 30\% of the visible light.
The spatial distribution of the SF-regions is compatible with recycled stellar ejecta.

%{\bf If 10=30\% of the light, how much relative mass in the young population? Is reprocessed stellar mass loss sufficient, or is external gas influx necessary? Enough reprocessed to make 3 solar masses/yr ?}

We calculate representative dynamical masses M$_{dyn}$ for the CHMDs ranging from $\sim10^{11}$ M$_{\odot}$ to $\sim7 \times 10^{11}$ M$_{\odot}$. The M$_{dyn}$ values are
larger by factors of 2.5 to 7.5 than the baryonic masses of the luminous stars and
gas but are affected by rather large errors. We test the Tully-Fisher relation for the
CHMDs and show that these lie below the relation defined by
lower mass galaxies, i.e., that their M$_{dyn}$ is lower than expected when extrapolating the relation from lower mass galaxies to higher HI line widths. %This indicates a fundamental difference between high-mass disk galaxies and their lower-mass cousins.

\end{abstract}

%%% ----------------------------------------------------------------------
\maketitle
%%% ----------------------------------------------------------------------

%%%%%%%%%%%%%%%%%%%%%%%%%%%%%%%%%%%%%%%%%%%%%%%%%%%%%%%%%%%%%%%%%%%%%%%%%%%
\section{Introduction}

Two basic cosmological questions are how do galaxies form and how do they achieve their present structure. The current approach to answer these questions involves the comparison   of structure-formation models to galaxy observations at low- and high-redshift. The stellar masses of galaxies can increase through major mergers, accretion of smaller satellite systems, and star formation by converting gas into stars; some of this gas can be externally accreted. Guo \& White (2008) showed that the relative importance of these three modes is a strong function of the stellar mass in the specific galaxy. Galaxy growth through major mergers depends strongly on stellar mass, minor mergers contribute more to galaxy growth than major mergers at all redshifts and stellar masses, and in galaxies significantly less massive than the Milky Way, star formation dominates the growth.

Gavazzi \& Scodeggio (1996) found that the star formation in late-type galaxies (Sa and later) is most probably regulated by the total mass of a galaxy. Their assumptions in reaching this conclusion were that galaxies develop in isolation, that they all have solar metallicity, that the star formation rate (SFR) decreases exponentially with time with an e-folding time depending on the galaxy mass, and that the initial mass function is ``Salpeter'', with stars from 0.1 to 120 M$_{\odot}$.

Balogh et al. (2004) identified two galaxy families in the colour distribution of Sloan Digital Sky Survey (SDSS) objects; a blue star-forming family and a red, passively-evolving population. They concluded that transitions between the two families can take place on short timescales of a few Gyr. Such transitions can be caused by mass influx from a merger with a small, gas-rich galaxy (Martin et al. 2007).

In normal, inactive galaxies, the stellar velocity dispersion of the bulge $\sigma_*$ tracks the maximal rotation velocity of the disk roughly as $v_m \simeq 1.7 \times \sigma_*$ (Whitmore et al. 1979). Although the theoretical basis of this correlation is still not clear, and the $v_m - \sigma_*$ relation is not as tight as has been claimed (e.g., Pizzella et al. 2005, specifically for LSB galaxies), nevertheless an empirical relation between $v_m$ and $\sigma_*$ does exist (Courteau et al. 2007). This implies that $\sigma_*$ can, in principle, be estimated from $v_m$. Since $v_m$ can be measured from HI observations for galaxies that are sufficiently gas-rich, in the absence of a resolved rotation curve $v_m$ can be estimated from a single-dish HI measurement while the inclination angle of the disk can be derived from optical observations. Given the rotation velocity, it is possible to deduce the total galaxy luminosity through the Tully-Fisher relation (Tully \& Fisher 1977) and the dynamical mass of the system given an estimate of the disk size from optical imaging. It is not clear, however, whether these assumptions can be applied also to any random sample of galaxies, although this seems to be the case at least for the ``All Digital HI catalog'' objects (ADHIC: Courtois et al. 2009).

The Arecibo Legacy Fast ALFA (ALFALFA) survey is an on-going, second generation, blind extragalactic HI survey using the seven-feed L-band focal plane array (ALFA) at the Arecibo Observatory (AO). The seven beams have sizes of 3\arcmin.3  along the azimuth direction and 3\arcmin.8  along the zenith angle direction. ALFALFA is performed by drift-scanning ribbons of the sky and repeating these after a few months. The combination of multiple scans with the fast one-second sampling (14 samples per source beam transit time) allows the centroiding of HI sources to much less than a  beam width (Giovanelli et al. 2005a, b).  ALFALFA will eventually survey more than 7000 square degrees of the high galactic latitude sky and is expected to detect more than 25000 extragalactic HI sources up to a redshift of 18000 km s$^{-1}$ exploiting AO's superior sensitivity, angular resolution and digital technology. The survey aims primarily to probe the faint end of the HI mass function (HIMF) and will provide a complete HI census in the surveyed sky area. The ALFALFA survey strategy was described by Giovanelli et al. (2005a) and survey results for selected sky regions were presented by Giovanelli et al. (2005b, 2007), Saintonge et al. (2007) and Kent et al. (2008). A revision of the HI mass function and a determination of $\Omega_{HI}$ from the 40\% of the total survey area with complete source extraction was published by Martin et al. (2010). The source detection algorithm and method were described by Saintonge (2007). ALFALFA is the best deep and unbiased HI survey now in existence. %{\bf check for new publications}

The ALFALFA survey yields a number of parameters for each detection. The ones most often used are the position and radial velocity (to characterize the location of each object in 3D space), and the total HI flux (to derive the HI mass). We  use here an additional parameter, $w(50)$ that is the width of the HI profile at 50\% of the peak flux density, to select a sample of objects with HI profiles wider than 550 km s$^{-1}$ from the published ALFALFA catalogs and associated data products, and study it. The  $w(50)$ parameter can be used to estimate the maximal rotation velocity of the gas in a galaxy (e.g., Courtois et al. 2009), since for disk galaxies $w(50)=2\times v_{max}sin(i)$.

There could be a number of reasons why a galaxy should show a very wide HI profile in a single-dish observation: it could have a high mass implying a fast asymptotic rotation velocity, it could originate in a galaxy showing not only regular rotation but also some chaotic dynamics, or it could result from the detection of a confused binary or multiple galaxy system where the entire HI profile in the relatively wide ALFALFA beam is contributed by two or more objects that may be strongly interacting (as found by e.g., Bothun et al. 1982). We expect to be able to distinguish between these two possibilities from inspecting the optical galaxy images and the galaxies' HI profiles, and from the characterization of the CHMD's galaxy neighbourhood, thus being able to select a clean, HI-based, sample of high-mass objects.

Trachternach et al. (2009) studied the baryonic Tully-Fisher (TF) relation in a sample of very low-mass dwarf galaxies. They concluded that the baryonic TF relation is followed by all rotationally-dominated galaxies. Since the relation was established using low-to-intermediate rotational velocities, it is interesting to see how well it reproduces the behaviour of high rotational velocities galaxies.

High-mass objects observed at present represent strong deviations of the initial density fluctuation field in the early Universe. It is possible that interacting galaxies also originate from such strong density fluctuations. In this case, the surroundings of galaxies with wide HI profiles should be characterized by a higher galaxy density than at random locations. This also is tested in our paper, where we adopt cosmology-corrected quantities: H$_0$ =  73 km s$^{-1}$ Mpc$^{-1}$, $\Omega_{matter}$=0.27 and $\Omega_{vacuum}$=0.73, as in NED.

The connection between HI gas and galaxy mass at the high-mass end was recently studied by Catinella et al. (2010). They found that the gas-to-stellar-mass ratio decreases with stellar mass and stellar surface mass density. Since we will be deriving both quantities here, it will be interesting to compare our results with those of Catinella et al.

%%%%%%%%%%%%%%%%%%%%%%%%%%%%%%%%%%%%%%%%%%%%%%%%%%%%%%%%%%%%%%%%%%%%%%%%%%%
%\begin{figure*}
%{\centering
% \includegraphics[clip=,angle=0,width=17.0cm]{Ring_gri_2.ps}
%}
% \caption{%
%Place holder: whatever we want to show
%    \label{Fig:RG1_pix}}
%\end{figure*}
%%%%%%%%%%%%%%%%%%%%%%%%%%%%%%%%%%%%%%%%%%%%%%%%%%%%%%%%%%%%%%%%%%%%%%%%%%%

%%%%%%%%%%%%%%%%%%%%%%%%%%%%%%%%%%%%%%%%%%%%%%%%%%%%%%%%%%%%%%%%%%%%%%%%%%%
\section{The sample}
\label{txt:sample}

The sample studied here consists of 28 objects selected from the ALFALFA observations.  We stress that the selection was done on the ALFALFA data set available only to the ALFALFA collaboration, including sources not yet published. All the objects with recession velocity smaller than 12000 km s$^{-1}$, HI profile width $w(50)\geq$550 km s$^{-1}$ and ALFALFA detection code 1 (implying high signal-to-noise and high confidence detection) were selected from the ALFALFA region that has been fully covered by the survey, i.e., 7$^h$.5$\leq$RA$\leq$16$^h$.5 and +4$^{\circ}\leq$Dec$ \leq +16^{\circ}$, which is $\sim$20\% of the final ALFALFA survey coverage. An ALFALFA detection code of 1 implies a S/N$\geq$6.5, but this cannot be translated into a flux density limit since the S/N depends, among others, on the HI line width (see e.g. Giovanelli et al. 2007). As an example, a source with a flux density of 0.72 Jy km sec$^{-1}$ would be detected at 5$\sigma$ if its line width would be 200 km sec$^{-1}$ (Giovanelli et al. 2005b, Figure 8).

The sample is therefore complete according to the selection criteria listed here and is unbiased regarding optical brightness or surface brightness criteria. The volume surveyed to detect these 28 objects is 7.2$\times10^5$ Mpc$^3$; the corresponding volume density of these very rare objects is 3.9$\times10^{-5}$ Mpc$^{-3}$. By selecting objects closer than 12000 km sec$^{-1}$ we avoid the outer edge of the ALFALFA survey, which is the velocity bin 120000$\leq v \leq$18000 km s$^{-1}$. %This velocity bin is strongly affected by RFI from the Federal Aviation Administration (FAA) radar at 1350 MHz, making the survey effectively ``blind'' to H I sources redshifted to the range 15,200$\leq$cz$\leq$16,000 km s$^{-1}$.
 The lowest recession velocity present in our sample is 4933 km s$^{-1}$ excluding also the low-velocity segment of each ALFALFA data cube that contains ``local'' objects, such as the Virgo Cluster.% was also ed. This segment .

The region has full SDSS coverage; it is thus possible to extract broad-band total magnitudes and colors as well as images with reasonable resolution for the morphological classification of the selected objects. SDSS also yields radial velocities for the sample objects and for some of their optical neighbours, while ALFALFA yields radial velocities  for the optically-faint but HI-rich neighbours. We aim to characterize the neighbourhood galaxy density of each object by the distance to its nearest catalogued neighbor and by the number of objects in a preset volume, say within 3h$^{-1}$ Mpc projected distance and 300 km s$^{-1}$ velocity difference. %{\bf Give value for cluster and for extreme field}

The galaxies selected from the ALFALFA survey are listed in Table 1. We give there the name of the object as used in the ALFALFA survey (the AGC designator), the derived HI position and the optical center of the galaxy from SDSS (both in J2000 coordinates), its heliocentric recession velocity and error (in parentheses) in km s$^{-1}$, the full width at half-maximum of the HI line profile and its error (in parentheses) in km s$^{-1}$, and the HI flux integral FI and its error (in parentheses) in Jy km s$^{-1}$. %The last column lists Sintmap {\bf what is this?}.
The heliocentric velocity is measured as the midpoint between the channels where the flux density drops to 50\% of each of the two peaks (or of one, if only one is present) at each side of the spectral feature and is measured in km s$^{-1}$, as explained e.g., in Giovanelli et al. (2007). The velocity width of the source line profile, w(50), is measured at the 50\% outer level of each of the two peaks and is corrected for instrumental broadening.

We checked carefully all 28 objects in the original selection to reject those that could have a wide HI profile by being a confused pair or multiple object, or by being affected by interactions to show a disturbed appearance. This was done by retaining only galaxies where a two-horned profile could clearly be seen, and by inspecting the SDSS image and radial velocity data to identify morphological disturbances or possible nearby companions. Information about the inspection of SDSS images and spectral data is given in Appendix A; the HI profiles are shown in Appendix B. The galaxies identified here as having an intrinsic wide HI profile, and retained as such, are marked in bold font in Table 1.

\begin{table*}
\caption{ALFALFA galaxies with wide HI profiles}
\label{t:sample}
{\small
\begin{tabular}{cccccc} \hline																			
Object	&	HI ellipse 	&	Optical 	&	v$_{50}$	& w(50) &	FI %&	Sintmap	
\\
AGC name & (J2000) $\alpha, \delta$ & (J2000) $\alpha, \delta$ & km s$^{-1}$  & km s$^{-1}$ & Jy km s$^{-1}$ %& %SintMap
\\ \hline
181756	&	082656.0+073011	&	082654.6+072953 & 9323(43)   & 685(86) & 4.26(0.13)  \\
192281	&	095900.0+130305	& 095859.9+130309	&	10470(3) &	885(5) & 3.62(0.17)  \\
{\bf 205190}  &	104801.9+142431	& 104802.2+142429	& 9683(4)    &	 698(7) &	 2.6(0.12)  \\
{\bf 200589}  &	104837.6+125846	& 104834.1+125859	& 10768(3)   &	553(7) &	 1.73(0.11)  \\
006042	&	105611.8+094544	& 105615.4+094515   &	9921(6)  &	568(12) &	 1.61(0.1)  \\
201697	&   105744.9+151755	&	105744.8+151827	&	10955(3) &	650(7) &	 1.48(0.1) \\
{\bf 006066}  &	105859.7+063141	&	105859.3+063121	&	11807(2) &	667(4) & 6.25(0.14)  \\
{\bf 222042}  & 120305.6+051023 &	120305.0+051028	&	11569(5) &	555(9) & 2.37(0.14)  \\
226077  &   122518.4+160740	&	122514.2+160713	&	9250(2)	 & 582(4) & 2.1(0.12) 	 \\
226111  &	125934.9+150248	&	125936.0+150257	&	10678(1) &	566(3) &	 2.38(0.1) \\
233609	&   130711.6+133918 &	130711.6+133933	&	8143(6)  &	624(11) &	 2.53(0.13) \\
{\bf 008375}  &	131954.6+155105 &	131956.3+155101	& 7007(34)   &	696(68) &	 2.39(0.14) \\
{\bf 008379}  &	132021.8+062019	&	132020.3+062010	&	11967(6) & 569(11) &	 2.93(0.12) \\
{\bf 008475}  &	132925.7+110015	&	132925.8+110028	&	6835(1)  &	591(2) &	 15.71(0.14) \\
{\bf 008488}  &	133003.3+132511	&	133002.3+132458	&	7363(5)  &	596(11) &	 4.03(0.13) \\
008559	&	133453.9+135008	&	133455.9+134956	&	7025(12) & 572(23) &	 4.31(0.12) 	 \\
008766	&	135133.1+140515	&	135130.9+140530	&	6996(8)  &	622(16) &	 3.1(0.13) 	 \\
008902	&	135903.1+153411	&   135902.8+153357	&	7559(3)  &	691(6) &	 20.14(0.13) 	 \\
{\bf 230914}  &	140004.6+120630	&	140004.4+120641	&	11848(5) &	580(11) &	 3.89(0.13) 	 \\
008943	&	140213.5+080235	&	140213.1+080212	&	4933(49) &	669(98) &	 1.67(0.1) \\
{\bf 009031}  &	140740.4+145204	&	140739.3+145151	&	11864(3) &	709(6) &	 2.76(0.16) \\
{\bf 248977}  &	144110.9+145327	&	144112.3+145324	&	9153(19) &	560(38) &	 2.13(0.13)  \\
{\bf 009624}  &	145735.0+081726	&	145735.8+081706	&	11068(18) &	599(36) &	 2.74(0.12) 	 \\
009788	&	151536.1+081735	&	151538.7+081803	&	10178(9) &	564(17) &	 3.77(0.14) \\
009794	&	1516 9.1+103033	&	151610.8+103034	&  6406(2)    &	630(5) &	 22.03(0.13) \\
009838	&	152512.3+070823	&   152512.2+070916 &	10267(9) &	575(19) &	 2.59(0.15) \\
{\bf 260110}  & 160448.5+140741	&	160448.1+140744	&	10160(8) &	610(15) &	 5.99(0.16) \\
{\bf 010272}  &	161257.3+095143	&	161256.9+095201	&	5124(31) &	580(62) &	 1.72(0.1) \\

\hline
\end{tabular}
}
\end{table*}

We identified 14 galaxies with wide HI profiles that did not show signs of interaction nor had nearby ``significant'' neighbour galaxies; these might be truly high-mass objects. The objects chosen from the ALFALFA selection as candidate high-mass galaxies (CHMGs, see below) were observed in the R-band and in rest-frame H$\alpha$, using the Wise Observatory narrow-band filter set. These observations %, combined with the SDSS $r$-band image,
 allow the determination of the H$\alpha$ equivalent width. This, together with the absolutely calibrated $ugriz$ magnitudes from SDSS %, with GALEX FUV and NUV and 2MASS
and other public-domain data where available, allows the determination of the total H$\alpha$ emission line flux and of an approximate star formation rate (SFR) and star formation history (SFH) for each object. %We link the results with the neighborhood galaxy density to characterize the nature of the sources with wide HI profiles. %If the results will be promising, we would like to extend this analysis to other survey regions once they approach full coverage.

%We are aware of the team project proposed by Salucci et al. (Discovery and basic HI properties of the most luminous late-type spiral galaxies in the local Universe) where they want to follow-up optically and with HI kinematic observations ALFALFA sources with HI line widths of 500 km s$^{-1}$ and wider. Since they aim to study about 15 objects with optical rotation curves, BVRI photometry, and HI synthesis observations, we do not see a conflict between our proposal and theirs. If, however, they feel that there is a possible conflict with the spiral galaxies that would show up in our sample, we are willing to collaborate to achieve both goals while optimizing the observing time in the follow-ups.

%--------------------------------------------------------------------------
\subsection{H$\alpha$ observations}

H$\alpha$ imaging of 14 CHMDs was carried out at the Wise Observatory on 8 nights, from March 2009 to April 2010, using the 40-inch telescope with the PI CCD camera (pixel scale of 0.6 arcsec$\cdot$pixel$^{-1}$) and narrow-band H$\alpha$ filters centered
approximately on the wavelength of the redshifted H$\alpha$ line of each galaxy. At least three 20-minute dithered exposures in the narrow-band H$\alpha$ filter which best fits its redshift and three 5-minute dithered exposures in R were obtained for each galaxy. %in both R and. %The H$\alpha$ information was collected from three 20-minute dithered exposures of each galaxy obtained with the Wise Observatory's PI CCD through .
The images in the same band were debiased, flat-fielded, sky-subtracted, and combined into a final image for each filter.

In order to derive the net line emission contribution (that includes also that of the
[NII] lines), we subtracted the continuum contribution from a properly scaled R-band image. The scaling was done using stars, assuming that stars would not contribute
specific emission or absorption features in the narrow redshifted band, and that their contribution in R would only be continuum. This is justified, since even the lowest redshift H$\alpha$ filter we used is significantly distant from the zero-redshift H$\alpha$ line that could appear in the spectra of the scaling stars.

%The images were reduced and combined using IRAF standard procedures. Images of the net H$\alpha$ emission (including the adjacent [NII] emission lines) were obtained by subtracting a scaled R image from the H$\alpha$ image. The scaling was such that on average the counts of foreground stars in both images would be equal.
H$\alpha$ equivalent widths (EW) were obtained using uncalibrated flux measurements of the galaxies in the R and net-H$\alpha$ images and the transmission curves of the H$\alpha$ and R filters. Calibrated H$\alpha$ flux measurements were obtained using the measured EW values and the continuum flux per frequency at the H$\alpha$ redshifted wavelength, estimated by linearly interpolating between the SDSS measured $r$ and $i$ fluxes.

%--------------------------------------------------------------------------
\subsection{SDSS and 2MASS data}

Catalog data were extracted for each sample galaxy from the SDSS %{\bf [Adi: which SDSS magnitudes are listed?]}
and are listed in Table~\ref{t:sample2}. The table lists a shortened name for each object obtained by truncating the position of the object from column 1 of Table~\ref{t:sample}  to retain the first four right ascension digits, the declination sign and two declination digits, and the total $ugriz$ magnitudes from SDSS. % and 2MASS.
 Table~\ref{t:sample3} %{\bf Check the last table from Oded}
lists the H$\alpha$ emission line equivalent width in \AA\, and flux in erg cm$^{-2}$ sec$^{-1}$ determined as explained above, the GALEX FUV and NUV magnitudes with errors for the few objects where these data are available, and the 2MASS J, H, and K$_s$ magnitudes with errors. Given the lack of approximately uniform coverage by GALEX of the objects in our sample, we decided to only list the FUV and/or NUV magnitudes without using them for further investigations.

%\label{t:sdss}
\begin{table*}
\caption{SDSS total photometry for the sample galaxies}
\label{t:sample2}
\begin{tabular}{cccccc} \hline
ACG &	$u$	& $g$	& $r$	& $i$	& $z$  \\ \hline
181756	&17.42$\pm$0.02	&15.42$\pm$0.00	&14.48$\pm$0.00	&14.05$\pm$0.00&13.66$\pm$0.00 \\
192281	&17.95$\pm$0.02	&16.35$\pm$0.00	&15.51$\pm$0.00&15.05$\pm$0.00&14.71$\pm$0.01\\
205190	&17.09$\pm$0.02	&14.93$\pm$0.00&13.82$\pm$0.00&13.52$\pm$0.00&12.92$\pm$0.00\\
200589	&16.76$\pm$0.01	&14.77$\pm$0.00&13.87$\pm$0.00&13.42$\pm$0.00&13.10$\pm$0.00\\
006042	&16.35$\pm$0.01	&14.40$\pm$0.00&13.54$\pm$0.00&13.15$\pm$0.00&12.81$\pm$0.00\\
201697	&17.82$\pm$0.03	&16.07$\pm$0.00&15.22$\pm$0.00&14.79$\pm$0.00&14.44$\pm$0.00\\
006066	&16.98$\pm$0.02	&15.02$\pm$0.00&14.16$\pm$0.00&13.75$\pm$0.00&13.38$\pm$0.00\\
222042	&17.00$\pm$0.02	&15.00$\pm$0.00	&14.12$\pm$0.00&13.68$\pm$0.00&13.35$\pm$0.00\\
226077	&17.10$\pm$0.02	&15.59$\pm$0.00	&14.85$\pm$0.00&14.47$\pm$0.00&14.17$\pm$0.00\\
226111	&17.07$\pm$0.01	&15.27$\pm$0.00&14.40$\pm$0.00&13.97$\pm$0.00&13.59$\pm$0.00\\
233609	&17.71$\pm$0.02	&16.71$\pm$0.00&16.41$\pm$0.00&16.22$\pm$0.01&15.92$\pm$0.01\\
008375	&16.11$\pm$0.01	&14.16$\pm$0.00&13.32$\pm$0.00&12.87$\pm$0.00&12.54$\pm$0.00\\
008379	&16.68$\pm$0.02	&14.91$\pm$0.00&14.08$\pm$0.00&13.66$\pm$0.00&13.33$\pm$0.00\\
008475	&15.74$\pm$0.01	&13.70$\pm$0.00&12.74$\pm$0.00&12.33$\pm$0.00&11.93$\pm$0.00\\
008488	&16.20$\pm$0.01	&14.15$\pm$0.00&13.25$\pm$0.00&12.80$\pm$0.00&12.49$\pm$0.00\\
008559	&16.11$\pm$0.01	&14.05$\pm$0.00&13.07$\pm$0.00&12.58$\pm$0.00&12.23$\pm$0.00\\
008766	&15.66$\pm$0.01	&13.78$\pm$0.00&12.90$\pm$0.00&12.41$\pm$0.00&12.01$\pm$0.00\\
008902	&16.10$\pm$0.01	&14.09$\pm$0.00&13.21$\pm$0.00&12.81$\pm$0.00&12.41$\pm$0.00\\
230914	&16.57$\pm$0.01	&14.71$\pm$0.00&13.89$\pm$0.00&13.48$\pm$0.00&13.18$\pm$0.00\\
008943	&15.25$\pm$0.01	&13.30$\pm$0.00&12.45$\pm$0.00&12.03$\pm$0.00&11.74$\pm$0.00\\
009031	&16.83$\pm$0.01	&14.98$\pm$0.00&14.07$\pm$0.00&13.58$\pm$0.00&13.13$\pm$0.00\\
248977	&17.77$\pm$0.02	&15.94$\pm$0.00&15.14$\pm$0.00&14.74$\pm$0.00&14.40$\pm$0.00\\
009624	&16.22$\pm$0.01	&14.32$\pm$0.00&13.45$\pm$0.00&12.99$\pm$0.00&12.64$\pm$0.00\\
009788	&16.83$\pm$0.02	&15.13$\pm$0.00&14.22$\pm$0.00&13.70$\pm$0.00&13.28$\pm$0.00\\
009794	&19.14$\pm$0.03	&17.13$\pm$0.00&16.18$\pm$0.00&15.62$\pm$0.00&14.96$\pm$0.00\\
009838	&16.73$\pm$0.02	&14.82$\pm$0.00&13.89$\pm$0.00&13.42$\pm$0.00&13.04$\pm$0.00\\
260110	&17.13$\pm$0.01	&15.10$\pm$0.00&14.18$\pm$0.00&13.72$\pm$0.00&13.33$\pm$0.00\\
010272	&15.79$\pm$0.01	&13.78$\pm$0.00&12.86$\pm$0.00&12.34$\pm$0.00&11.98$\pm$0.00\\ \hline
\end{tabular}

Note: Magnitude errors smaller than 0.01 are reported here as 0.00.
\end{table*}

\begin{table*}
\caption{H$\alpha$, GALEX, and 2MASS total photometry for the galaxies in the sample}
\label{t:sample3}
\begin{tabular}{cccccccc} \hline
ACG    & EW H$\alpha$ & F$_{H\alpha}$  & FUV & NUV & J	& H	& K$_s$  \\
 & \AA\ & $\times10^{-13}$  & AB mag & AB mag & mag & mag & mag  \\ \hline
181756& - & - & - & - & 12.39$\pm$0.03 & 11.66$\pm$0.04 & 11.51$\pm$0.06 \\
192281 & - & - & - & 17.79$\pm$0.02 & 13.57$\pm$0.05 & 12.72$\pm$0.06 & 12.31$\pm$0.06 \\
205190 & 14$\pm$02 & 1.1$\pm$0.2 & - & - & 11.70$\pm$0.02 & 10.99$\pm$0.03 & 10.66$\pm$0.04 \\
200589 & 19$\pm$03 & 1.6$\pm$0.3 & - & - & 11.84$\pm$0.02 & 11.15$\pm$0.03 & 10.85$\pm$0.04 \\
006042 & - & - & - & - & 11.29$\pm$0.03 & 10.60$\pm$0.03 & 10.46$\pm$0.06 \\
201697 & - & - & - & - & 11.29$\pm$0.03 & 11.00$\pm$0.03 & 10.46$\pm$0.06 \\
006066 & 19$\pm$06 & 1.2$\pm$0.4 & - & - & 11.82$\pm$0.04 & 11.18$\pm$0.05 & 10.96$\pm$0.08 \\
222042 & 21$\pm$04 & 1.4$\pm$0.2 & - & - & 12.04$\pm$0.04 & 11.50$\pm$0.05 & 11.09$\pm$0.07 \\
226077 & - & - & - & - & 13.44$\pm$0.08 & 12.67$\pm$0.10 & 12.23$\pm$0.09 \\
226111 & - & - & - & - & 12.43$\pm$0.03 & 11.63$\pm$0.04 & 11.38$\pm$0.04 \\
233609 & - & - & - & - & 14.12$\pm$0.11 & 13.59$\pm$0.17 & 13.10$\pm$0.16 \\
008375 & 10$\pm$04 & 1.3$\pm$0.6 & - & 19.30$\pm$0.06 & 11.26$\pm$0.02 & 10.60$\pm$0.02 & 10.30$\pm$0.03 \\
008379 & 25$\pm$08 & 1.7$\pm$0.5 & - & 17.56$\pm$0.05 & 12.30$\pm$0.05 & 11.53$\pm$0.06 & 11.23$\pm$0.08 \\
008475 & 34$\pm$06 & 7.8$\pm$1.5 & 19.17$\pm$0.14 & - & 10.29$\pm$0.03 & 9.63$\pm$0.03 & 9.33$\pm$0.04 \\
008488 & 16$\pm$18 & 2.3$\pm$2.7 & 20.42$\pm$0.30 & - & 11.12$\pm$0.02 & 10.56$\pm$0.03 & 10.24$\pm$0.04 \\
008559 & - & - & - & - & 10.88$\pm$0.02 & 10.26$\pm$0.03 & 10.05$\pm$0.03 \\
008766 & - & - & - & - & 10.75$\pm$0.02 & 10.00$\pm$0.02 & 9.67$\pm$0.03 \\
008902 & - & - & - & - & 11.09$\pm$0.02 & 10.23$\pm$0.02 & 9.95$\pm$0.03 \\
230914 & 29$\pm$06 & 2.3$\pm$0.5 & 20.99$\pm$0.32 & - & 12.01$\pm$0.03 & 11.26$\pm$0.04 & 10.97$\pm$0.04 \\
008943 & - & - & - & - & 10.30$\pm$0.02 & 9.63$\pm$0.03 & 9.37$\pm$0.03 \\
009031 & 32$\pm$08 & 2.3$\pm$0.6 & 22.34$\pm$0.58 & - & 11.96$\pm$0.03 & 11.13$\pm$0.04 & 10.79$\pm$0.04 \\
248977 & 09$\pm$02 & 0.2$\pm$0.05 & - & 20.69$\pm$0.32 & 13.02$\pm$0.05 & 12.65$\pm$0.10 & 12.18$\pm$0.09 \\
009624 & 26$\pm$04 & 3.2$\pm$0.5 & - & 18.28$\pm$0.06 & 11.46$\pm$0.03 & 10.73$\pm$0.03 & 10.56$\pm$0.05 \\
009788 & - & - & - & - & - & - & - \\
009794 & - & - & - & - & 10.68$\pm$0.02 & 9.80$\pm$0.03 & 9.48$\pm$0.03 \\
009838 & - & - & - & 19.21$\pm$0.12 & 11.82$\pm$0.03 & 11.23$\pm$0.04 & 10.90$\pm$0.05 \\
260110 & 27$\pm$05 & 1.7$\pm$0.3 & - & - & 12.11$\pm$0.02 & 11.42$\pm$0.03 & 11.10$\pm$0.03 \\
010272 & 23$\pm$03 & 5.1$\pm$0.6 & 21.28$\pm$0.31 & 19.05$\pm$0.09 & 10.72$\pm$0.02 & 9.98$\pm$0.02 & 9.74$\pm$0.03 \\ \hline
\end{tabular}

Notes:\\
(1) Magnitude errors smaller than 0.01 are reported here as 0.00. \\
(2) 201697 and 009788 have no 2MASS data. \\
(3) 2MASS data for ACG 248977 are those listed for 2MASX J14411233+1453242. \\
(4) Only the candidate high-mass galaxies have H$\alpha$ measurements. The fluxes are given in units of erg cm$^{-2}$ s$^{-1}$.
\end{table*}

%--------------------------------------------------------------------------
%\subsection{Neighbourhood of the galaxies}
%\label{txt:surroundings}

%In this section we discuss the large-scale neighborhood of each galaxy. In the next section we shall discuss the more immediate neighborhood of each object in the context of discussing the findings from optical data bases.

\section{Results}
\label{txt:results}

%We present in Table 3 various derived properties for the sample galaxies.

%Dynamical mass from HI w(50)

%Photometric mass from M$_r$ and luminosity

%Table 3: object, L, D (major aXis), dynamic mass, stellar mass

%\begin{table*}
%\caption{Derived properties for the sample galaxies}
%\label{t:properties}
%\begin{tabular}{cccccccccc} \hline
%Name    &	L 	& $a$   & $i$	& M$_{dyn}$	& M$_*$	& M$_{HI}$ & J & H & K \\
% & L$_{\odot}$ & kpc    & $^{\circ}$ & M$_{\odot}$ & M$_{\odot}$ & M$_{\odot}$ & \\ \hline
%0826+07	&       & 29	&14.48$\pm$0.00	&14.05$\pm$0.00&13.66$\pm$00 \\
%192281	&	& 32 &12.86$\pm$0.00&12.34$\pm$0.00&11.98$\pm$0.00\\
%205190 & & 39 \\
%1046+12 & & 53 \\
%006042 & & 50 \\
%201697 & & 36 \\
%006066 & & 81 \\
%222042 & & 56 \\
%226077 & & 23 \\
%226111 & & 44 \\
%233609 & & 15 \\
%008375 & & 51 \\
%008379 & & 65 \\
%008475 & & 76 \\
%008488 & & 55 \\
%008559 & & 67 \\
%008766 & & 65 \\
%008902 & & 36 \\
%230914 & & 65 \\
%008943 & & 35 \\
%009031 & & 54 \\
%248977 & & 6 \\
%009624 & & 42 \\
%009788 & & 36 \\
%009794 & & 78 \\
%009838 & & 51 \\
%260110 & & 28 \\
%010272 & & 33 \\
%\hline
%\end{tabular}
%\end{table*}

%{\bf Do these galaxies harbor AGNs? check SDSS line profiles

% Bulge; is it massive?}

The basic question stated above is whether the wide HI profiles detected by the ALFALFA survey are produced by single, non-interacting galaxies or by two or more galaxies that are interacting and are too close together to be resolved by the rather wide ALFALFA beam. This has been shown to be the case for the galaxy with the widest HI profile reported by Bothun et al. (1982).

The relevant information is contained in the ALFALFA HI profile.  We expect this profile to have a clean two-horned shape for non-interacting disky galaxies. Courtois et al. (2009) explained this in their description of some ``good, bad, and ugly'' profiles in the ADHIC: PGC 71392, in their Figure 6, shows a clean profile with $w(50)$=989 km sec$^{-1}$ and is a high inclination S0/Sa (see also Giovanelli et al. 1986); PGC 10314 has $w(50)$=33 km sec$^{-1}$ and appears to be face-on, and the two nearby galaxies PGC 68870 and PGC 68878 show contamination of the HI profile of one galaxy by that of the other.

\begin{figure*}
{\centering
 \includegraphics[clip=,angle=0,width=11.3cm]{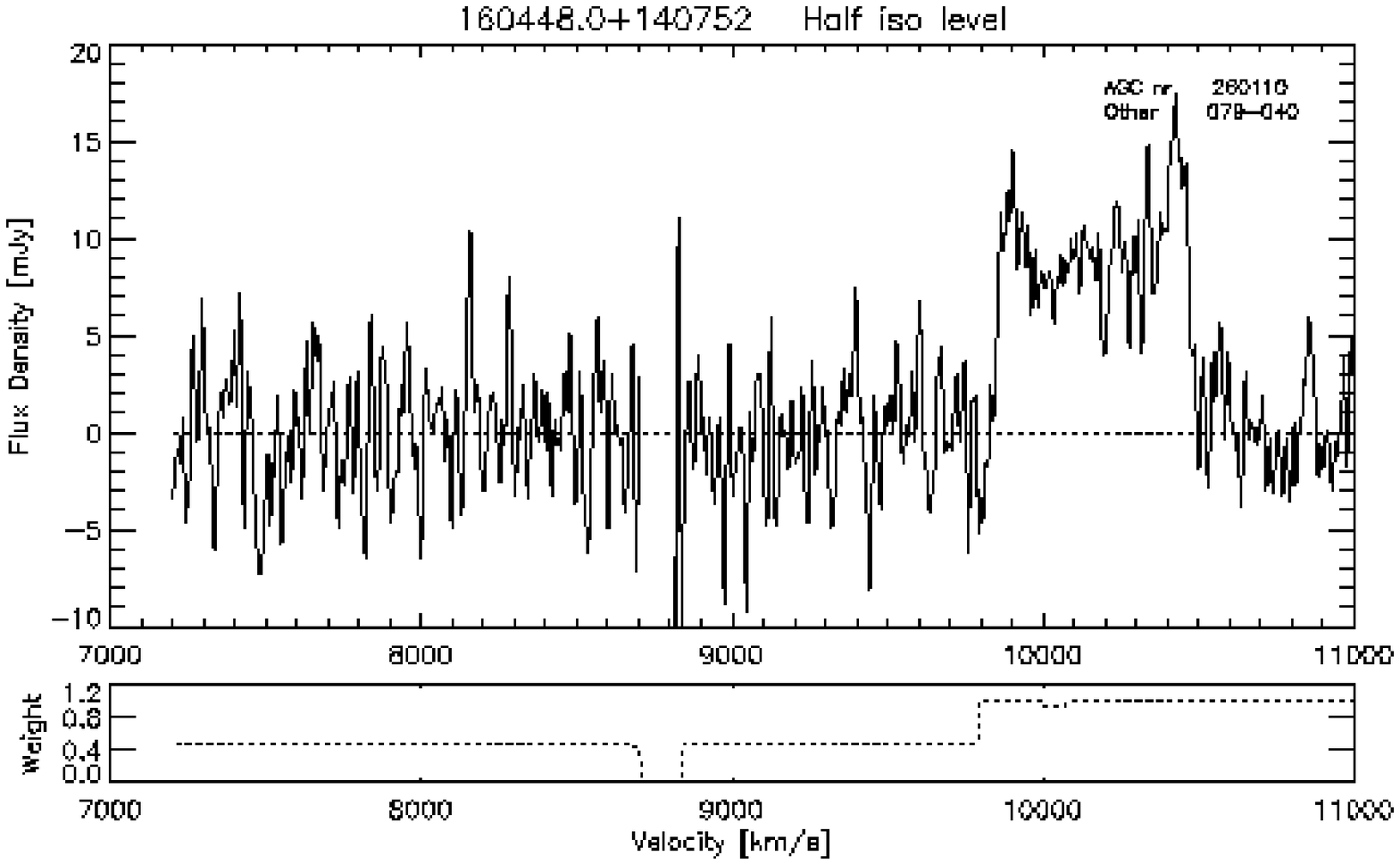}
 \includegraphics[clip=,angle=0,width=11.3cm]{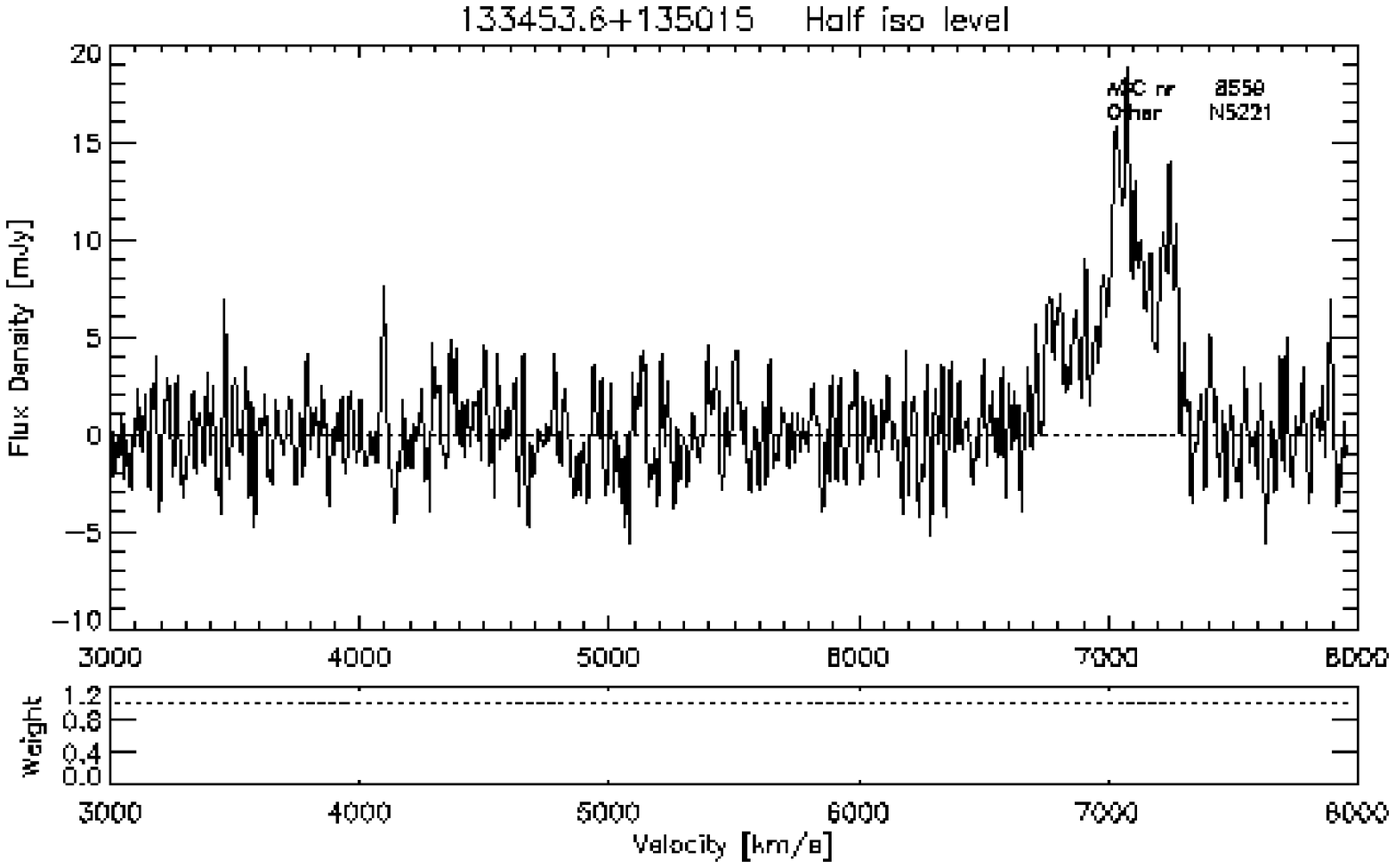}
 \includegraphics[clip=,angle=0,width=11.3cm]{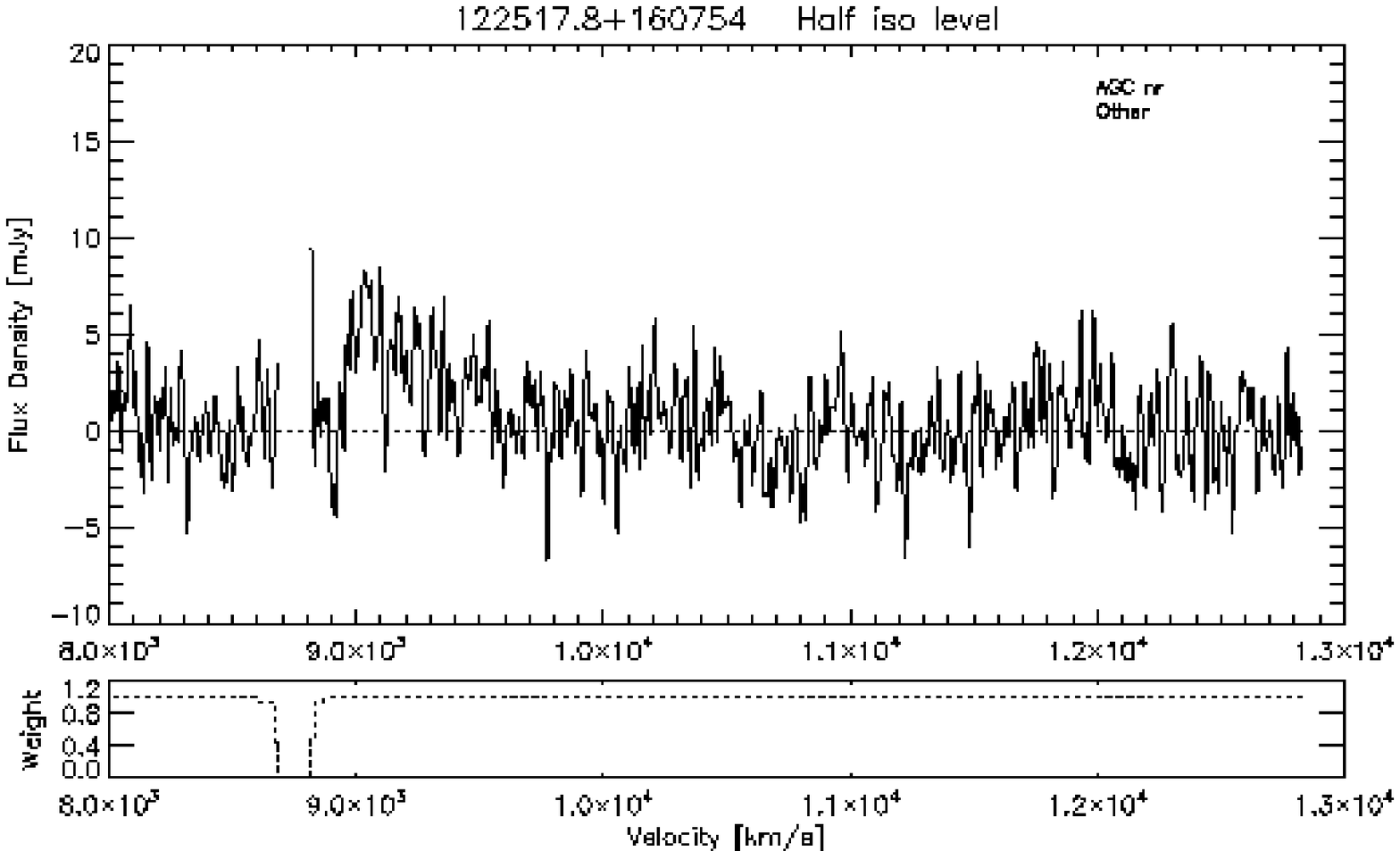}
}
 \caption{%
{\it Top panel:} A ``good'' HI profile; ACG 260110, a fairly isolated galaxy.
{\it Middle panel:} A ``bad'' profile; ACG 008559, an interacting galaxy with a strong warp or tilt of the outer disk.
{\it Bottom panel:} An ``ugly'' profile; ACG 226077, a galaxy whose HI profile is affected not only by the nearby neighbour that is included in the ALFALFA beam, but also by radio interference for velocities below $\sim$9000 km sec$^{-1}$.
}
    \label{Fig:Good-bad-ugly}
\end{figure*}

We checked the ALFALFA profiles of all galaxies in Table~\ref{t:sample} and %confirmed that this indeed is the case. We
show in Figure~\ref{Fig:Good-bad-ugly} examples of the three kinds of profiles from among the objects in our sample.

%%%%%%%%%%%%%%%%%%%%%%%%%%%%%%%%%%%%%%%%%%%%%%%%%%%%%%%%%%%%%%%%%%%%%%%%%%%
%%%%%%%%%%%%%%%%%%%%%%%%%%%%%%%%%%%%%%%%%%%%%%%%%%%%%%%%%%%%%%%%%%%%y%%%%%%%

\subsection{High-mass candidate galaxies and their neighbourhoods}

We found here that, in general, the 28 sample galaxies have large physical sizes and are very luminous. The largest object is 82 kpc wide while the smallest is only $\sim$12 kpc. In many cases the target galaxy had nearby objects that would have been included in the ALFALFA beam. In cases where redshifts were available, some objects were shown to be physical companions of the target galaxy. We also found that some companions showed emission-line spectra and blue continua, indicating fairly recent star formation possibly triggered by a past interaction. Other companions had spectra of early-type galaxies but showed also Balmer absorption lines; such cases could possibly be similar to E+A galaxies (Dressler \& Gunn 1983) indicating the presence of a relatively large population of A and B stars along with an old population dominated by G, K, and M spectral types, representing possibly a star formation episode about 1 Gyr ago, with the stellar bulk being much older.

Only 14 galaxies among the 28 selected from the ALFALFA data set appear to be fairly isolated, since they do not seem to have SDSS companions closer than $\sim$10 arcmin that could be confused in the derivation of the HI line width; these are candidates for being high-mass objects and are listed in Table 4. Here and in Appendix A we give additional details about each object and about their neighbourhood properties. In particular, we add in column 10 of Table 4 the number of identified neighbours in a 3h$^{-1}$  Mpc and 300 km sec$^{-1}$ volume around the galaxy; these could well be physical companions that may have interacted with the galaxy in the not-too-distant past. The data for the neighbourhood search are primarily from NED, supplemented by information from ALFALFA.

Table 4 lists the luminosity of the candidate galaxies in the $r$-band using M$_{\odot}$(r)=+4.52 mag
(from www.ucolick.org/$\sim$cnaw/sun.html), %and the luminosity-distance from NED,
the size of
their major axes in kpc using the angular size and the luminosity-distance from NED, the inclination $i$ calculated from the SDSS isophotal axes a and b
and cos(i)=b/a, the maximal rotation velocity corrected for inclination from v$_{max} = \frac {w(50)}{2 \times sin(i)}$, where $w(50)$ is the HI profile width listed in Table 1, the dynamical mass from
M$_{dyn} = \frac {v^2_{max} R}{G}$,
where R is the maximal extent of the visible galaxy with the power-of-ten of the value
shown in parentheses, M$_*$ an indicative stellar mass derived from the $r$-band luminosity by assuming all stars are K5V (this estimate will be refined below), and M$_{HI}$, the mass in atomic hydrogen from the ALFALFA flux integral and the luminosity distance listed in NED, using M$_{\rm HI}$=2.36$\times 10^5$ D$^2_{\rm Mpc} \times$FI,
where D$_{\rm Mpc}$ is the distance to the object in Mpc. Note that all estimates were made disregarding Milky Way and internal extinction, and also not correcting M$_{\rm HI}$ for the helium content to arrive at an estimate of the diffuse baryon masses. Also, Bernardi et al. (2010) list the $r$-band absolute magnitude of the Sun as +4.67; this would imply a 15\% brighter luminosity for all the objects. Column 9 of the table shows
F=$\frac{M_{dyn}}{M_{\rm gas}+M_{\rm stars}}$, the ratio of the dynamical mass to the total baryonic mass (stars and gas, now correcting for the helium content with M$_{gas}$=1.3$\times$M$_{\rm HI }$). Figures~\ref{f:candidates0},~\ref{f:candidates1},~\ref{f:candidates2},
and~\ref{f:candidates3} in Appendix B show the HI profiles of these CHMD.
 The table also lists the error in the derived parameters; in some cases e.g. ACG 009624, the error in the derived dynamical mass is similar to the value itself. In most cases, the rather large errors are dominated by the error in the inclination.

We stress that the calculated value for M$_{dyn}$ is a lower limit, since it assumes that the HI distribution extends only as far as the 25 mag arcsec$^{-1}$ isophote of the optical image; in most disk galaxies this is not the case and the HI extends further out of the luminous disk by a factor of a few. On the other hand, this calculation also assumes that the gas dynamics are dominated by circular rotation. M$_*$ is also a lower limit, since we did not correct the photometry for Milky Way and internal extinction.

%%%%%%%%%%%%%%%%%%%%%%%%%%%%%%%%%%%%%%%%%%%%%%%%%%%%%%%%%%%%%%%%%%%%%%%%%%%

\section{Discussion}
\label{txt:discussion}

\subsection{Simplified star formation history for CHMDs}

The availability of the integrated SDSS colours allows a comparison of the 14 candidate high-mass galaxies with the relations found by Wu et al (2007) between the SDSS colour indices and the absolute magnitudes of galaxies. This indicates that, in most cases, the integrated colours can be produced by evolved stellar populations (SPs). In disk galaxies such colours correspond to high-luminosity systems, as their Figure 3 shows. The CHMDs are thus definitely on the ``red sequence'' of galaxies.

The present star formation rate (SFR) can be derived from the H$\alpha$ luminosity, following e.g., Pflamm-Altenburg, Weidner, \& Kroupa (2007). They give a formula to convert the H$\alpha$ luminosity to an SFR (in M$_{\odot}$ yr$^{-1}$) valid for L(H$\alpha)\geq2.8\times10^{36}$ erg s$^{-1}$:
\begin{equation}
SFR=L(H\alpha)/X
\end{equation}
where $X$=1.89$\times10^{41}$ erg s$^{-1}$ for a canonical (broken) IMF, or 3.3$\times10^{41}$ erg s$^{-1}$ for a Salpeter IMF.

The resultant SFRs for the objects with H$\alpha$ line emission, listed in the last column of Table 4 for a canonical IMF, are rather modest with the median SFR=2 M$_{\odot}$ yr$^{-1}$ and the highest value at 4.3 M$_{\odot}$ yr$^{-1}$ (for 009624), or $\sim$40\% lower for a Salpeter IMF. We are, therefore, not witnessing very strong SF events but rather mild ones, implying a possibility to maintain the SF process at its present level, given the detected total HI content, over many Gyr by having many and frequent such SF events. In fact, among all galaxies we find the shortest HI consumption  time (for 009624) to be $\sim$4 Gyr.

Given this result, we decided to test first a possibility that the CHMDs have been forming stars at a constant rate since their formation. Our data set, in particular the availability of SDSS and 2MASS colours, allows a rough estimation of the global star formation history (SFH) in the candidate high-mass disk galaxies. This procedure aims to best-fit the measured global colours of a CHMD with colours predicted by accepted population synthesis programs and has been explained in Zitrin \& Brosch (2008) and in Zitrin, Brosch  \& Bilenko (2009). The baseline synthesized colours originate from the Bruzual \& Charlot (2003) GALAXEV library. To allow for slight variations in the simplest scenario, we modeled a composite of two constant SFR models where we allowed each SF process to start at a different time and to produce different fractions of the light detected at present from each galaxy. Note that a single stellar population (SP) formed with a constant SFR would result automatically from such a procedure as one of the two SPs, where the second population would not contribute to the present-day light. The criterion by which the proper population mix was chosen was the lowest reduced-$\chi^2$ of the fit. We can reject this constant SFR scenario, based on the resultant $\chi^2$ values, when comparing with the $\chi^2$ values of the other tested scenarios.

Since we could not reproduce the colours with a constant SFR, even when involving two SPs formed at different times with constant SFRs, we tested two other possible SF scenarios; scenario A combining a continuous star formation process with a single $\delta$-function SF burst while allowing any metallicity value for each of the two SPs, and scenario B assuming two ``instantaneous'' $\delta$-function bursts occuring at different times, but fixing the metallicity value to either solar or 2.5 times higher. The parameters that change are the start times of the two processes and their relative proportions in the present-day light collected from the CHMD. %We also tried two metallicity recipes: one with solar metallicity and the other .

%{\bf Adi results: one very old SB and one a few 100 Myr old}

We found that scenario A produced exclusively ``old'' galaxies, where only $\sim$1/3 of the objects showed a relatively young population that was nevertheless $\sim$3 Gyr old and contributed up to 34\% of the present light. Scenario B, on the other hand, allowed the presence of a reasonable fraction of young to very young stars in most objects, less than 10 Myr old, producing $\sim$10-30\% of the observed light and yielded $\sim3\times$ lower reduced $\chi^2$ values than scenario A. %These results are valid for most of the 12 objects where we could form seven colour indices, including (H$\alpha$-V), and also for the two objects lacking H$\alpha$ lin emission, where only six colour indices are available. {\bf TBD, after Adi runs new models, maybe a combination of exponential decay with $\tau\simeq$1-3 Gyr and a delta burst}

Based on the resultant minimal $\chi^2$ we can, therefore, reject  the scenario of a SP formed with a constant SFR combined with an SP formed by a recent SF burst, We thus adopt a scenario where most of the stars in a CHMD were formed a long time ago, perhaps close to a Hubble time ago,  during a short period but a small fraction of their SP formed in a recent SF event. This fraction of young stars now produces 10-30\% of the light collected from the galaxy.

It is tempting at this point to speculate what could be the possible very recent SF trigger, a few tens of Myr at most, in a galaxy that seems to have been simply aging with time. One possibility is that the SF process is fed by stellar ejecta and not by externally accreted material. This recycling of material through stars can be done by gentle mass loss via stellar winds and planetary nebulae formation, with $\sim$40-50\% of the stellar mass returned to the ISM and only a small fraction of ISM ($\sim$10\% of the total) coming from supernovae (e.g., Pozzetti et al. 2007, Matig \& Bournaud 2010).

Another option could be a minor merger with a gas-rich companion that rejuvenated the old disk, since our optical survey and the shapes of the HI profiles do not seem to support major mergers. We point out that signs of a past but recent interaction could possibly be detected with 2D kinematic mapping, either optically or by mapping the HI with a synthesis telescope. It is also possible that some of the other 14 galaxies not identified as CHMDs could be transition objects, following the recent accretion of a gas-rich object, indicating that with this sample we are seeing ``real-time'' evolution of massive disks from the red to the blue sequence, or at least into the green valley.

%{\bf NB still to do: in this case the SF should be more central-check the H$\alpha$ images vs the blue ones}
These two options can be checked against the location of the star-forming regions. If the SF trigger is accumulation of stellar ejecta, then the SF should be more concentrated that the general galaxy light. If the trigger is external gas accumulation by IGM accretion (e.g., Salim \& Rich 2010) then the young stars are expected form outside (most of) the disk and not show a particular central condensation.

We tested this by blinking the net-H$\alpha$ images against the R-band images and indicate the results in the last column of Table~\ref{t:massive}. Code C implies that the line emission is centrally concentrated while code D indicates that it is diffuse. The results show that in most cases the H$\alpha$ emission is centrally concentrated, supporting the possibility that the source of material for the formation of the young stellar population is recycling of internal ISM produced by stellar ejecta. We cannot rule out influx of fresh material, at least for the objects with a mixed classification (C/D). However, it is significant that in no case have we observed a galaxy where the HII regions were exclusively relegated to the periphery of the object.

\subsection{Stellar mass estimate}
We estimated above the stellar mass by assuming that all stars were K5V; this is certainly not the case, since the stellar population models showed the best fits to be a composite of a population almost a Hubble time ago with a small fraction formed at most a few tens of Myr ago. The young population contributes 10-30\% of  the visible light. We can use this result to refine the estimate of the stellar mass of each galaxy. We shall first explain the method, then apply it to each CHMG.

The mass estimate uses a value for the M/L to convert from measured luminosity to mass; above we used the value for a K5V star. However, the M/L values for different populations e.g., by using the top panel from Figure 5 of Anders et al. (2009), indicate that M/L$_V\simeq2\times10^{-2}$ M$_{\odot}$L$^{-1}_{\odot}$ for a stellar population 10 Myr old, while it is $\sim$5 M$_{\odot}$L$^{-1}_{\odot}$ for a population $\sim10^{10}$ yr old.

If the stellar composition of a galaxy of total luminosity $L$ is produced by only two stellar populations, say population 1 (young) and population 2 (old), which have (M/L)$_1$ and (M/L)$_2$ respectively, and if each population produces a fraction X or (1-X) of the total luminosity recorded for the galaxy, then the mass of first population is:
\begin{equation}
M_1=X \times L \times (M/L)_1
\end{equation}
whereas for the second population
\begin{equation}
M_2=(1-X) \times L \times(M/L)_2
\end{equation}
The mass ratio of the two stellar populations, for the CHMGs studied here, is
\begin{equation}
M_1/M_2=\frac{X}{1-X} \times \frac{(M/L)_1}{(M/L)_2} \simeq\frac{0.2}{0.8} \times \frac{0.02}{5}
\end{equation}
The young population contributes $\sim$0.1\% to the total mass, thus assuming that all the mass is in the old stars is certainly not a bad approximation. Since the total stellar mass in the CHMGs is of order 9$\times10^{10}$ M$_{\odot}$, the mass of the newly formed stellar population should be $\sim10^8$ M$_{\odot}$ and the time to form it, at a typical rate of 2 M$_{\odot}$ yr$^{-1}$ would be some 50 Myr. This rough estimate justifies our conclusion that in these CHMGs we see a recent SF event in an old galaxy.

\subsection{CHMDs and the Tully-Fisher relation}
Catinella et al. (2010) studied a sub-sample of an unbiased collection of galaxies with stellar masses M$_*\geq$10$^{\rm 10}$ M$_{\odot}$ at nearby redshifts (0.025$\leq$z$\leq$0.05). The objects were primarily selected from the GALEX and SDSS data sets; those analyzed  in the paper, some 200 galaxies, have HI measurements from Arecibo. The preliminary results of Catinella et al. show a strong decrease in the M$_{HI}$/M$_*$ ratio with stellar mass and with the stellar surface density in M$_{\odot}$ kpc$^{-2}$, $\mu_*$=M$_*$/(2$\pi R^2_{50,z})$, with $R_{50,z}$ being the radius %in arcsec
 containing 50\% of the Petrosian flux in the z-band. Specifically, M$_{HI}$/M$_*$ decreases by about one order of magnitude for 10$\leq$log M$_*\leq$11.3 and by two orders of magnitude for 8.0$\leq$log $\mu_* \leq$9.5 (see their Figure 10).
The CHMD behaviour shows a similar trend of the HI-to-stellar mass ratio to decrease with the stellar mass increase, except for the aberrant object 248977 discussed below.

We found that in all cases M$_{dyn}$ was 2.5--7.5$\times$ larger than the sum of M$_*$ and
M$_{\rm HI}$ corrected for the helium content albeit with rather large errors (see Table 4).%; correcting for helium would not change much the situation, since even for the most gas-rich object, 006066, the atomic gas makes up only 1/3 of the baryonic mass, which by itself is only a small fraction of the total mass of a CHMD.
 This seems to indicate that all galaxies studied here have significant but not excessive amounts of dark matter. The stellar mass to gas mass ratios found here are similar to those of field Sa and Sab galaxies (Read \& Trentham 2005).

%Simple interpretation of disk rotational properties under the assumption of Newtonian gravity implies that M $\propto$ Rv$^2$, thus log(M) $\propto \, 2 \times $log(v)+log(R). A MOND-like gravity law, where the gravitational attraction depends linearly on distance, would predict log(M) $\propto 2 \times$log(v). The relation found here, log(M) $\propto$ log(v), fits neither of the relations.

Another interesting result concerns the immediate neighbourhood of the high-mass candidate galaxies. One would expect massive galaxies to reside in high galaxy density regions yet, as column 10 of Table 4 shows, these neighbourhoods are relatively sparse. In fact, some of the more massive candidates (e.g., 006066 or 009624) have very few possible neighbours within 3h$^{-1}$ Mpc and 300 km sec$^{-1}$.

The shape of HI profiles themselves is an accepted criterion to indicate possible signs of interaction (e.g., Haynes et al. 1998). Although the ALFALFA profiles shown in Figures~\ref{f:candidates0},~\ref{f:candidates1},~\ref{f:candidates2},
and~\ref{f:candidates3} of Appendix B are not as clean and ``high-quality'' as those shown by Haynes et al., we can state with reasonable confidence that the profiles appear to be symmetric in the areal symmetry criterion. We thus identified a small population of high-mass disk galaxies that are fairly isolated and do not appear to have had
interactions in the last few Gyr.%; this is a strong constraint on the formation mechanisms of massive galaxies.

\begin{figure}
\label{f:log(vmax)_vs_Log(DynMss)}
\includegraphics[clip=,angle=-90,width=14cm]{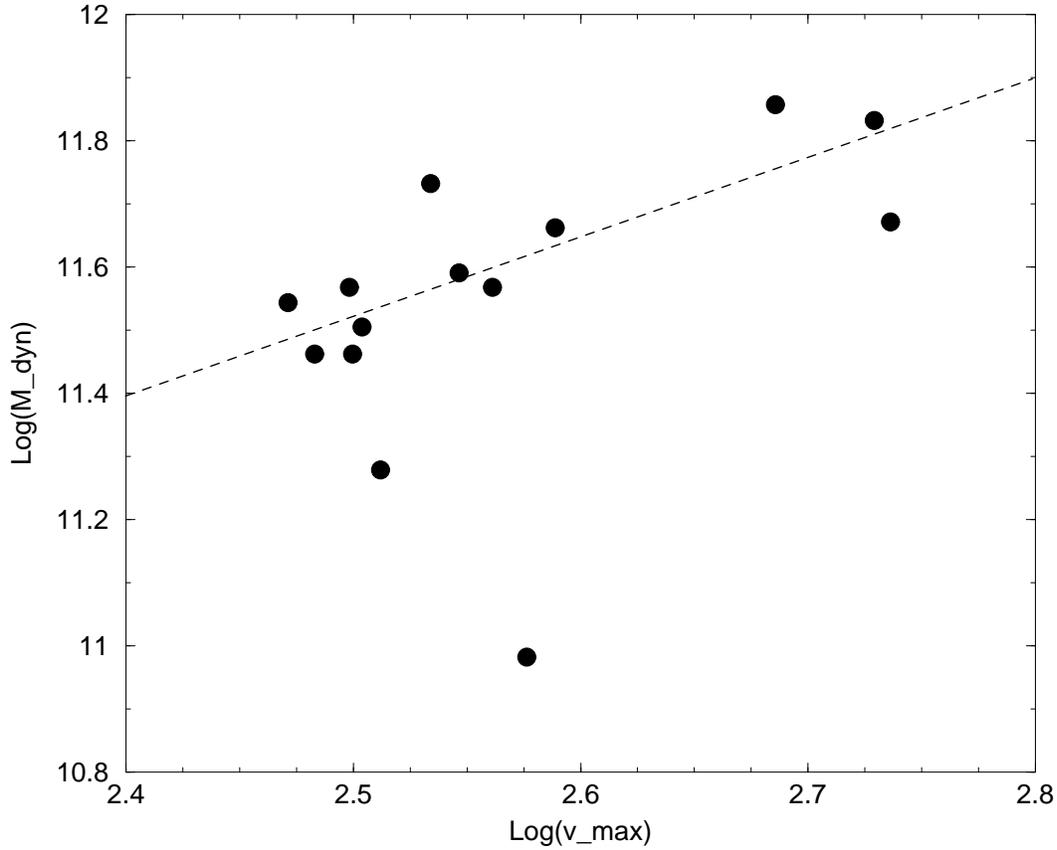}
\caption{Logarithmic plot of the maximal rotational velocity vs. the total dynamical mass for the CHMDs. The general trend is for a relatively flat relation, i.e. the maximal rotational velocity being $\sim$independent of galaxy dynamical mass, with one object (ACG 248977) deviating considerably. The error bars are not plotted, for clarity, but the errors are listed in Table 4. The relation derived in equation 5 is plotted as a dashed line.}
\label{f:HiMass_plot}
\end{figure}

The sub-sample of candidate high-mass objects that do not appear to show obvious signs of interaction offers a possibility to check the mass vs. rotational velocity at the high-v$_{max}$ end. %Their HI profiles are shown in Appendix B, Figures~\ref{f:candidates0},~\ref{f:candidates1},~\ref{f:candidates2}, and~\ref{f:candidates3}.
 The mass vs. rotational velocity (a.k.a., the Tully-Fisher=TF) relation, was shown e.g., in Figure 4 of Giovanelli et al. (1986) and is shown here for the CHMDs in Figure~\ref{f:HiMass_plot}.

ACG 248977, a small galaxy extending less than 10 kpc as shown in the SDSS images, yet with
a very wide HI profile (Figure~\ref{f:candidates2}, 3$^{\rm rd}$ row-right), is one of the objects that deviates significantly from the Giovanelli et al. (1986) relation and from the relation suggested here for the high v$_{max}$ galaxies, rotating $\sim3\times$ faster than what its $r$-band luminosity would allow, or by being fainter by one order of magnitude than its v$_{max}$ would predict. It is possible that we are not observing pure rotation in this object and that HI synthesis observations, or a single-dish observation with significantly higher S/N will indicate chaotic motion of HI clouds. Note also that the apparent size of HI14110.4+145339, as inspected by us, seems to be about three times larger overall than the listed SDSS major axis (Appendix A). If this is true, then its M$_{dyn}$ would be larger by the same factor alleviating the discrepancy. On the other hand, we also note that the HI profile is affected by RFI, as the plotted baseline shows. In adopting the listed w(50) we have, therefore, accepted essentially a lower limit to the profile width implying a lower limit to M$_{dyn}$.%; this would have the opposite effect to the location of the object in Figure~\ref{f:HiMass_plot}.}

Eliminating ACG 248977, and fitting a linear regression to the log(M$_{dyn}$) vs. log(v$_{max}$) distribution, yields a correlated set (correlation coeffcient 0.73) with
\begin{equation}
log(M_{dyn})=(8.4\pm0.9)+(1.25\pm0.35)\times log(v_{max})
\end{equation}
We plotted this relation with a dashed line in Figure~\ref{f:HiMass_plot}.

%We found many cases where the target galaxy had nearby objects (within 100 kpc {\bf TBD}) that would have been included in the ALFALFA beam. In cases when redshifts were available, some objects were shown to be physical companions of the target galaxy. We also found that in some cases, the companions showed emission-line spectra and blue continua. Other companions had spectra of early-type galaxies but showed also Balmer absorption lines; these cases may possibly be similar to E+A galaxies (Dressler \& Gunn 1983) and could indicate the presence of a relatively large population of A and B stars together with an old population dominated by G, K, and M spectral types, representing possibly a star formation episode about 1 Gyr or less ago, while the stellar bulk is much older.

Many of the 14 galaxies discussed here appear to fit well the upper-right part of the
Giovanelli et al. (1986) plot but the three objects with the highest v$_{max}$ values lie below the extrapolation of the Giovanelli et al. relation. This was also the case for UGC 12591 (see Giovanelli et al. 1986). We note in this context that the study of the high-mass end of the TF relation (Noordermeer \& Verheijen 2007) found a similar result,
that galaxies with v$_{max} \geq$200 km sec$^{-1}$ were rotating faster than expected (or were less luminous). However, Meyer et al. (2008) found slopes of $\sim$4 for lower mass galaxies (v$_{max}\leq$300 km sec$^{-1}$) measured by HIPASS, as also found by Trachternach et al. (2009) for 11 very low-mass dwarf galaxies for which they have HI synthesis maps. Our result, a slope of approximately unity, together with the results of Giovanelli et al. and of Noordermeer \& Verheijen, indicate a fundamental difference between the high-mass and lower mass galaxies in the definition of the TF relation at the high-mass end of the disks.

\begin{table*}
\caption{Candidate high-mass galaxies}
\label{t:massive}
\begin{tabular}{ccccclcccccc} \hline
ACG &	L$_r$ 	  & $a$ & $i$        & v$_{max}$     & M$_{dyn}$   & M$_*$	     & M$_{HI}$    & F & N & SFR & HII\\
     & L$_{\odot}$& kpc & $^{\circ}$ & km s$^{-1}$ & M$_{\odot}$ & M$_{\odot}$ & M$_{\odot}$ &   & 3h$^{-1}$Mpc & M$_{\odot}$yr$^{-1}$ & \\ \hline
205190 & 3.8(10) & 25 & 46 & 485 & 7.2(11) & 8.1(10) & 1.2(10) & 7.5 & 24 & 1.1 & C \\
        &       &       & 14    & 115   & 2.4 &     & 0.06    & 2.5   &   & 0.2 \\ \hline
200589 & 4.5(10) & 26 & 61 & 316 & 2.9(11) & 9.6(10) & 1.0(10) & 2.7 & 14 & 2.0 & C \\
        &       &       & 10    & 29    & 0.4 &     & 0.06    & 0.4   &    & 0.4  \\ \hline
006066 & 4.3(10) & 41 & 77 & 342 & 5.4(11) & 9.1(10) & 4.4(10) & 3.6 & 3 & 1.8 & C \\
        &       &       & 07    & 11    & 0.4 &     & 0.09    & 0.3   &     & 0.6 \\ \hline
222042 & 4.1(10) & 28 & 66 & 304 & 2.9(11) & 8.7(10) & 1.6(10) & 2.7 & 18 & 2.1 & C \\
        &       &       & 09    & 22    & 0.3 &     & 0.10      & 0.3   &   & 0.3 \\ \hline
008375 & 3.1(10) & 25 & 73 & 364 & 3.7(11) & 6.6(10) & 5.9(09) & 7.3 & 6 & 0.7 & C \\
        &       &       & 07    & 37    & 0.6 &     & 0.35       & 1.3   &   & 0.3 \\ \hline
008379 & 4.4(10) & 33 & 74 & 296 & 3.5(11) & 9.4(10) & 2.1(10) & 2.9 & 7 & 2.7 & C \\
        &       &       & 07    & 46    & 0.7 &     & 0.08       & 0.6   &   & 0.8 \\ \hline
008475 & 5.1(10) & 28 & 57 & 352 & 3.9(11) & 1.1(11) & 3.7(10) & 2.5 & 60 & 4.0 & D \\
        &       &       & 12    & 46    & 0.7 &     & 0.04      & 0.4   &   & 0.8 \\ \hline
008488 & 3.7(10) & 28 & 69 & 319 & 3.2(11) & 7.9(10) & 1.1(10) & 4.0 & 48 & 1.4 & C \\
        &       &       & 09    & 25    & 0.4 &     & 0.03      & 0.6   &   & 1.6 \\ \hline
230914 & 5.2(10) & 33 & 67 & 315 & 3.7(11) & 1.1(11) & 2.7(10) & 2.5 & 14 & 3.6 & C \\
        &       &       & 09    & 11    & 0.4 &     & 0.08      & 0.3   &   & 0.8 \\ \hline
009031 & 4.4(10) & 27 & 66 & 388 & 4.6(11) & 9.4(10) & 1.9(10) & 3.9 & 8 & 3.6 & C/D \\
        &       &       & 09    & 30    & 0.6 &     & 0.11      & 0.6   &   & 0.9 \\ \hline
248977 & 1.0(10) &  6 & 48 & 377 & 9.6(10) & 2.1(10) & 9.0(09) & 2.9 & 5 & 0.2 & C \\
        &       &       & 14    & 85    & 3.1 &     & 0.54      & 0.9   &   & 0.05 \\ \hline
009624 & 7.0(10) & 21 & 34 & 536 & 6.8(11) & 1.5(11) & 1.7(10) & 4.0 & 5 & 4.3 & C \\
        &       &       & 20    & 270   & 4.9 &     & 0.07      & 2.9   &   & 0.7 \\ \hline
260110 & 2.9(10) & 14 & 34 & 545 & 4.7(11) & 6.2(11) & 2.9(10) & 4.7 & 19 & 1.9 & C/D \\
        &       &       & 20    & 273   & 3.3 &     & 0.09      & 3.3   &   & 0.3 \\ \hline
010272 & 2.5(10) & 16 & 63 & 325 & 1.9(11) & 5.3(10) & 2.2(09) & 3.4 & 14 & 1.5 & C \\
        &       &       & 10    & 43    & 0.4 &     & 0.13      & 0.7   &   & 0.2 \\
\hline
\end{tabular}
\\ Notes: \\
(a) Powers-of-ten are given in parentheses. \\
%(b) The SFR refers to those objects that had measured H$\alpha$ emission.\\
(b) The last column, HII, gives the classification of the HII region distribution: C=concentrated and D=diffuse. \\
(c) The second line for each entry gives the error in the parameter listed immediately above it, in the same units and with the same power-of-ten as in the first line.
\end{table*}

%We single out 248977, a very small galaxy covering only a few kpc in the optical image, yet with a very high spin or exceedingly low brightness. This is one of the objects that also deviates significantly from the Giovanelli et al. (1986) relation and from the relation suggested here for the high v$_{max}$ galaxies, rotating faster by $\sim3\times$ than what its $r$-band luminosity would allow, or being fainter by one order of magnitude than its v$_{max}$ would predict. Eliminating 248977 fo which we have no reasonable explanation, and fitting a linear regression line for M$_{dyn}$ vs. v$_{max}$, indicates a highly correlated set with a correlation coefficient of 0.69, with: \\ log(M$_{dyn})=9.1(\pm0.88)+1.09(\pm0.34)\times{\rm log \, (v}_{max}$).

 We stress that in Figure 2 we have not plotted the error bars; these are relatively large and their inclusion (in a log-log plot) would result in a figure that would not be understandable. However, as mentioned above, the formal errors are listed in Table 4.

%%%%%%%%%%%%%%%%%%%%%%%%%%%%%%%%%%%%%%%%%%%%%%%%%%%%%%%%%%%%%%%%%%%%%%%%%%%
\section{Summary}
\label{txt:summ}

\begin{enumerate}

\item We described a complete sample of 28 galaxies with wide HI profiles selected from the ALFALFA HI survey that could be candidate high-mass objects.

\item  Using on-line data bases, we showed that some of the objects could have been confused by nearby neighbour galaxies that could have been included in the radio telescope beam producing the wide HI profiles.

\item  We eliminated those, as well as objects showing obvious signs of interaction in SDSS images, and reduced the sample to 14 objects lacking visible signs of interaction and/or very close neighbours. We propose these as candidate high-mass disk galaxies (CHMDs).

\item  We checked the degree of isolation of the CHMDs and found that some may indeed be located in low galaxy density neighbourhoods, contrary to expectations that high-mass galaxies would be found in high galaxy density regions.

%We studied a complete sample of 28 objects selected from the ALFALFA survey to have 21-cm line profiles wider than 550 km sec$^{-1}$ using archival optical data. We found that half of the objects could have been influenced by interactions with neighbours, compromising the interpretation of wide HI profiles as indicating high mass. However, the other half of galaxies did not show any signs of external influences, thus we conclude that they have intrinsically wide neutral hydrogen profiles and call this sub-sample "candidate high mass galaxies".

%The sub-sample allows the testing of the mass-rotation velocity relation at the high v$_{max}$ end. We find, as already suggested by Giovanelli et al. (1986) that the seemingly linear trend defined by the low v$_{max}$ galaxies breaks at the high v$_{max}$ end. Specifically, we find a $\sim$flat relation of dynamical mass with v$_{max}$

\item  We investigated simple models for the star formation histories of the CHMDs and found that, in most cases, these objects show signs of young and massive star formation. Tests with different star formation scenarios indicated that most of the stars formed about a Hubble time ago, with 10-30\% of the light produced by very young stars formed at most a few 10s of Myrs ago.

\item We found that the mass of the young stellar component is a small fraction of the total stellar mass; the CHMGs are thus showing signs of rejuvenation of their old stellar populations. With the current star formation rate, the young stellar population can be produced in a few tens of Myr; the HI reservoir detected by ALFALFA suffices to maintain the star formation for about a Hubble time.

\item  We calculated dynamical masses from the HI profile width, the optical inclination, and the optical radius, and compared those with the HI and stellar masses, finding that the luminous matter makes up only $\sim$30\% of the dynamical mass.

\item  We investigated the location of these CHMDs in the Tully-Fisher relation and found that they deviate significantly from the relation defined by lower-mass galaxies.

    \end{enumerate}

\section*{Acknowledgments}
The authors would like to acknowledge the work of the entire ALFALFA collaboration team in observing, flagging, and extracting the catalog of galaxies used in this work. We are grateful for the selection of galaxies from the ALFALFA data set provided by Martha Haynes and Riccardo Giovanelli. We acknowledge some constructive remarks from an anonymous referee.
%during the PV phase to complete this project.

Funding for the Sloan Digital Sky Survey (SDSS) has been provided by the Alfred P. Sloan Foundation, the Participating Institutions, the National Aeronautics and Space Administration, the National Science Foundation, the U.S. Department of Energy, the Japanese Monbukagakusho, and the Max Planck Society. The SDSS Web site is http://www.sdss.org/.

The SDSS is managed by the Astrophysical Research Consortium (ARC) for the Participating Institutions. The Participating Institutions are the University of Chicago, Fermilab, the Institute for Advanced Study, the Japan Participation Group, the Johns Hopkins University, Los Alamos National Laboratory, the Max-Planck-Institute for Astronomy (MPIA), the Max-Planck-Institute for Astrophysics (MPA), New Mexico State University, University of Pittsburgh, Princeton University, the United States Naval Observatory, and the University of Washington.

This publication makes use of data products from the Two Micron All Sky Survey, which is a joint project of the University of Massachusetts and the Infrared Processing and Analysis Center/California Institute of Technology, funded by the National Aeronautics and Space Administration and the National Science Foundation.

This research has made use of the NASA/IPAC Extragalactic Database (NED) which is operated by the Jet Propulsion Laboratory, California Institute of Technology, under contract with the National Aeronautics and Space Administration.

% References

\section{Appendix A: Information on individual objects}
\label{txt:singles}

This Appendix presents an independent estimate of each object's morphology derived from the SDSS images, and reports results of a search for possible neighbours using primarily SDSS information. In the individual comments below we adopt SDSS isophotal radii $a$ and $b$ as listed in SDSS DR7; these are defined at the 25 mag arcsec$^{-1}$ isophote level. As second choice, when SDSS isophotal data are not available, we adopt the RC3 blue major and minor axes at the 25 mag arcsec$^{-1}$ isophote level. We also adopt for each object the NED conversion from angular size to linear size, called here ``the plate scale''. The object identifier is from Table 1, and we repeat next to the name the HI redshift from column 4 of the table.% 1.

{\bf 181756} (9323 km s$^{-1}$):
The SDSS $r$ image has isophotal radii a=24\arcsec.20 and b=10\arcsec.48 with a major axis position angle PA=59$^{\circ}$. With the NED plate scale of 1\arcmin$\simeq$36 kpc and the major axis size is 29 kpc. The SDSS image shows a featureless disk with a central bulge. A number of nearby bluish diffuse features  may, or may not, be connected with the galaxy. The morphological impression is of an S0 galaxy, since no arms or dark lanes are seen. The SDSS spectrum shows nebular emission lines with strong H$\alpha$+[NII] and [OII] on top of a typical absorption spectrum.

This object has a nearby neighbour, SDSS J082656.50+073053.0 about one arcmin and 300 km sec$^{-1}$ away, considered here to be a physical companion. This is the brightest and bluest of the nearby diffuse blobs. Its SDSS spectrum is that of a very late-type galaxy, with a strong blueward sloping continuum, strong emission lines of H$\alpha$, [NII], [OIII], [SII], and [OII]. H$\beta$ shows a narrow emission line on top of a wide absorption feature.

{\bf 192281} (10470 km s$^{-1}$):
The SDSS image shows a disk with possible isophote twisting outside the brighter central part.
With a cosmology-corrected plate scale of 42.34 kpc arcmin$^{-1}$ and isophotal radii of $a$=22\arcsec.86 and $b$=6\arcsec.95, the major axis is 32 kpc at position angle 6$^{\circ}$.5. With $g$=16.2 this is a fairly bright galaxy, but not the brightest in the neighbourhood. The SDSS spectrum exhibits strong emission lines, the strongest being H$\alpha$+[NII] and [OII], and weak absorptions. H$\beta$ shows narrow emission in a  wider absorption profile.

This object has a close neighbour, 2MASX J09590327+1302211 = SDSS J095903.29+130221.0, which at 10689 km s$^{-1}$ is 1\arcmin.2 and  220 km s$^{-1}$ away. This is an $r$=15.2 elliptical galaxy, with weak H$\alpha$ and [OII] emission and H$\beta$ and higher Balmer lines in absorption. There are other similar galaxies in the neighbourhood, within a few galactic radii,  but they lack SDSS redshifts thus it is not possible at this time to evaluate whether this is a compact galaxy group or a projection effect.

{\bf 205190} (9683 km s$^{-1}$):
This object is known also as MCG +03-28-009=2MASXi J1048021+142428 and is a NVSS radio source. The major axis of 2$a$=77.4 arcsec (using the 2MASS Ks measure) and the plate scale of 38.50 kpc arcmin$^{-1}$ imply a 50 kpc wide galaxy.
The SDSS image shows an elliptical or lenticular galaxy with a strong dark lane that has a pronounced warp at its outer half-radius. The extensions of the dark lane appear as luminous ansae on both sides of the galaxy. The galaxy appears to have undertaken a major merger, or to have accreted recently a dust-and-gas rich galaxy, with the accreted material not yet relaxed to the equatorial plane of the galaxy. Note that the bluish object visible on the SDSS image near the northern end of the dark lane is a foreground star. The object does not have an SDSS spectrum. The few diffuse objects within 10 arcmin are all background galaxies.

{\bf 200589} (10768 km s$^{-1}$):
This galaxy is also known as CGCG 066-023. Its SDSS image shows a spiral disk with a strong bulge, implying an Sa classification. Dust lanes trace possible spiral arms both near the galaxy periphery and in its inner parts.
The cosmologically-corrected plate scale is 42.43 kpc arcmin$^{-1}$. The isophotal axes are $a$=75\arcsec.6 and $b$=36\arcsec.1 with a major axis position angle of 109$^{\circ}$, implying a really large galaxy with a physical major axis of 53 kpc. The SDSS spectrum shows  weak H$\alpha$ in emission and H$\beta$ in absorption. With $r$=13.87, at this distance the object is very luminous, M$_r \simeq$--22 mag.

A possible neighbour is SDSS J104831.76+125844.9, a bluish galaxy with with $r$=18.61  and  ($g-r$)=0.47 some 20\arcsec to the SW. Since it has no redshift, it is not possible at present to decide whether it is a real neighbour or a projected one. %{\bf maybe with H$\alpha$?}

{\bf 006042} (9921 km s$^{-1}$):
This galaxy, classified as SB(s)b:, is also NGC 3466=UGC 06042=PGC 32872. With a plate scale of 39.01 kpc arcmin$^{-1}$ and a major axis of 76\arcsec.59 at a position angle of 50$^{\circ}$, this is 50 kpc wide spiral. The minor axis isophotal radius is 19\arcsec.7. The SDSS image shows a bluish spiral arm in N3466 winding from the location of the superposed companion. UGC notes a 0\arcmin.15x0\arcmin.10 companion superposed at a distance of 0.35 arcmin to PA=85$^{\circ}$,  another companion at two arcmin away at PA=83, which is a 0\arcmin.4x0\arcmin.3 galaxy, a 15.7 mag 0\arcmin.8x0\arcmin.1 Sc at 10 53.2 +10 00 at a distance of 2.9 arcmin at PA=253$^{\circ}$, and UGC 06045=NGC 3467 at a distance of 7.0 arcmin away at PA=87.

The very close superposed UGC galaxy, known also SDSS J105616.79+094516.7, is an $r$=15.81 mag lenticular with a typical early-type galaxy spectrum. With a redshift of 9960$\pm$60 km sec$^{-1}$, this is definitely not a projected companion. A dark patch, visible on the SDSS finding chart image of the companion, appears to be connected with the bluish arm in N3466, indicating that the smaller lenticular might be interacting N3466.
N3466 was previously measured in HI and is included in the 21-cm spectral line digital archive (Springob et al. 2005) with $w(50)$=728$\pm$16 km s$^{-1}$. %The detected HI line width there was narrow, only 568$\pm$12 km s$^{-1}$.

NED notes that a poor galaxy cluster, WBL 291 (White et al. 1999), whose center is 2.4 arcmin away from the object, is at essentially the same redshift, 9707 km s$^{-1}$. This
indicates that the immediate neighbourhood is relatively dense.

{\bf 201697} (10955 km s$^{-1}$):
This object is also CGCG 095-077 NED02=MCG +03-28-040=PGC 032951. The plate scale of 43.09 kpc arcmin$^{-1}$ and the $\sim$50\arcsec major axis imply a 36 kpc galaxy. The SDSS finding chart shows a strongly distorted spiral with a warped disk pointing a southward appendage toward a nearby lenticular 20\arcsec to the south. The SDSS spectrum shows strong H$\alpha$+[NII] and [SII] emission, with weak emission components of H$\beta$ within an absorption profile, and [OII]. The lenticular neighbour is SDSS J105743.81+151740.2, with no SDSS redshift but with a size similar to that of the target galaxy, hinting that it might be a physical and interacting neighbour. Other galaxies in the neighbourhood, also lacking SDSS spectra, appear to form a chain of galaxies. SDSS J105741.28+152003.2, a tiny 10 arcsec bluish spiral at 11040$\pm$30 km s$^{-1}$, is a physical companion and shows a blueward-sloping spectrum with strong H$\alpha$, H$\beta$, [NII], [SII], [OIII] and [OII] emission lines. The Balmer lines from H$\beta$ and higher are narrow and visible in wide absorption troughs.

{\bf 006066} (11807 km s$^{-1}$):
This galaxy, classified in NED as SAab: sp, is also UGC 06066. The plate scale at its redshift is 46.25 kpc arcmin$^{-1}$ and the isophotal major axis $a$=105\arcsec.15 implies a physical size of 81 kpc at PA=41$^{\circ}$. The SDSS spectrum of the central region shows only absorption features, including fairly pronounced Balmer lines. The SDSS finding chart shows a disk galaxy with tightly wound spiral arms and with no obvious signs of interaction. No immediate neighbours were identified in the SDSS data.

{\bf 222042} (11569 km s$^{-1}$):
This is CGCG 041-030. The SDSS finding chart shows a disk galaxy with an inner dark ring encircled by a luminous outer ring. The inner regions show that this might be a barred galaxy with the arms emerging from the ends of the bar, with a possible SBb morphological classification. A bright nucleus is visible at the center of the bar. The SDSS spectrum shows a narrow H$\alpha$ in emission and higher Balmer lines in absorption. The plate scale of 45.35 kpc arcmin$^{-1}$ and the major axis of 74\arcsec.42 imply a 56 kpc size. The object shows no obvious signs of interaction.

{\bf 226077} (9250 km s$^{-1}$):
The plate scale at the redshift of this object is 36.53 kpc arcmin$^{-1}$; with a major axis of 36\arcsec.99 at a position angle of 46$^{\circ}$ this is not a very big galaxy, only 23 kpc wide. The SDSS image shows a spiral galaxy with an inner ring, possibly warranting an Sa(r) classification. The spectrum shows strong H$\alpha$+[NII] and [SII] emissions on a typical absorption spectrum, with H$\beta$ and higher Balmer lines also in absorption. This galaxy has a nearby neighbour, SDSS J122512.55+160536.0 about 1.5 arcmin to the south at 9180$\pm$60 km s$^{-1}$. The companion also shows H$\alpha$+[NII] and [SII] emissions, although these are weaker than in the primary target.

{\bf 226111} (10678 km s$^{-1}$):
This galaxy, also known as CGCG 100-019 and MCG +03-33-020, The plate scale of 41.98 kpc arcmin$^{-1}$ and the major axis of 59\arcsec.83 at PA=69$^{\circ}$ imply a size of 44.2 kpc. The SDSS image shows a highly inclined disk with dark truncations of the inner disk with a possible morphological classification of Sc. A nearby physical neighbour, SDSS J125932.41+150413.5=2MASX J12593238+1504138 at 10920$\pm$60 km s$^{-1}$ is visible 1.2 arcmin to the north-west.  The main object shows weak H$\alpha$ and [SII] emission on a spectrum with strong absorption lines. The SDSS companion has relatively strong H$\alpha$+[NII], [SII] and [OII] emissions on a spectrum with shallow absorption lines. Another physical companion is SDSS J125928.98+145958.1, a blue 10 arcsec disk at 10800$\pm$60 km s$^{-1}$ and 3\arcmin.3 away, with a blueward-sloping continuum and H$\alpha$+[NII], [OII] and [SII] emission lines.

{\bf 233609} (8143 km s$^{-1}$):
The SDSS finding chart shows a featureless bluish disk, possibly face-on and devoid of spiral arms, with a superposed star. The plate scale is 32.57 kpc arcmin$^{-1}$. It, and the isophotal $r$ major axis of 28\arcsec.48 at PA=150$^{\circ}$, imply a smallish 15 kpc galaxy. No SDSS spectrum exists in the archives, but the object has as real neighbour SDSS J130717.44+133847.8 1.5 arcmin to the south-east as an amorphous disk at 8070$\pm$60 km s$^{-1}$. SDSS J130712.74+133848.2 about 50\arcsec to the south is a background object at 30030$\pm$60 km s$^{-1}$ with a blueward-sloping continuum and H$\alpha$+[NII], [SII] and [OII] emissions, and another background object, SDSS J130709.55+133804.6, is about 1.5 arcmin to the south south-west and at 43110$\pm$60 km s$^{-1}$.

{\bf 008375} (7007 km s$^{-1}$):
This galaxy is IC 0881=UGC 08375=CGCG 101-025, classified Sa in NED, with an isophotal major axis of 2a=108\arcsec.33 at PA=11$^{\circ}$. With the plate scale of 28.17 kpc arcmin$^{-1}$, this galaxy has a semi-major axis of 25 kpc. The SDSS finding chart shows an edge-on disk with a strong bulge, warranting the Sa classification. Dust lanes are visible in the disk. The SDSS spectrum shows strong absorption lines, including Balmer absorptions. A physical neighbour is SDSS J132006.93+155353.2=IC 0882, about four arcmin to the north-east at 6930$\pm$60 km s$^{-1}$, with an absorption SDSS spectrum. This is a face-on disk with a bright bulge, perhaps classifiable as S0. Other galaxies within a few arcmin are all in the background.

{\bf 008379} (11967 km s$^{-1}$):
This object is also UGC 08379=CGCG 044-071, classified Sbc in NED. The plate scale is 46.49 kpc arcmin$^{-1}$ and the major axis being 84\arcsec.09 at PA=79$^{\circ}$ implies that the galaxy has a 33 kpc radius. The SDSS spectrum shows weak H$\alpha$ in emission along with [SII] and higher Balmer lines in absorption. The SDSS finding chart shows a two-arm spiral that appears to be fairly isolated.

{\bf 008475} (6835 km s$^{-1}$):
This object is NGC 5162=NGC 5174=UGC 08475=CGCG 072-087=MCG +02-34-018, an Scd:, according to NED. The double NGC assignment is mentioned in UGC as `Listed in CGCG as ``double nebula'', which is incorrect. Star superimposed south center.' The SDSS isophotal major axis is 123\arcsec.16 at PA=165$^{\circ}$; at a plate scale of 27.63 kpc arcmin$^{-1}$ this implies a 76 kpc wide galaxy. The SDSS spectrum shows weak H$\alpha$+[NII], [SII] and [OII] emission lines over a strong, absorption-feature continuum.

A physical companion is some eight arcmin to the south-west: this is SDSS J132902.43+105455.7, an $r$=17.71 mag bluish disk or flat irregular with a blueward-sloping continuum and strong H$\alpha$, H$\beta$, [NII], [SII], [OIII] and [OII] emission lines. A more distant physical neighbour is SDSS J132956.72+110419.5, about seven arcmin to the north-east at 6750$\pm$30 km s$^{-1}$; the SDSS spectrum shows H$\alpha$, [NII] and [SII] emissions on a blueward sloping continuum. A foreground object, some eight arcmin to the north-west and at 4290$\pm$30 km s$^{-1}$, is SDSS J132852.20+110549.2, a bluish disk with blue continuum and  H$\alpha$, H$\beta$, [NII], [SII], and [OIII] emission lines. Another bluish disk, SDSS J132848.66+110156.6, is eight arcmin to the west and at 6060$\pm$30 km s$^{-1}$. All these possible real neighbours appear to be forming stars intensively and in the same region of space; yet the neighbours are between one tenth and two tenths the size of N5162 and would not be considered ``significant neighbours'', according to the Karachentseva (1973) isolation criteria. For this reason we include it in our selection of possible high-mass galaxies.

{\bf 008488} (7363 km s$^{-1}$):
This object is NGC 5185=UGC 08488=CGCG 072-104=MCG +02-34-025, an Sb galaxy according to NED, with a major axis of 111\arcsec.70 (from RC3) at PA=58$^{\circ}$; with the plate scale of 29.62 kpc arcmin $^{-1}$, this implies a galaxy with a semi-major axis of 28 kpc. The SDSS finding chart shows a tilted spiral,  possibly barred, where the north-east side is more open than the south-west side.

SDSS J133001.26+131839.4 is a physical companion; this is a blue disky $\sim$10 arcsec wide object, $r$=17.88 mag galaxy some five arcmin south and 7350$\pm$30 km s$^{-1}$ with a blueward-sloping continuum, H$\alpha$, [NII], and [SII] emission lines. A more distant companion, some 10 arcmin to the south south-east, is SDSS J133025.61+131148.2 at 7380$\pm$60 km s$^{-1}$ with the appearance of a nucleated dE but whose spectrum shows weak H$\alpha$, strong [NII], and even stronger [OIII] emission lines. Another emission-line object is SDSS J133011.83+133623.2, some 10 arcmin to the north north-east at 7560$\pm$60 km s$^{-1}$, showing H$\alpha$, [NII] and [SII] emission lines on a blueward-sloping continuum. The appearance of this $\sim$10 arcsec blue disk is of a nucleated dE object. 008488 is another case of  a large galaxy with very small, star-forming galaxies in its immediate surroundings.

{\bf 008559} (7025 km s$^{-1}$):
This galaxy is NGC 5221=UGC 08559=ARP 288=VV 315b=VIII Zw 325=CGCG 073-040=MCG +02-35-006 classified as Sb: in NED. The SDSS finding chart shows a disky galaxy with a strong bulge and a circum-nuclear dark lane. The outer parts of the object are strongly warped with bright bluish features to the west and south. To the north-east the warp appears softer and less bright and blue than the other parts of the warped disk. A reddish diffuse concentration is visible in the outermost northward part of the warp. The plate scale is 28.12 kpc arcmin $^{-1}$ and, together with the isophotal major axis of 143\arcsec.90 (from RC3), implies a 67 kpc object.

Possible neighbours include SDSS J133503.61+135519.7 a nucleated dE five arcmin to the north north-east at 7350$\pm$60 km s$^{-1}$ with the spectrum of an early-type galaxy, the larger SDSS J133455.94+134431.7=NGC 5222, a seemingly elliptical with a superposed distorted spiral, four arcmin south at 6960$\pm$60 km s$^{-1}$, with H$\alpha$, [NII], strong [SII], and [OII] emission and and with whom the primary galaxy is assumed to have interacted, and SDSS J133430.04+134814.8, another possible nucleated dE some five arcmin to the west south-west at 7290 km s$^{-1}$. SDSS J133448.16+134439.4, a diffuse bluish disk, is two arcmin to the west of N5222; it is at 6630$\pm$30 km s$^{-1}$ and shows an early-type galaxy spectrum, while SDSS J133443.53+134507.3, some five arcmin to the south-west and at 7590$\pm$60 km s$^{-1}$, shows Balmer absorptions. The region seems to have more than its share of objects at redshifts similar to that of the primary target, indicating that some of the small galaxies might be fragments created by the galaxy interaction.

{\bf 008766} (6996 km s$^{-1}$):
This galaxy is IC 0944=UGC 08766=CGCG 073-085=MCG+02-35-019, classified in NED as Sa, with an isophotal $r$-band major axis of 138\arcsec.55 at PA=106$^{\circ}$. Since the plate scale is 28.13 kpc arcmin$^{-1}$, this implies a 65 kpc galaxy. The SDSS spectrum shows H$\alpha$ in emission and higher Balmer lines in absorption on a continuum with shallow absorption lines. The SDSS finding chart shows the nearby physical companion, SDSS J135132.95+140639.0, which is
 an SBb (or SAB(s)b:, as listed in the UGC note) galaxy at 7380$\pm$30 km s$^{-1}$ about one arcmin away,%{\bf [how far away?]},
 with H$\alpha$, [NII] and [SII] in emission, H$\beta$ emission with narrow emission in a wide absorption trough and the higher Balmer lines in absorption, on a blueward-sloping continuum. Smaller objects in the same neighbourhood are SDSS J135117.73+140630.8 3.5 arcmin to the west north-west and at 6180$\pm$30 km s$^{-1}$ being more than 500 km s$^{-1}$ away, with a blueward-sloping continuum and strong  H$\alpha$, H$\beta$, [NII], [OIII] and [SII], and SDSS J135211.55+135959.6,  a spiral at 7350$\pm$30 km s$^{-1}$ some eight arcmin to the south-west. Another spiral, SDSS J135219.28+141618.8 about 12 arcmin to the north-east and at 7260$\pm$60 km s$^{-1}$, shows an absorption spectrum including Balmer lines.

{\bf 008902} (7559 km s$^{-1}$):
This object, classified in NED as Sb:,  has a 70\arcsec.50 major axis (from RC3 at 25 mag arcsec$^{-1}$). With a plate scale of  30.60 kpc arcmin$^{-1}$, this makes it a 36 kpc object. The SDSS finding chart shows a spiral galaxy very close to a bright star, and a number of galaxies of similar size nearby. The galaxy itself shows H$\alpha$ and [SII] in emission on an absorption spectrum, whereas SDSS J135906.56+153434.2, a diffuse irregular neighbour one arcmin to the north-east at 7260$\pm$30 km s$^{-1}$, shows a  blueward-sloping continuum with strong Balmer, [NII], [SII], and [OIII] emission lines. Another physical companion, SDSS J135909.06+153336.9, an edge-on Sb with a prominent equatorial dark lane 1.5 arcmin east and at 7440$\pm$30 km s$^{-1}$, shows H$\alpha$ and [SII] in emission and other Balmer lines in absorption. A more distant object, SDSS J135855.46+153808.7, a bluish flocculent spiral at 7650$\pm$30 km s$^{-1}$, shows H$\alpha$, [OII] and [SII] in emission and other Balmer lines in absorption. H$\beta$ shows a narrow emission component within the absorption profile, all on a fairly flat (in F$_{\lambda}$) continuum.

{\bf 230914} (11848 km s$^{-1}$):
This galaxy is also CGCG 074-030 and has an 85\arcsec.06 major axis at PA=115$^{\circ}$. With a  plate scale of 45.96 kpc arcmin$^{-1}$, this is a 65 kpc galaxy that can be classified as Sb. SDSS shows an absorption spectrum. Two galaxies are within three arcmin; one is SDSS J135955.08+120319.2, a 20 arcsec face-on spiral at 11670$\pm$30 km s$^{-1}$ with Balmer, [NII], [SII], [OII] and [OIII] emission lines on a blueward-sloping continuum. Others are SDSS J140031.27+120451.8, a face-on Sc at 11610$\pm$30 km s$^{-1}$ also showing Balmer, [NII], [SII], [OII] and [OIII] emission lines, and SDSS J140025.49+120848.6, a nucleated dE some four arcmin to the north-east at 11820$\pm$60 km s$^{-1}$ with H$\alpha$ and [NII] emission on a general absorption spectrum.

{\bf 008943} (4933 km s$^{-1}$):
This object is NGC 5417=UGC 08943=CGCG 046-039, an Sa galaxy with 105\arcsec.81 major axis in PA=117. With a plate scale of 20.09 kpc arcmin$^{-1}$, this is 35 kpc object. The SDSS spectrum shows absorptions only. A physical companion, SDSS J140215.25+080022.9, a featureless 10\arcsec blob two arcmin south of N5417 and at 5190$\pm$60 km s$^{-1}$, shows Balmer lines in absorption on a blueward-sloping continuum.

{\bf 009031} (11864 km s$^{-1}$):
This galaxy is UGC 09031=CGCG 103-076=MCG +03-36-053, classified as a possible spiral (S?) in NED. With an isophotal $r$ major axis of 69\arcsec.80 at PA=40$^{\circ}$, and a plate scale of 46.03 kpc arcmin$^{-1}$, this is a 27 kpc (semi-major axis) disk. The SDSS finding chart shows an almost edge-on disk with a prominent equatorial dark lane. The general appearance is of an Sb galaxy. The SDSS spectrum shows an absorption line spectrum and emission at H$\alpha$ and [NII], and absorption at higher Balmer lines. A small galaxy about 20 arcsec to the north-west shows a warped disk; with no redshift it is not possible to establish whether this is a physical companion or is projected. There are other small galaxies at slightly larger distances, most lacking redshifts and those that have one, are in the background.

{\bf 248977} (9153 km s$^{-1}$):
The SDSS finding chart shows a $\sim$30 arcsec diffuse image with a very bright center reminiscent of a nucleated dE (ndE), significantly larger than the size given in SDSS. This object is SDSS J144112.34+145324.4 with an optical redshift of 9150$\pm$60 km s$^{-1}$ and with an absorption spectrum that includes also H$\alpha$ and $\beta$ in absorption. At this redshift, the 10\arcsec.55 size from SDSS translates into some 6 kpc; this is a relatively small galaxy that would be $\sim$20 kpc if the size shown in the SDSS image would be adopted. Two diffuse $\sim$5 arcsec blobs are nearby: SDSS J144110.38+145331.5 with $r$=15.6 and ($g-r$)=0.42 that might be a dark-lane galaxy, and SDSS J144110.79+145333.9 with $r$=16.81 and ($g-r$)=0.30. These do not have optical redshifts and there is a possibility they are associated with the main galaxy and may have been confused in the same ALFALFA beam.

{\bf 009624} (11068 km s$^{-1}$):
This object is NGC 5790=UGC 09624=CGCG 048-076=MCG +01-38-022. Classified in NED as (R)SA0/a and in UGC as S0/Sa, it shows an inner yellowish spiral with an outer bluish ring with dust lanes and spiral structure on the SDSS finding chart. The SDSS spectrum shows H$\alpha$, [NII], [SII], [OI], [OIII] and [OII] in emission, while H$\beta$ and higher Balmer lines are in absorption. There are a number of nearby galaxies without redshifts, but a physical companion is SDSS J145730.93+082326.3, a 10\arcsec nucleated dE or dwarf S0 about six arcmin north at 11130$\pm$60 km s$^{-1}$. The main galaxy has a major axis of 57\arcsec.30 (RC3, isophotal at 25  mag arcsec$^{-1}$) and, with a plate scale of 43.57 kpc arcmin$^{-1}$, is 42 kpc wide (21 kpc semi-major axis).

{\bf 009788} (10176 km s$^{-1}$):
The galaxy is UGC 09788=CGCG 049-079, classified Im? in NED. The plate scale of 39.68 kpc arcmin$^{-1}$ and major axis of 54\arcsec.70 (RC3) imply a 36 kpc object. The SDSS finding chart shows a disky galaxy with a number of dust lanes meandering along the equatorial plane. A faint and fuzzy extension is visible to the north, extending some 15 arcsec from the nucleus, and a bluish extension is visible at the north-west end of the equatorial dark lane. This extension points to a physical companion, SDSS J151534.35+081839.2, a flattened disk one arcmin to the north-west and at 10410$\pm$30 km s$^{-1}$, that shows a flat spectrum (in F$_{\lambda}$) with emission lines at H$\alpha$, [NII], [SII], [OIII] (weak), H$\beta$ in a wide absorption trough, and [OII].

Another physical companion is SDSS J151543.03+081710.0, a nucleated dE about 70\arcsec to the south-east and on the extension of the major axis of the primary galaxy at 10230$\pm$60 km s$^{-1}$, showing deep Balmer absorptions. SDSS J151543.06+082027.3 is further away about two minutes to the north-east. This is a ndE with a 15 arcsec size and a redshift of 10350$\pm$60 km s$^{-1}$. It has a similar spectrum, indicating a significant population of early-type stars. Even more distant physical companions can be found: SDSS J151522.00+082127.3 five arcmin to the north-west at 10260$\pm$60 km s$^{-1}$, with a similar morphology and spectrum as the two nearer companions, and SDSS J151540.51+081303.6, an edge-on 30 arcsec Sc that is five arcmin to the south at 9900 $\pm$60 km s$^{-1}$, has a blueward-sloping continuum, and shows H$\alpha$, H$\beta$, [NII], [SII], [OIII] and [OII] emission lines.

{\bf 009794} (6406 km s$^{-1}$):
This extremely disturbed object is UGC 09794=VIII Zw 461=CGCG 077--54=MCG +02-39-009. NED classifies it as SBdm: whereas UGC notes it as SBc-IRR. The $r$ band isophotal major axis quoted in NED, 7\arcsec.30, is an obvious mistake; the corresponding value for the B isophotal major axis in RC3 is 181\arcsec.20 at PA=42$^{\circ}$, which we adopt. With a plate scale of 25.75 kpc arcmin$^{-1}$, this is a 78 kpc galaxy. The SDSS finding chart shows a possible Sc with a strong warp or arm kink at the north-east end and a less pronounced warp at the opposite end. Dust lanes along the equatorial plane and parallel to it are visible.
The SDSS fiber was not set on the nucleus but a small distance to the south-west; the resultant spectrum shows a redward-sloping continuum with H$\alpha$, [NII], [SII] emission lines, and with H$\beta$ in narrow emission superposed on a wide absorption. An immediate physical neighbour is SDSS J151603.23+103023.4, a 15\arcsec nucleated dE about two arcmin to the west at 6630$\pm$60 km s$^{-1}$ and with an absorption spectrum. Other nearby galaxies are in the background.

{\bf 009838} (10267 km s$^{-1}$):
This galaxy is known also as UGC 09838=CGCG 049-158 and is classified Sab in NED and in UGC. The RC3 isophotal major axis is 77\arcsec.30; this translates into 51 kpc at a plate scale of 39.75 kpc arcmin$^{-1}$. The galaxy looks like an Sb with an inner ring about half-way from the center to the edge. SDSS J152502.75+071056.2 is a physical neighbour two arcmin to north-west
at 10380$\pm$60 km s$^{-1}$, with H$\alpha$, [NII], [SII] and [OII] emission lines on a slightly blueward-sloping continuum, and  SDSS J152506.35+070552.5, a 30" blue irregular galaxy three arcmin to the south-west at 10410$\pm$30 km s$^{-1}$ with H$\alpha$, H$\beta$, [NII], [SII], [OIII] and [OII] emission lines. Other physical companions are SDSS J152526.14+070504.9,  a 30\arcsec spiral at 10440$\pm$30 km s$^{-1}$, with H$\alpha$, [NII] and [SII] emission lines and with a number of condensations in the arms, and SDSS J152449.46+070845.2, a blue face-on 25\arcsec two-arm spiral with H$\alpha$, [NII] and [OII] emission on a blueward-sloping continuum five arcmin to the west.

{\bf 260110} (10160 km s$^{-1}$):
This galaxy is also CGCG 079-040, whose SDSS finding chart shows an elliptical featureless body.
With a  major axis of 43\arcsec.13 at PA=56$^{\circ}$ and a plate scale of 39.25 kpc arcmin$^{-1}$ this is a 28 kpc galaxy. The SDSS spectrum shows weak H$\alpha$, [NII], [SII] and [OII] emission on a general absorption spectrum.
A physical but distant companion is SDSS J160514.09+141320.1, an edge-on 30\arcsec disk seven arcmin to the north-east at 10410$\pm$30 km s$^{-1}$. Other nearby (projected) companions are SDSS J160457.91+140815.2, a galaxy classifiable as Sb(r) with two spiral arms emerging from the inner ring, is two arcmin to the east at 11160$\pm$30 km s$^{-1}$ and shows H$\alpha$, H$\beta$ (in an absorption trough), [NII], [SII] and [OII] emission, SDSS J160512.92+140910.6, an edge-on blue 25\arcsec disk four arcmin to the east north-east at 11430$\pm$30 km s$^{-1}$ with H$\alpha$, H$\beta$ (in an absorption trough), [NII], [SII], [OIII] and [OII] emission, and SDSS J160504.78+141005.9, a blue nucleated disk four arcmin to the north-east at 10770$\pm$30 km s$^{-1}$ with H$\alpha$, H$\beta$, [NII], [SII], [OIII] and [OII] emission on a blueward-sloping continuum; although more than 500 km s$^{-1}$ this might be a physical companion.

{\bf 010272} (5124 km s$^{-1}$):
This galaxy is NGC 6081=IC 1202=UGC 10272=CGCG 079-078=MCG +02-41-019, classified S0 in NED, with an isophotal $r$-band major axis of 96\arcsec.37 at PA=129$^{\circ}$, which at a plate scale of 20.54 kpc arcmin$^{-1}$ implies a 16 kpc semi-major axis galaxy. The object shows an SDSS spectrum with H$\alpha$, [NII], and [SII] in emission on a general absorption spectrum. A physical companion is SDSS J161311.43+094952.0, an elongated blue irregular four arcmin to the south-east at 4890 km s$^{-1}$ with strong H$\alpha$, H$\beta$,  [NII], [SII], and [OIII] emission on a blueward-sloping continuum.

%%%%%%%%%%%%%%%%%%%%%%%%%%%%%%%%%%%%%%%%%%%%%%%%%%%%%%%%%%%%%%%%%%%%%%%%%%%

\section{Appendix B: HI profiles of the candidate massive disks}
In this Appendix we show the HI profiles of all the CHMDs obtained by the ALFALFA survey. The ACG numbers of the CHMGs are shown in bold font.
\begin{figure}
\includegraphics[clip=,angle=0,width=7cm]{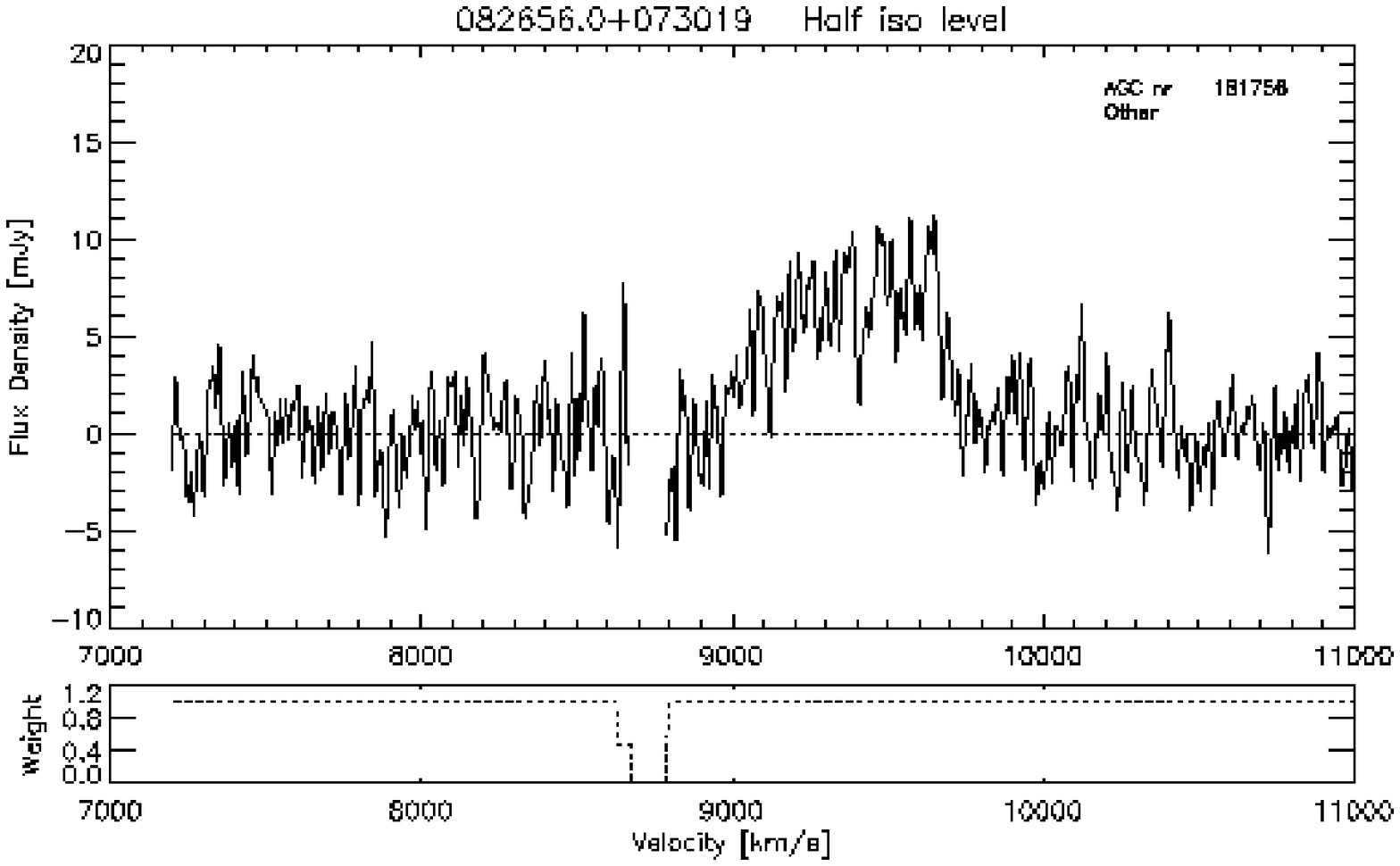}
\includegraphics[clip=,angle=0,width=7cm]{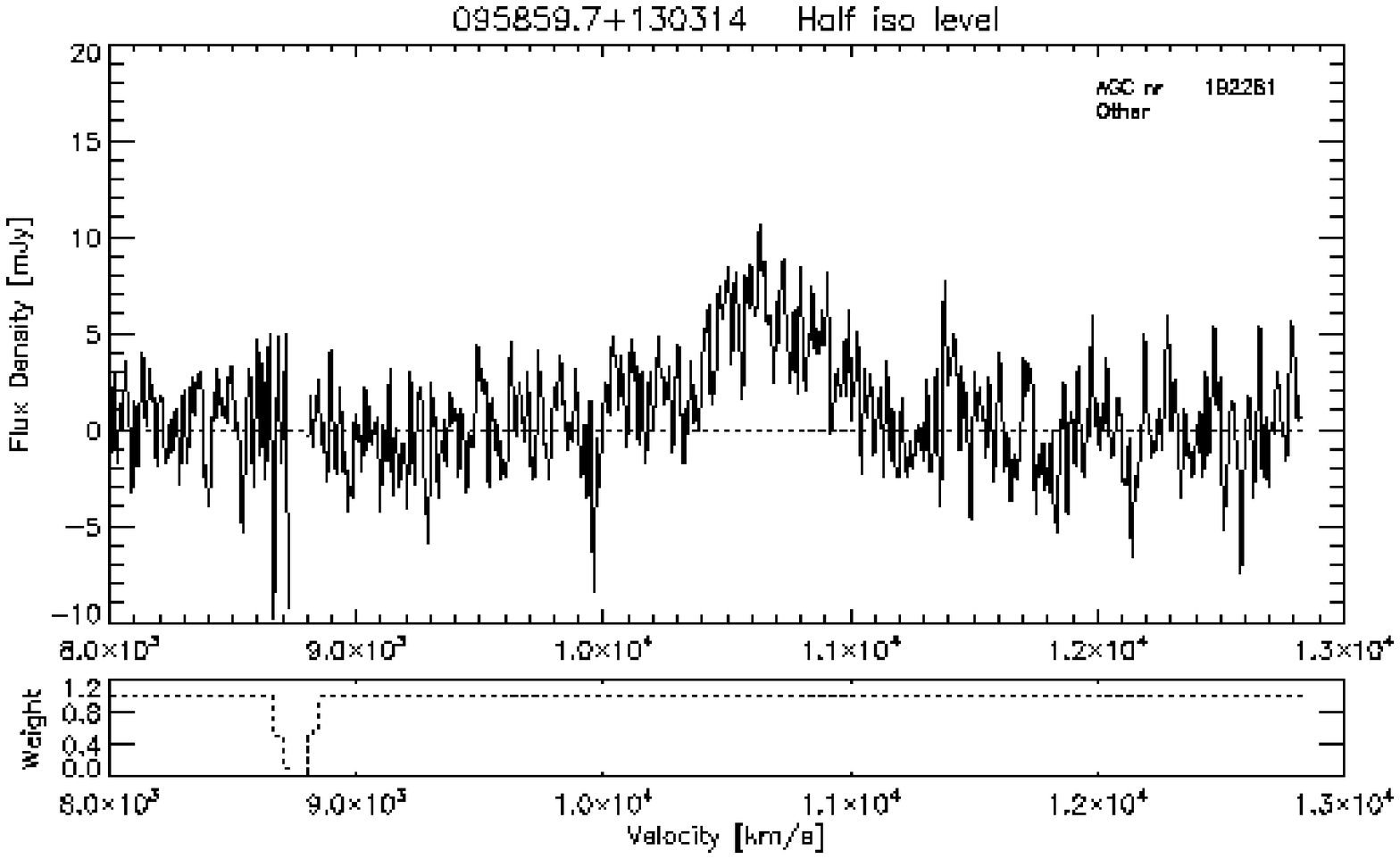}
\includegraphics[clip=,angle=0,width=7.0cm]{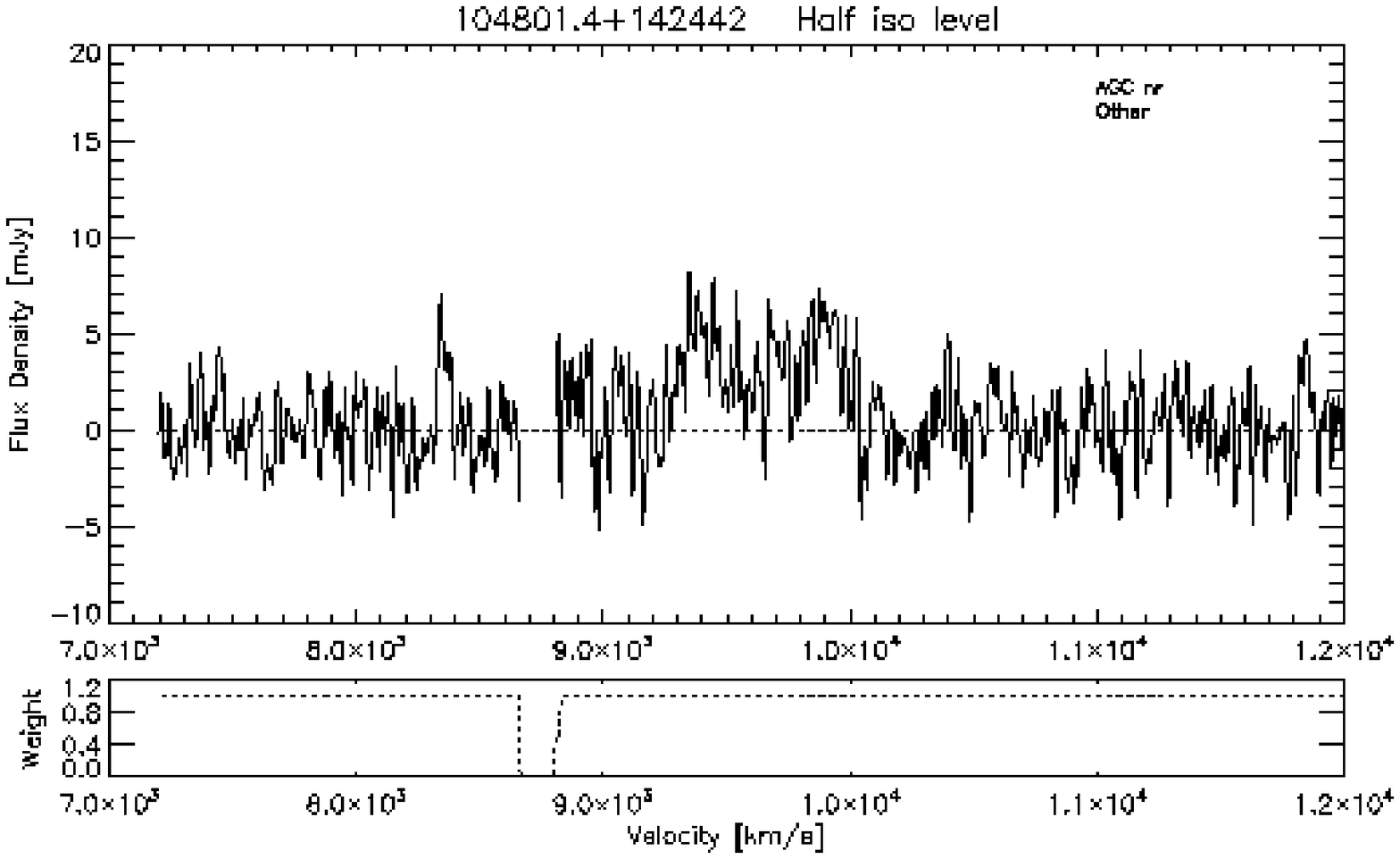}
\includegraphics[clip=,angle=0,width=7.0cm]{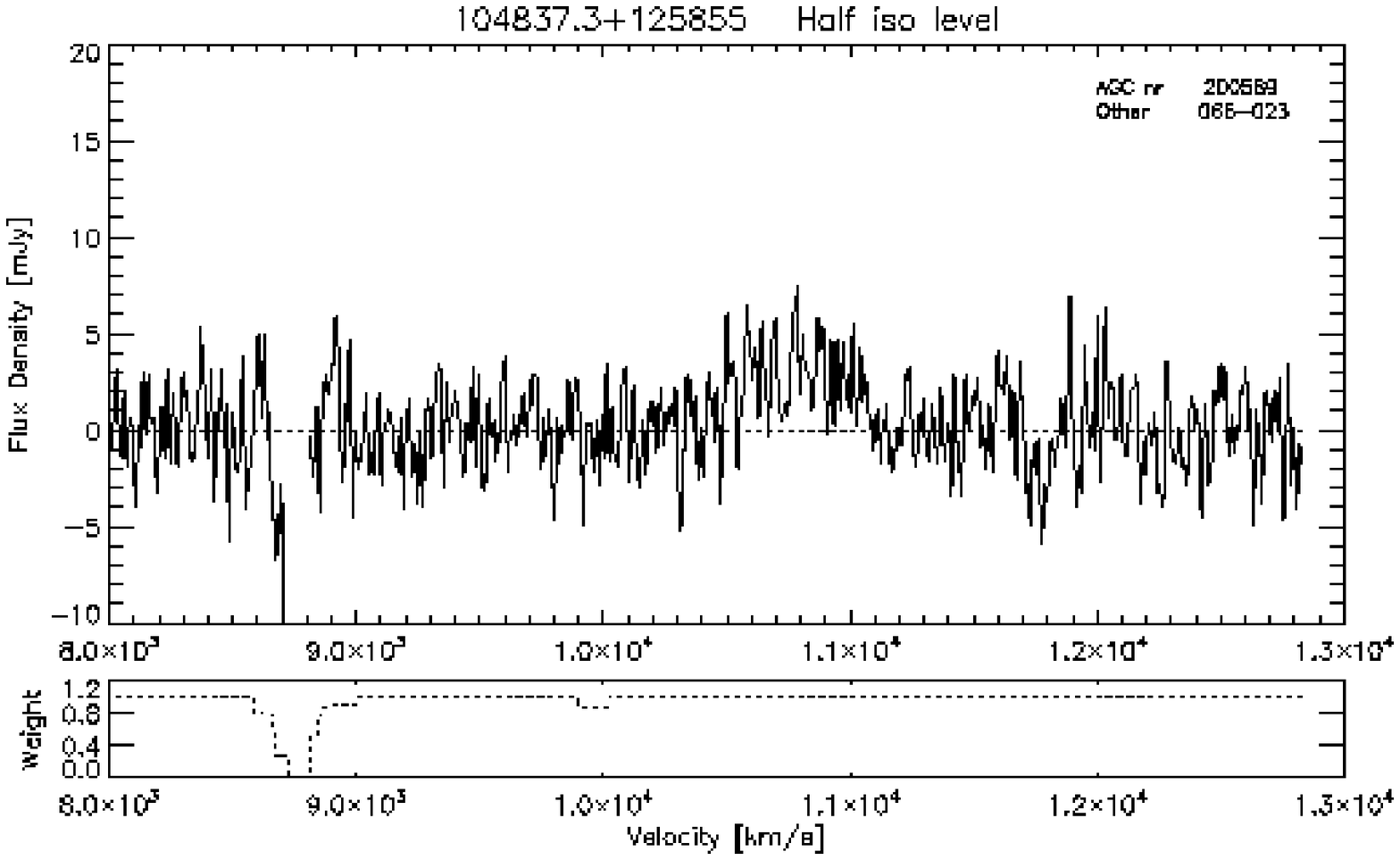}
\includegraphics[clip=,angle=0,width=7.0cm]{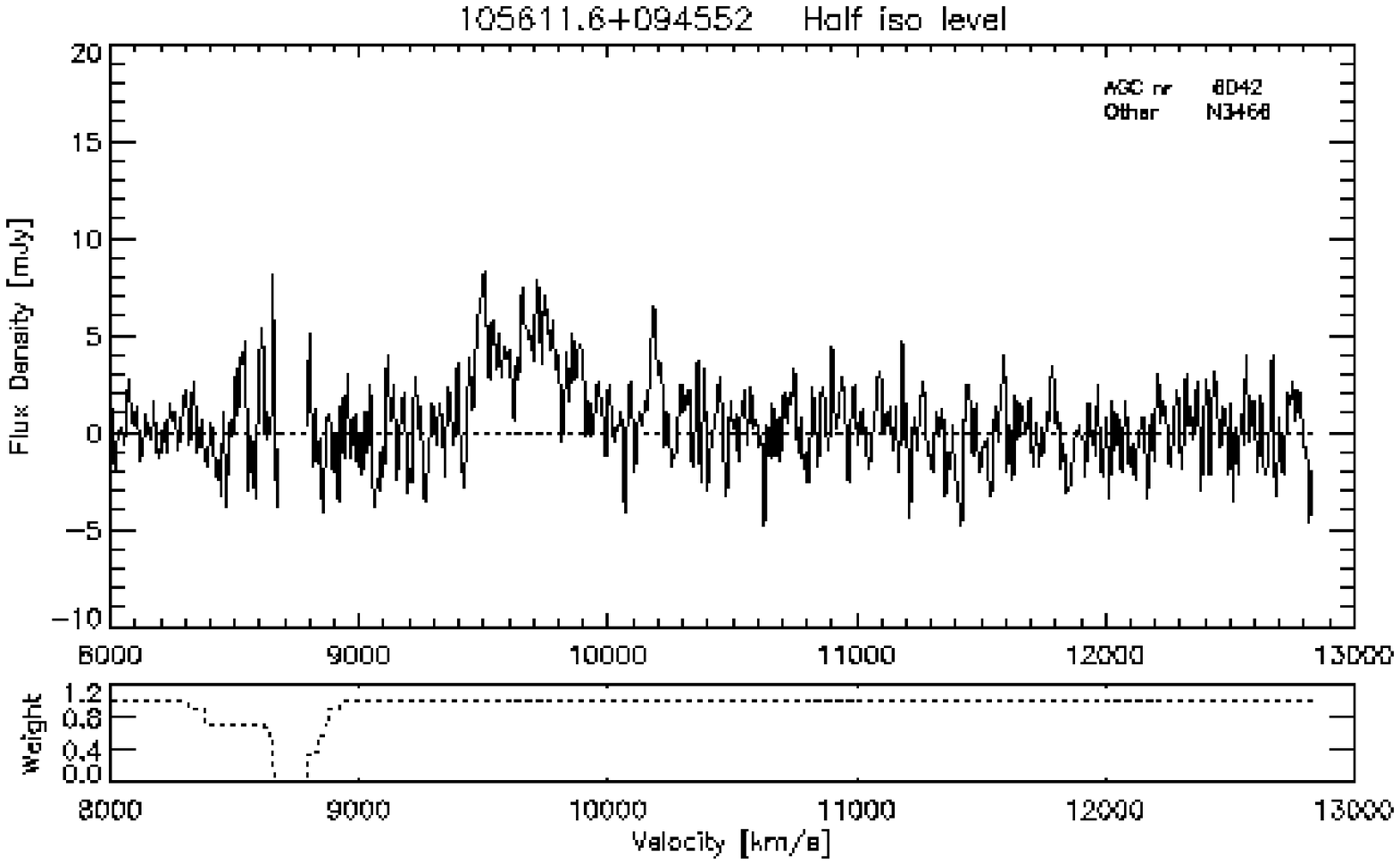}
\includegraphics[clip=,angle=0,width=7.0cm]{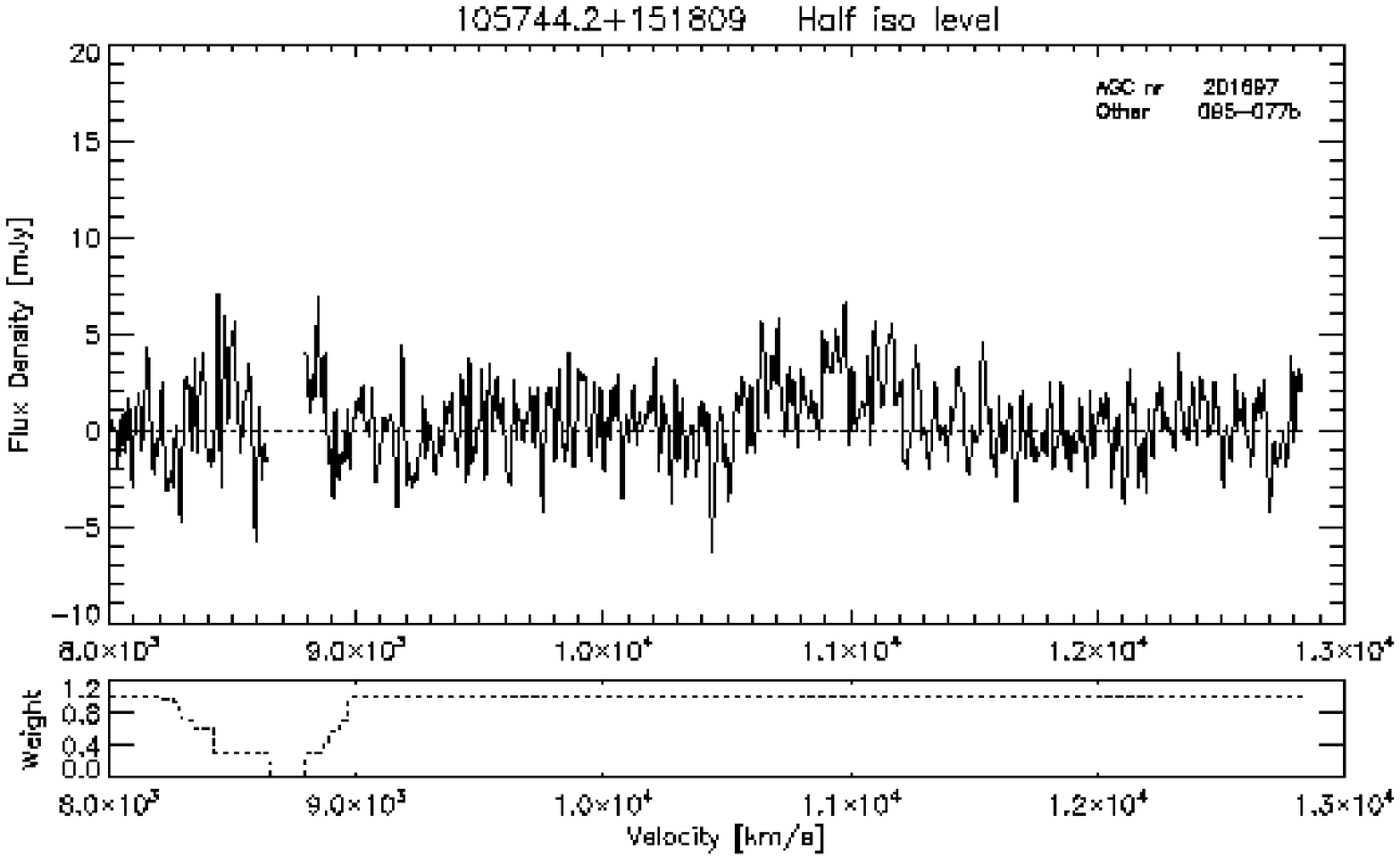}
\includegraphics[clip=,angle=0,width=7.0cm]{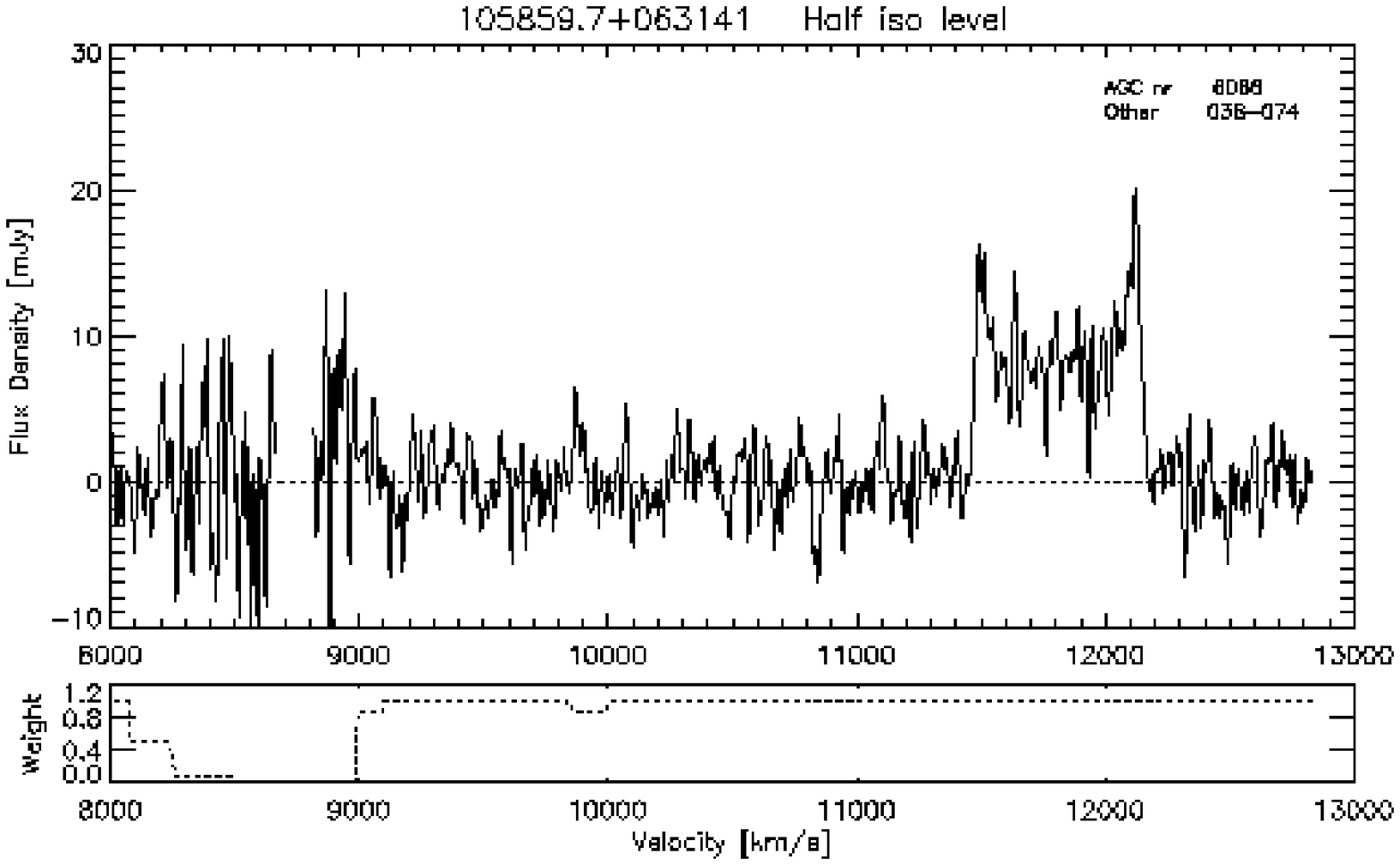}
\includegraphics[clip=,angle=0,width=7.0cm]{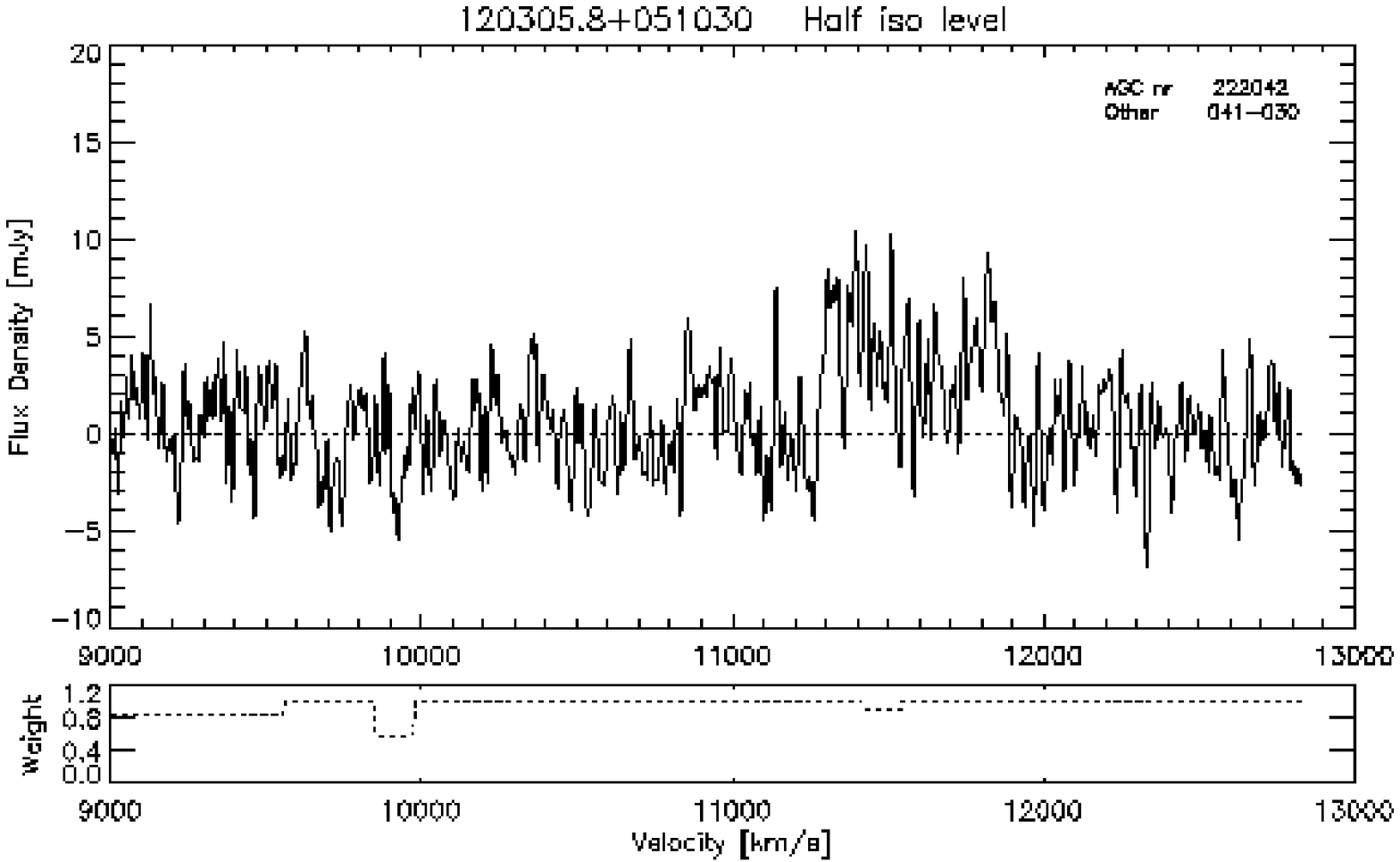}
\caption{Mosaic of HI profiles for possible high-mass galaxies. The bottom part of each panel shows the adopted baseline; segments excised from the spectrum and missing in the plot, mainly because of interference, are indicated. The plots belong to the following ACG objects, with those objects retained as showing intrinsic wide HI profiles given in bold fonts: 181756 (top-left), 192281 (top-right), {\bf 205190} (2$^{\rm nd}$ row-left), {\bf 200589} (2$^{\rm nd}$ row-right), 006042 (3$^{\rm rd}$ row-left), 201697 (3$^{\rm rd}$ row-right), 006066 (bottom-left), and 222042 (bottom-right).}
\label{f:candidates0}
\end{figure}

\begin{figure}
\includegraphics[clip=,angle=0,width=7cm]{HI122517.8+160754_small.EPS}
\includegraphics[clip=,angle=0,width=7cm]{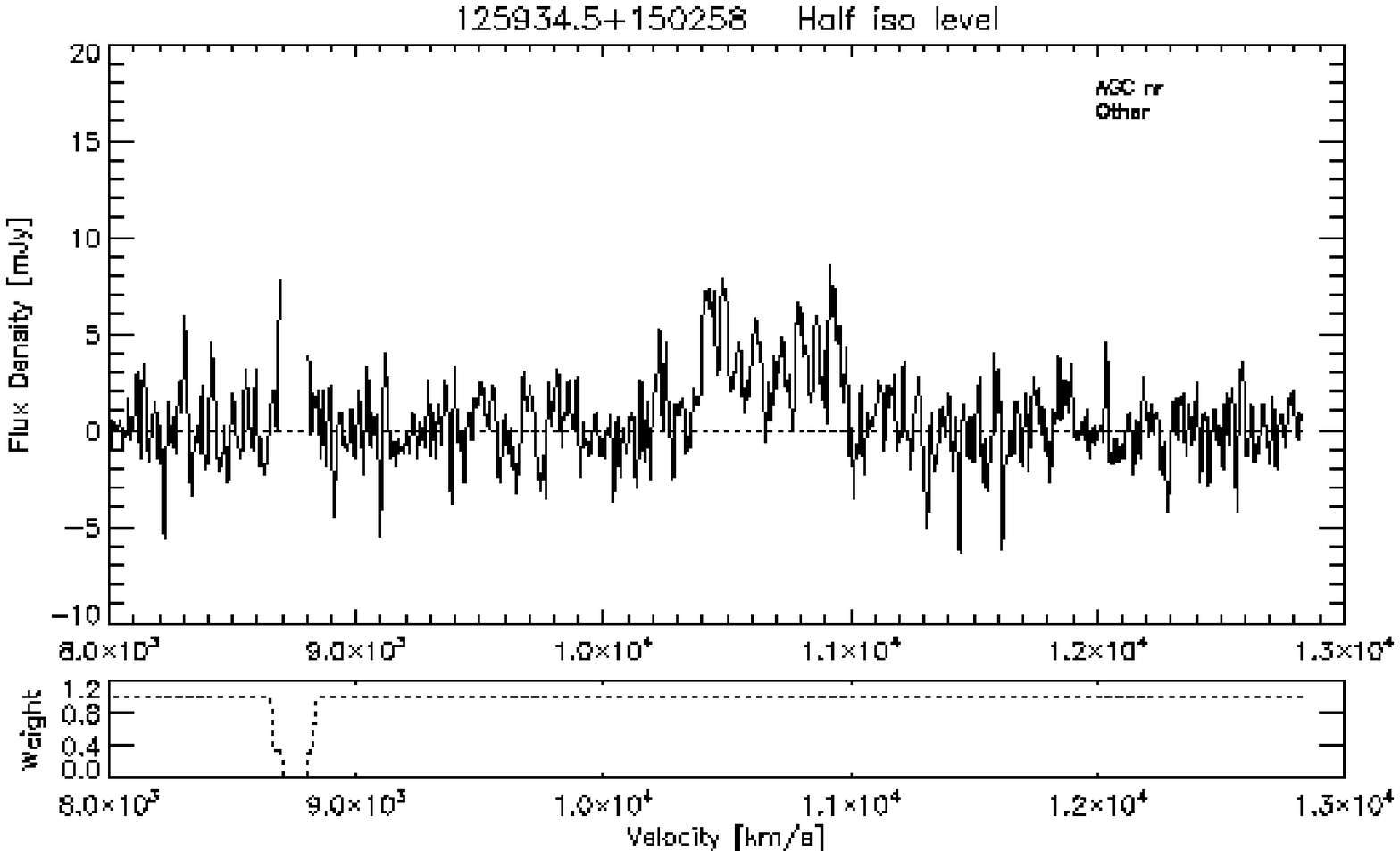}
\includegraphics[clip=,angle=0,width=7cm]{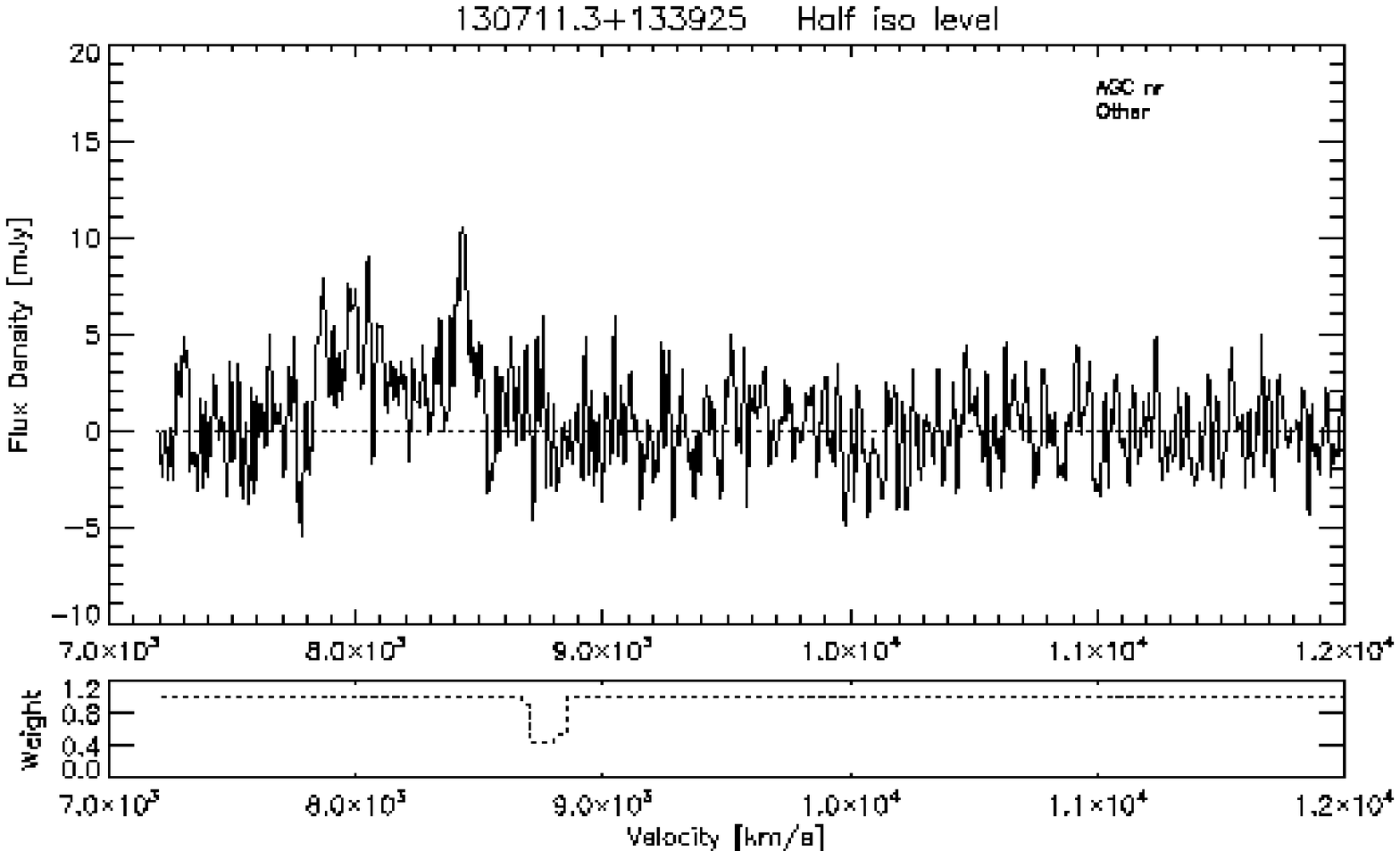}
\includegraphics[clip=,angle=0,width=7cm]{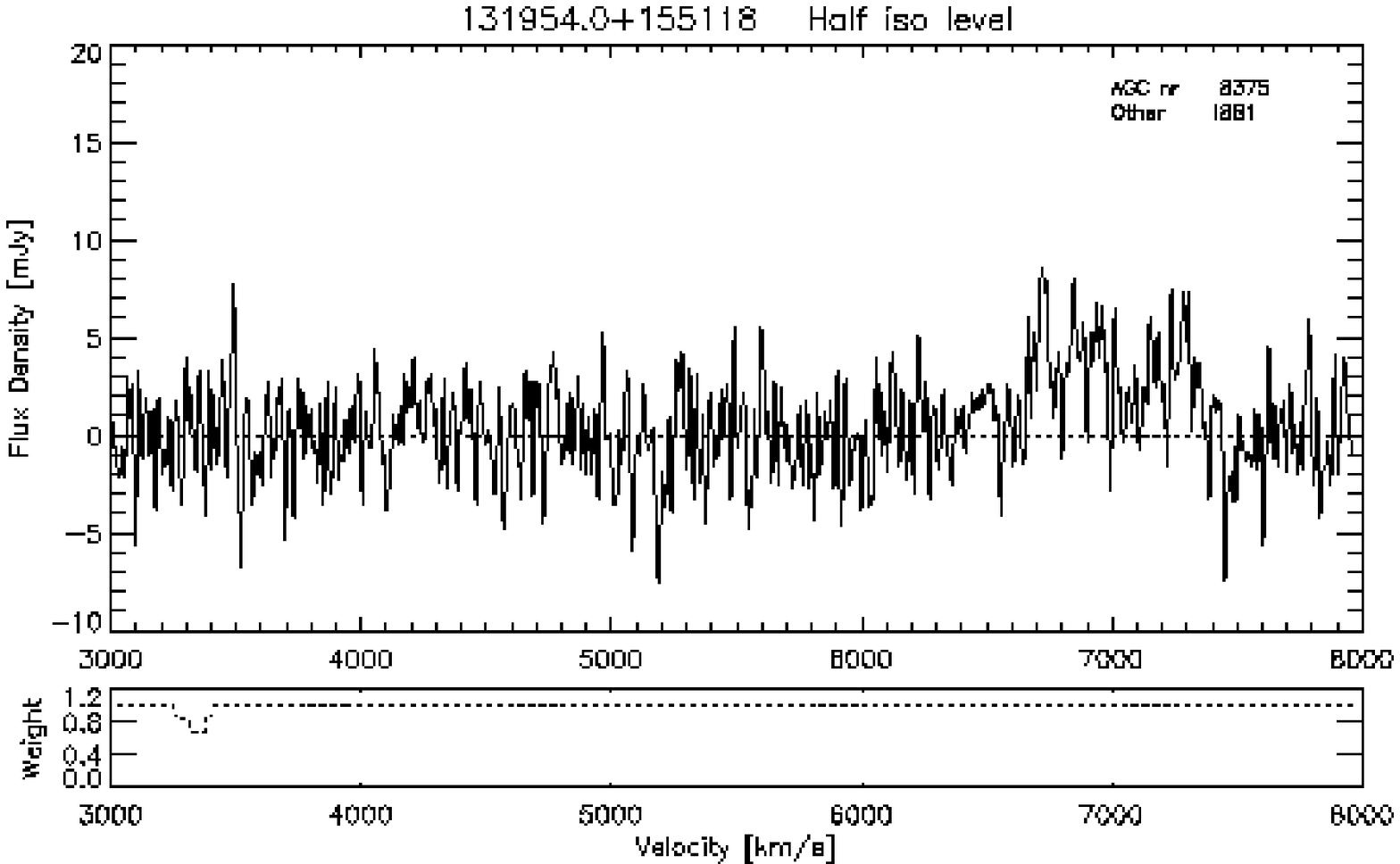}
\includegraphics[clip=,angle=0,width=7cm]{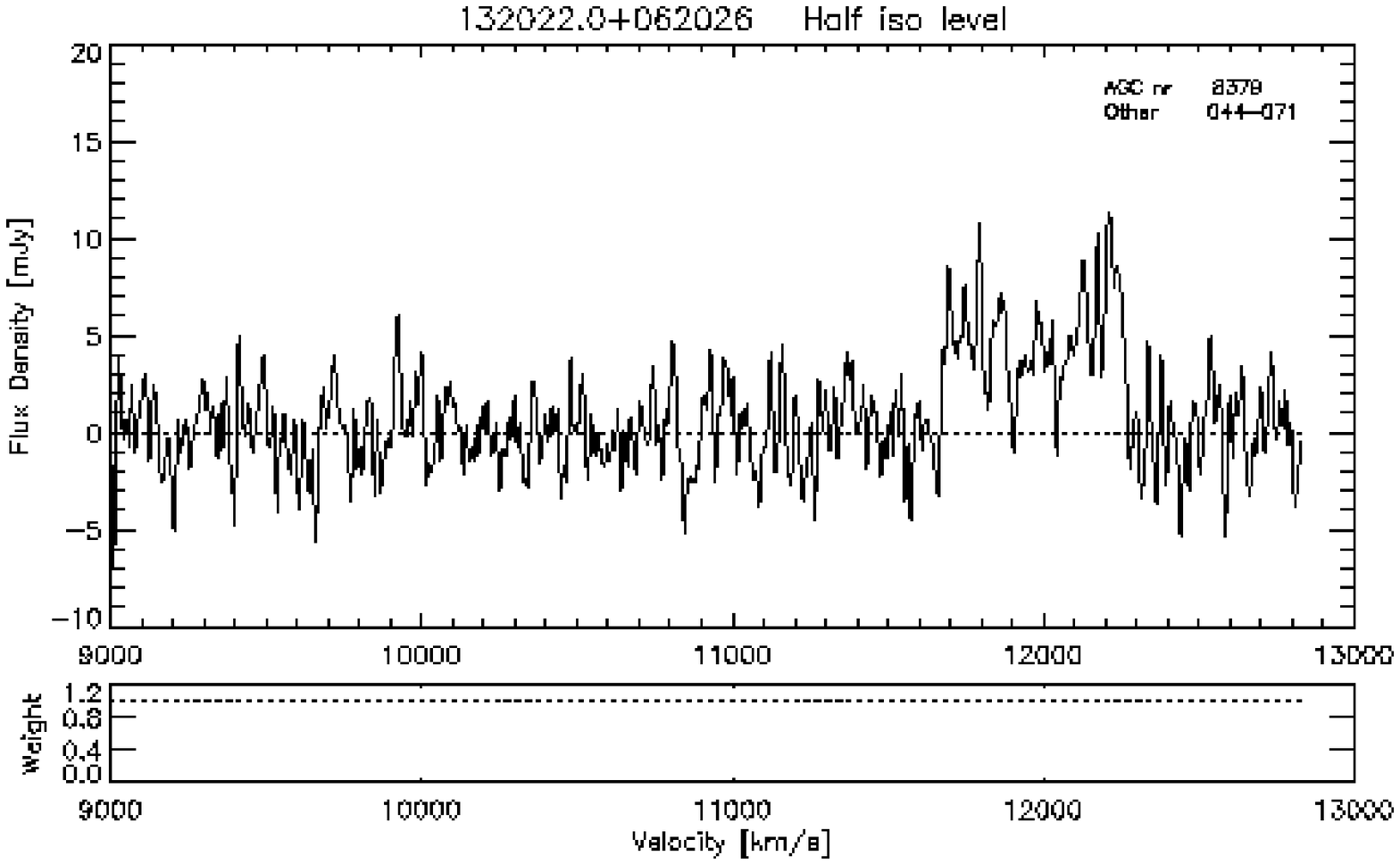}
\includegraphics[clip=,angle=0,width=7cm]{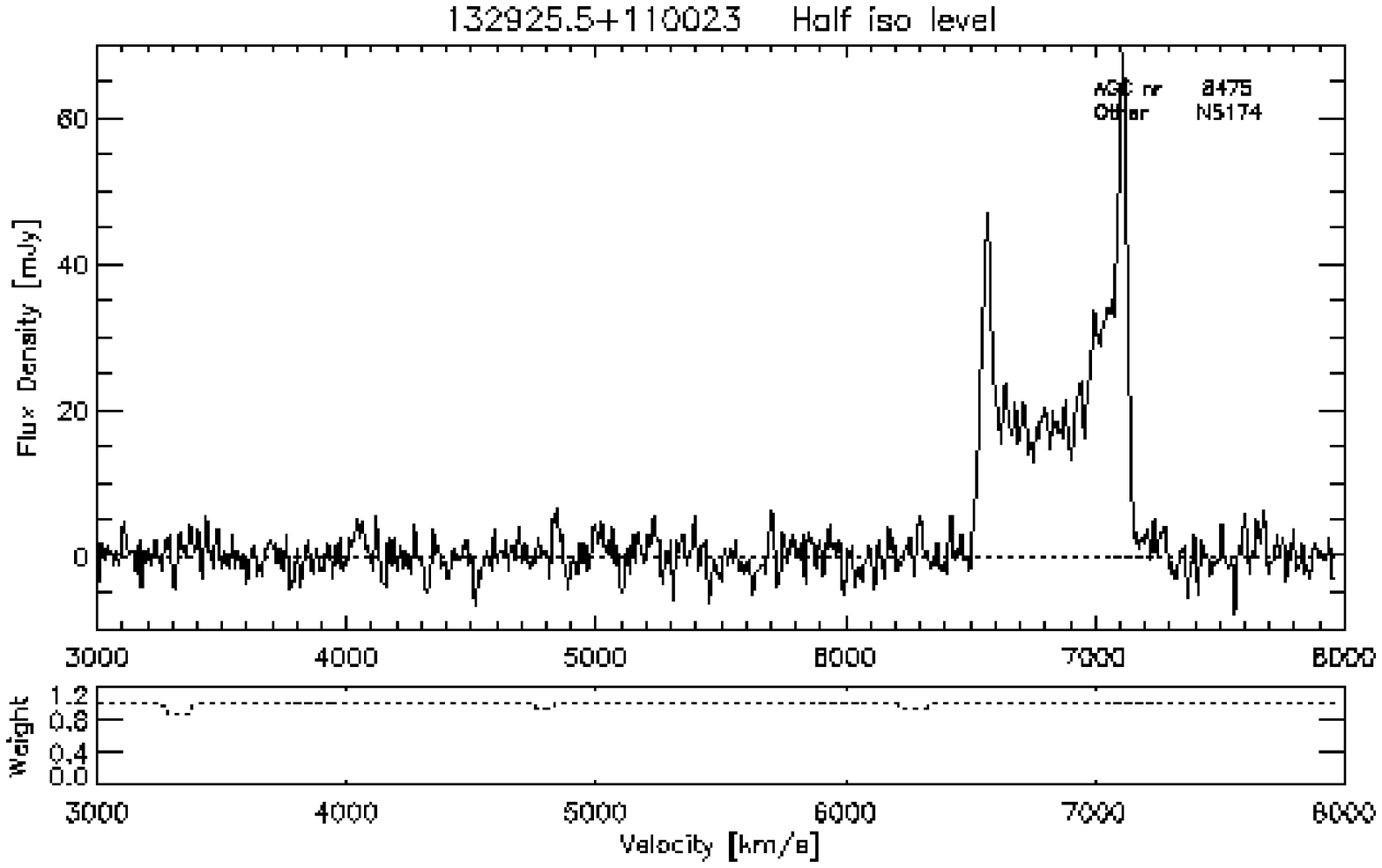}
\includegraphics[clip=,angle=0,width=7cm]{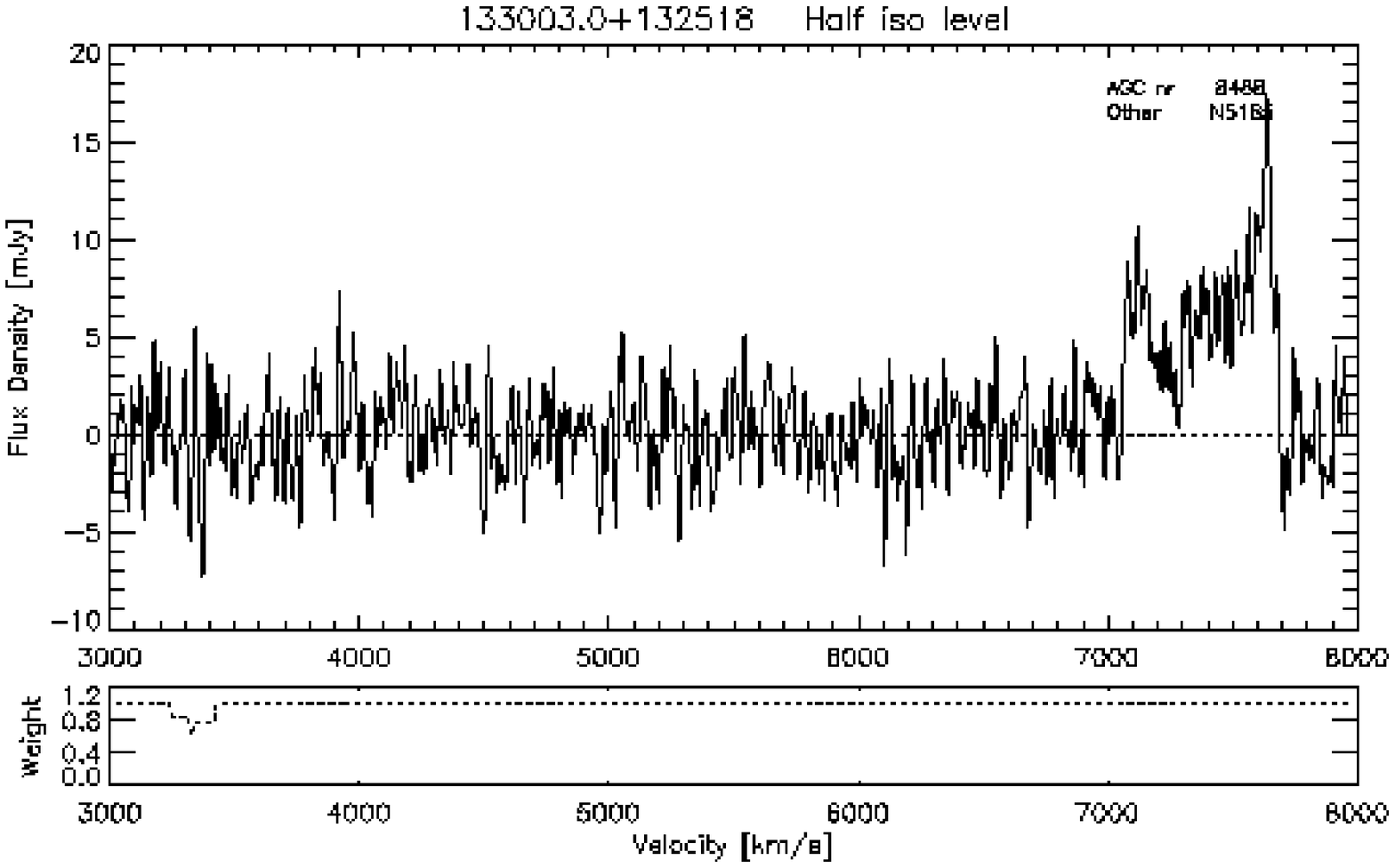}
\includegraphics[clip=,angle=0,width=7cm]{HI133453.6+135015_small.EPS}
\caption{Mosaic of HI profiles for possible high-mass galaxies (continued). The plots belong to the following ACG objects, with those objects retained as showing intrinsic wide HI profiles given in bold fonts: 226077 (top-left), 226111 (top-right),  223609 (2$^{\rm nd}$ row-left), {\bf 008375} (2$^{\rm nd}$ row-right), {\bf 008379} (3$^{\rm rd}$ row-left), {\bf 008475} (3$^{\rm rd}$ row-right), {\bf 008488} (bottom-left), and 008488 (bottom-right).}
\label{f:candidates1}
\end{figure}

\begin{figure}
\includegraphics[clip=,angle=0,width=7cm]{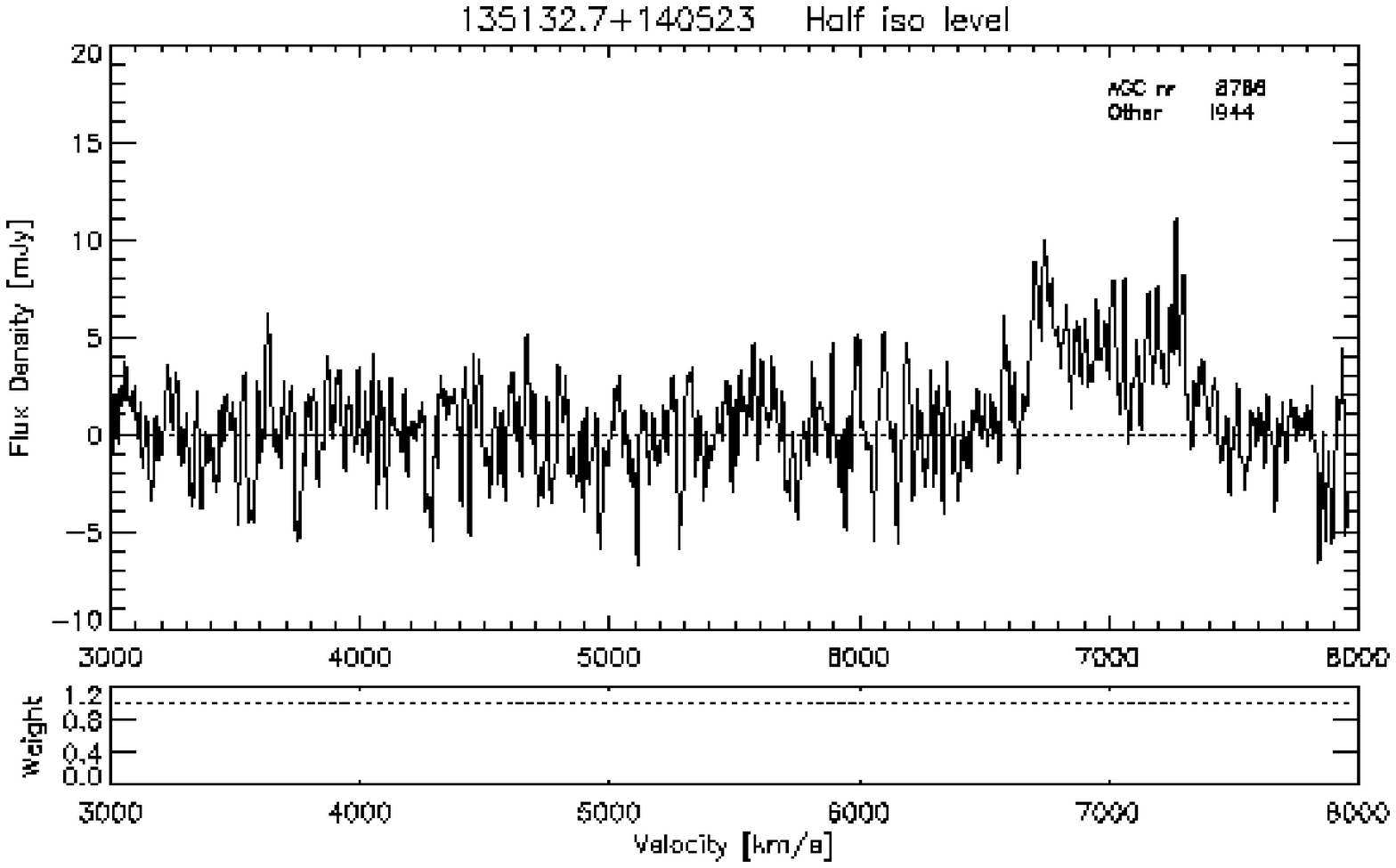}
\includegraphics[clip=,angle=0,width=7cm]{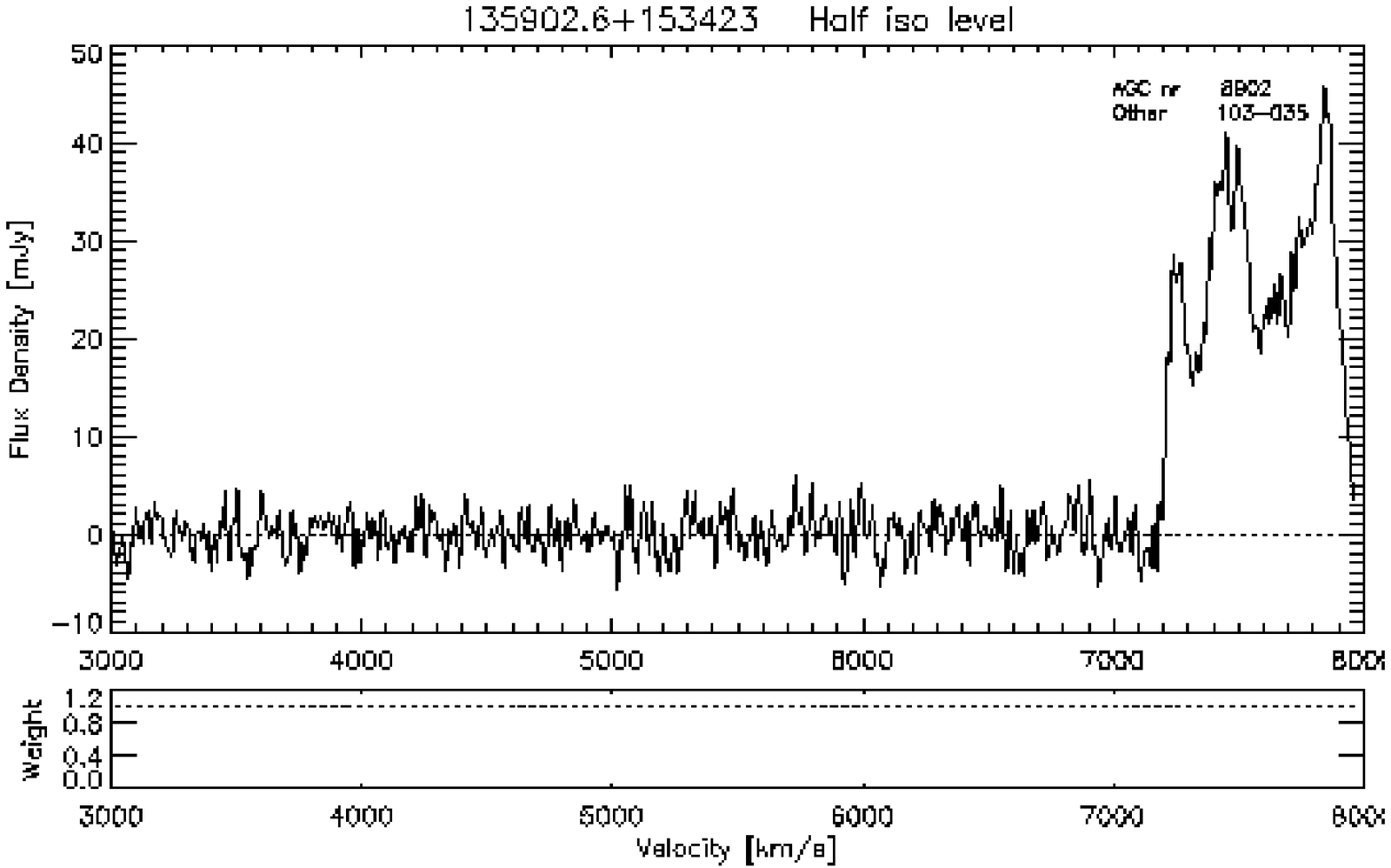}
\includegraphics[clip=,angle=0,width=7cm]{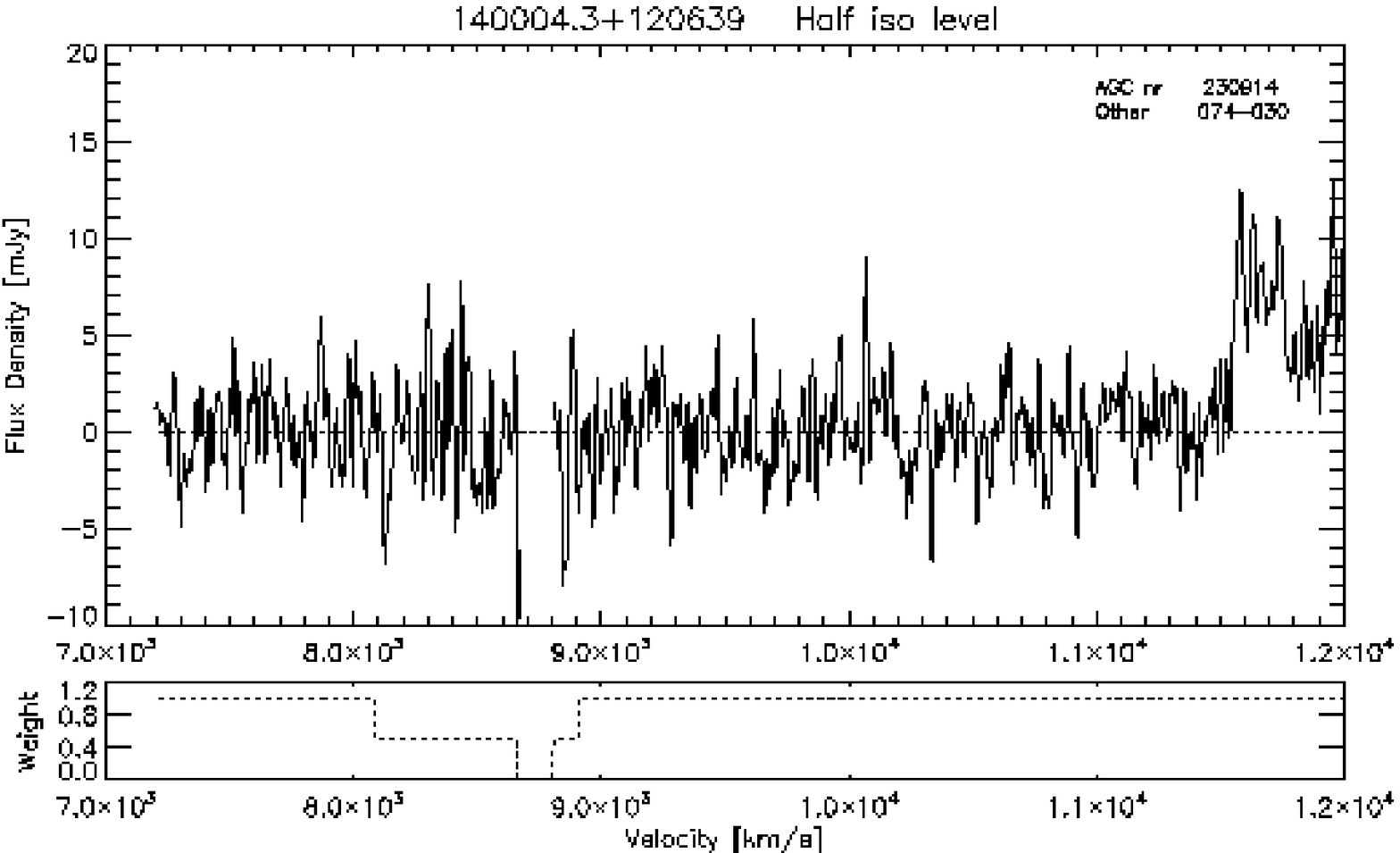}
\includegraphics[clip=,angle=0,width=7cm]{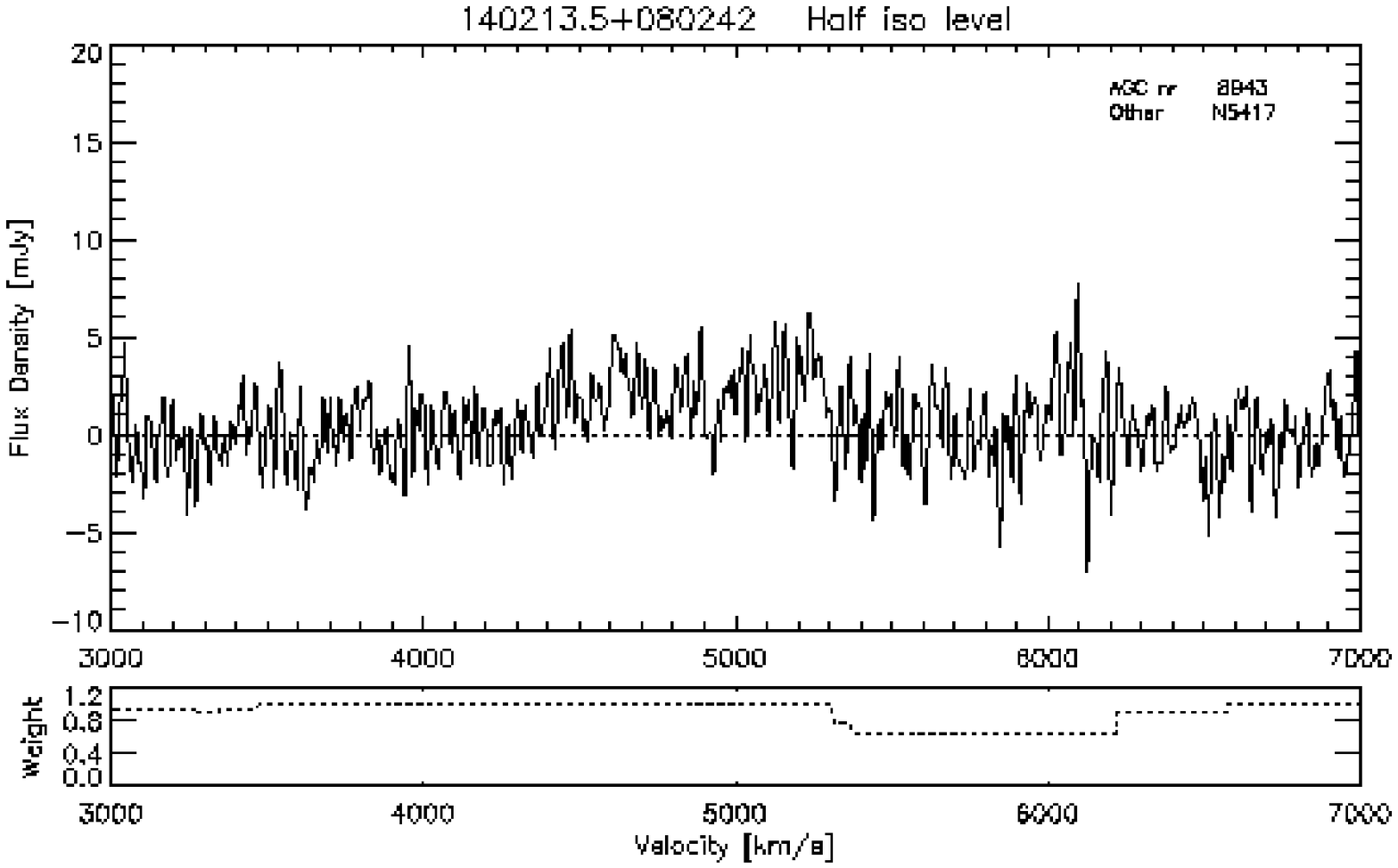}
\includegraphics[clip=,angle=0,width=7cm]{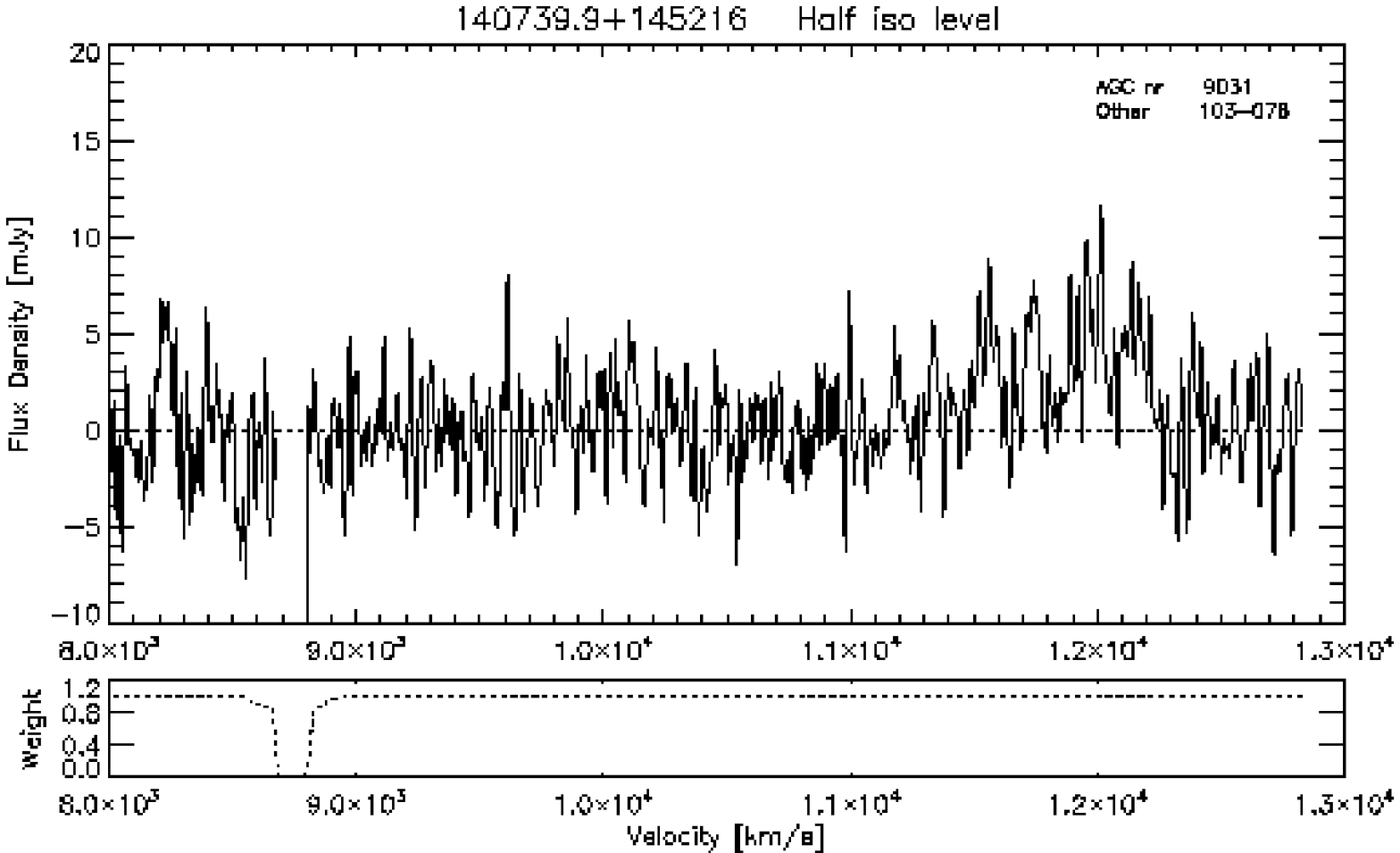}
\includegraphics[clip=,angle=0,width=7cm]{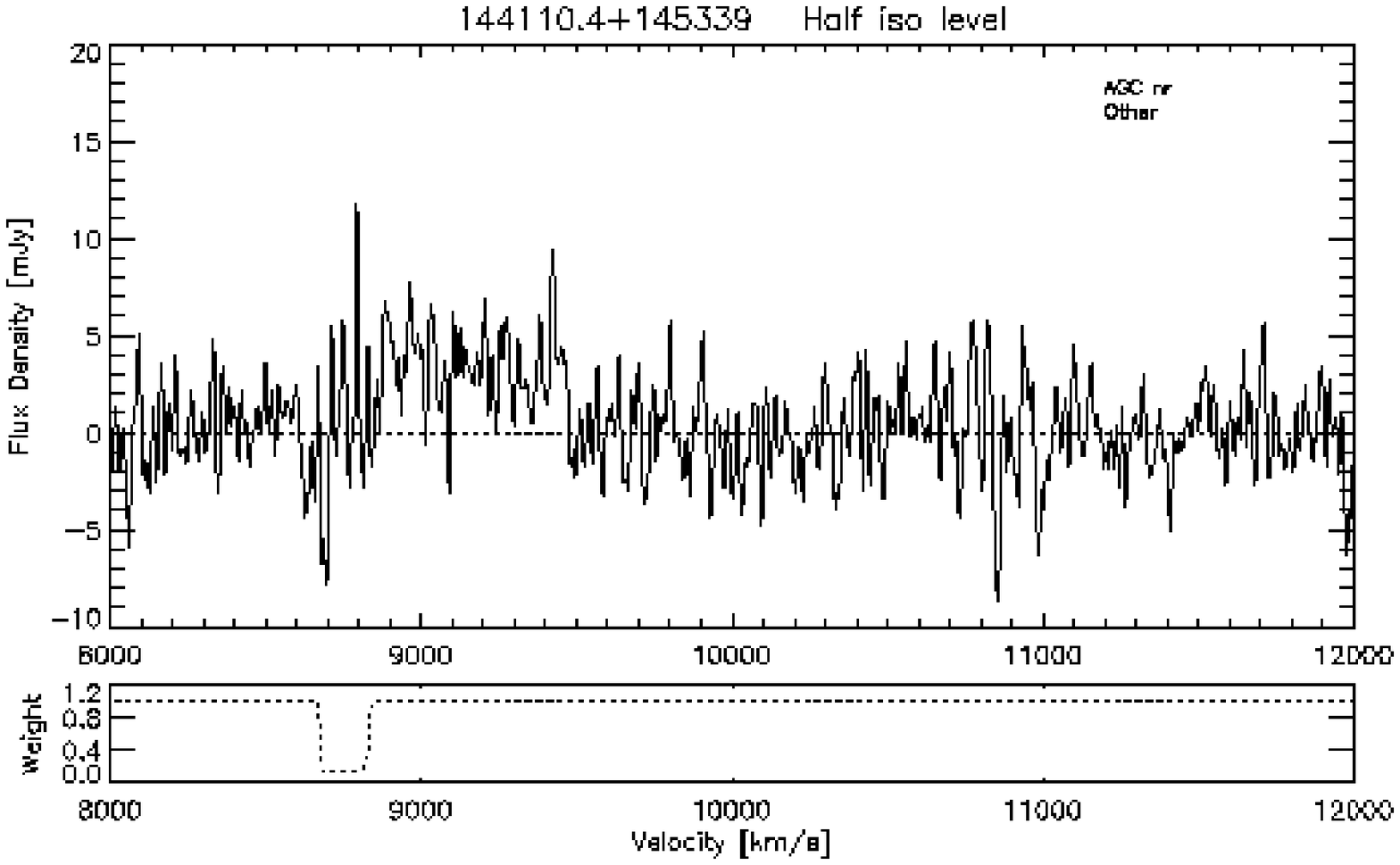}
\includegraphics[clip=,angle=0,width=7cm]{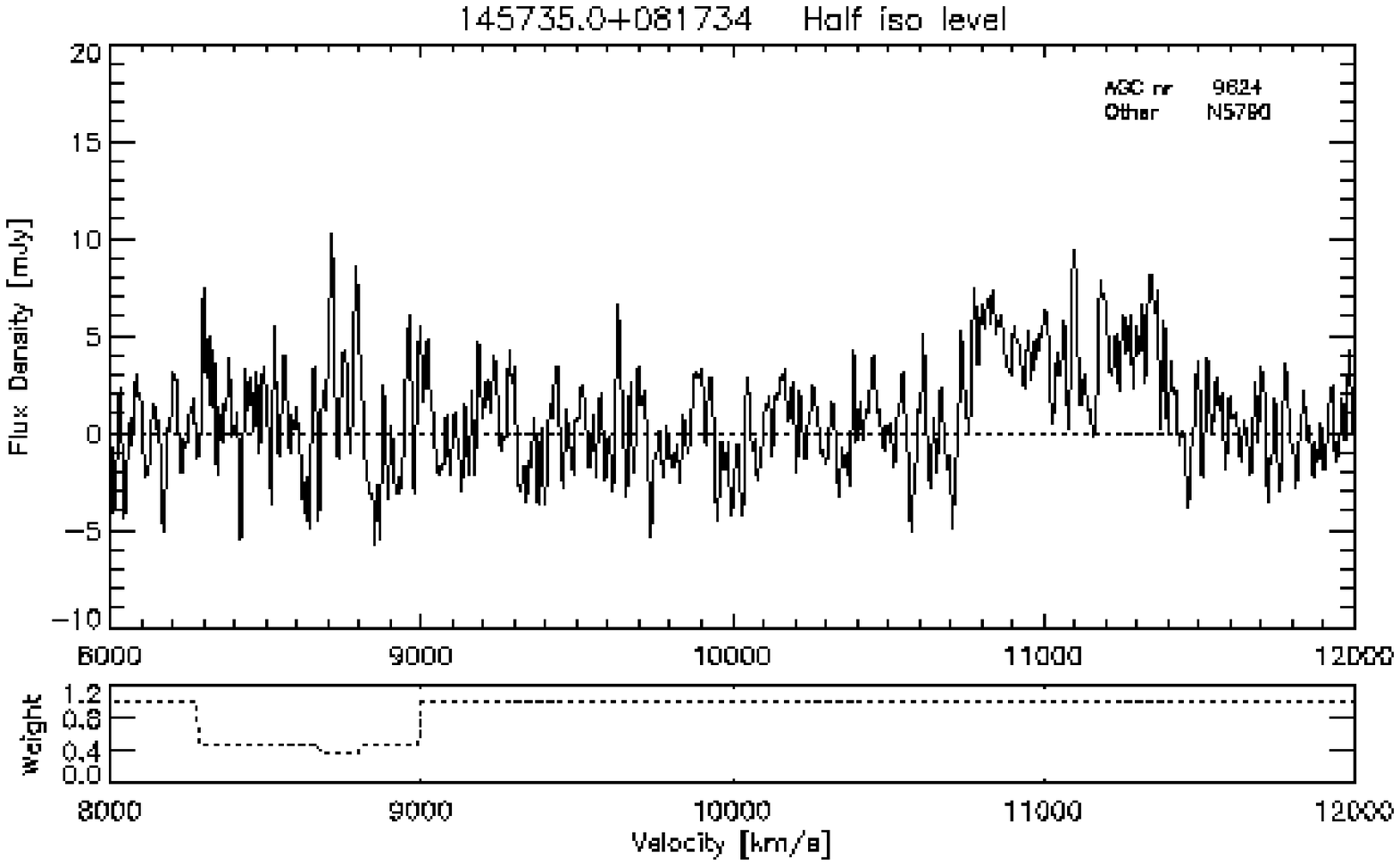}
\includegraphics[clip=,angle=0,width=7cm]{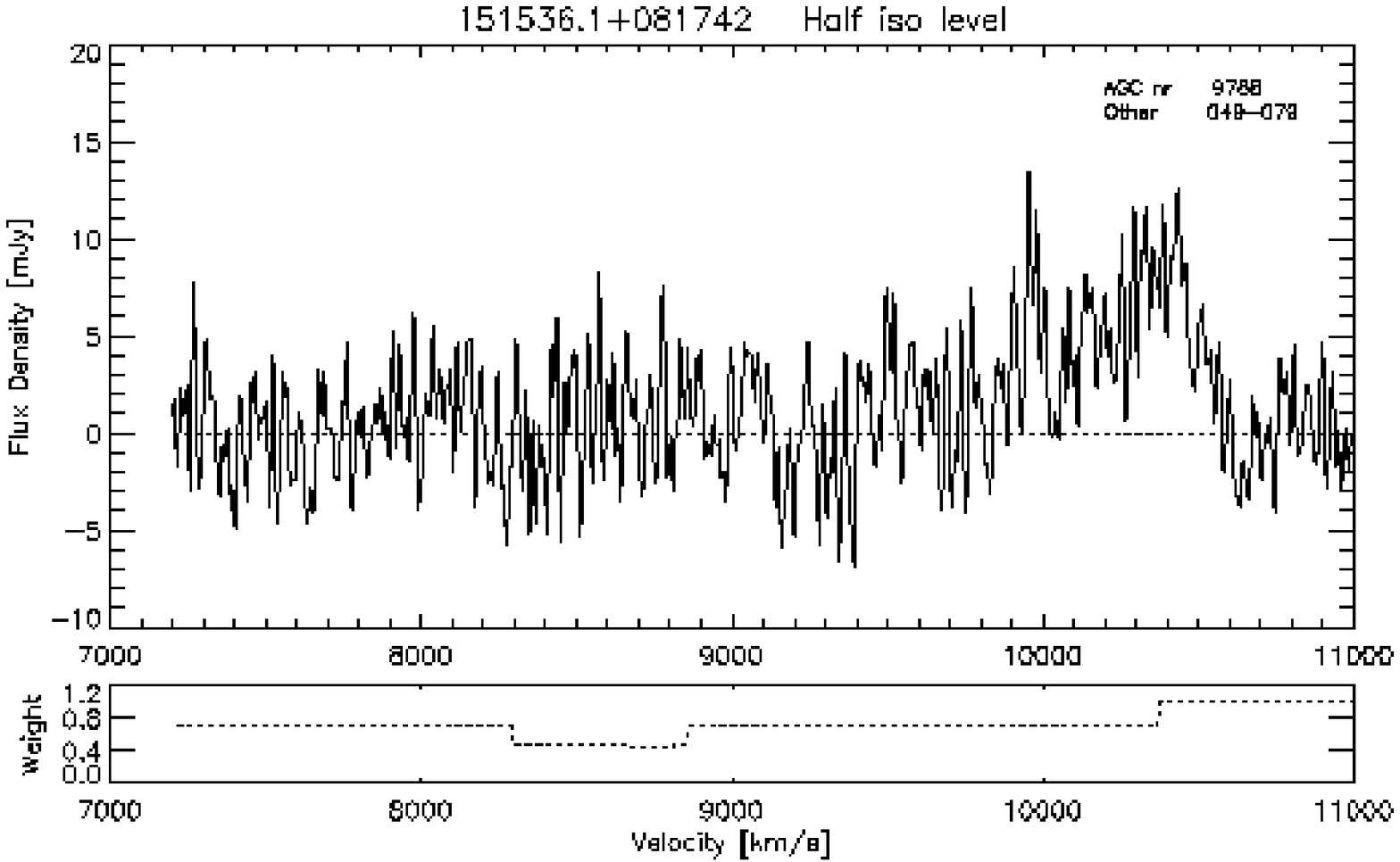}
\caption{Mosaic of HI profiles for possible high-mass galaxies (continued). The plots belong to the following ACG objects, with those objects retained as showing intrinsic wide HI profiles given in bold fonts: 008766 (top-left), 008902 (top-right), {\bf 230914} (2$^{\rm nd}$ row-left), 008943 (2$^{\rm nd}$ row-right), {\bf 009031} (3$^{\rm rd}$ row-left), {\bf 248977} (3$^{\rm rd}$ row-right), {\bf 009624} (bottom-left), and 009788 (bottom-right).}
\label{f:candidates2}
\end{figure}

\begin{figure}
\includegraphics[clip=,angle=0,width=7cm]{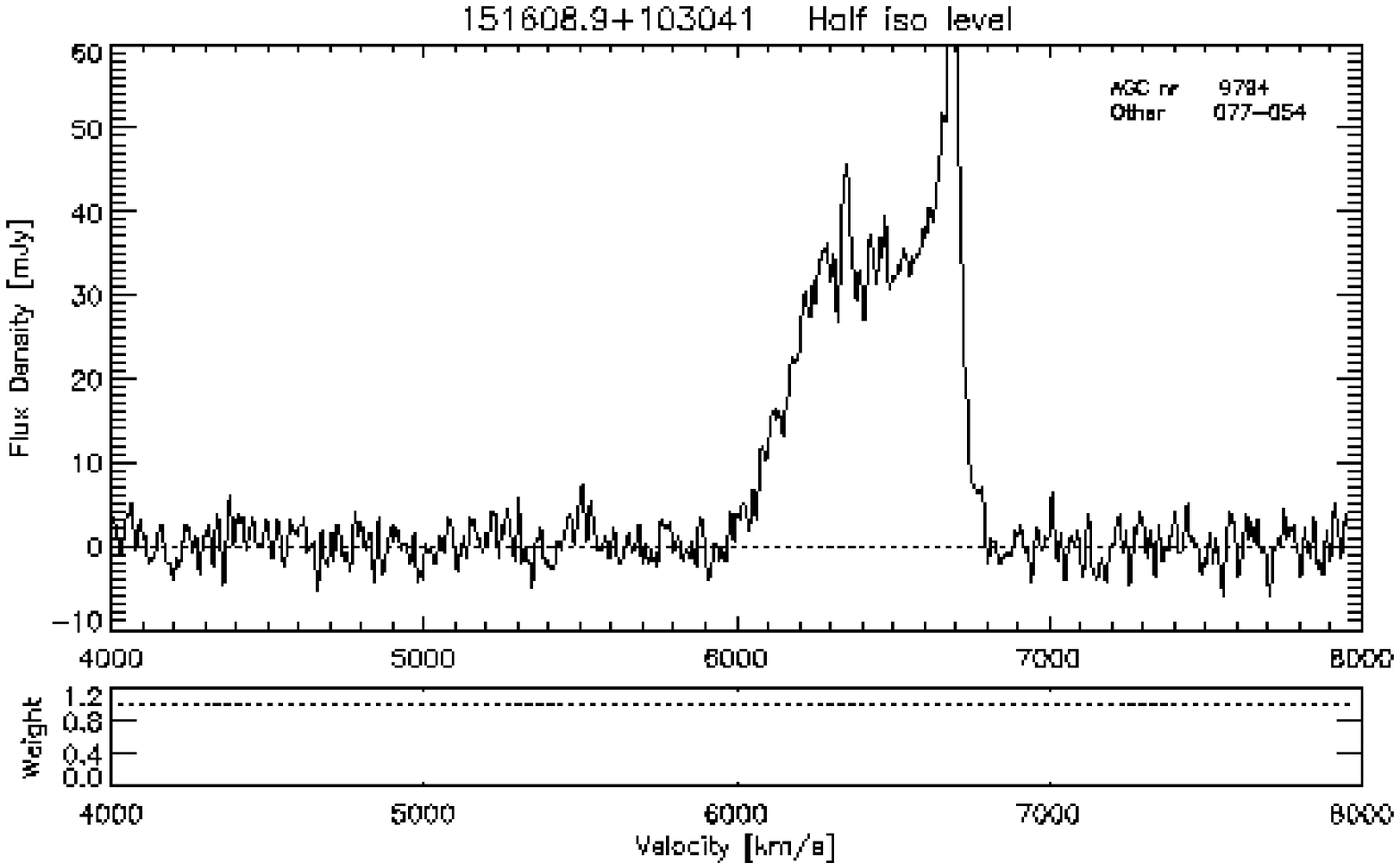}
\includegraphics[clip=,angle=0,width=7cm]{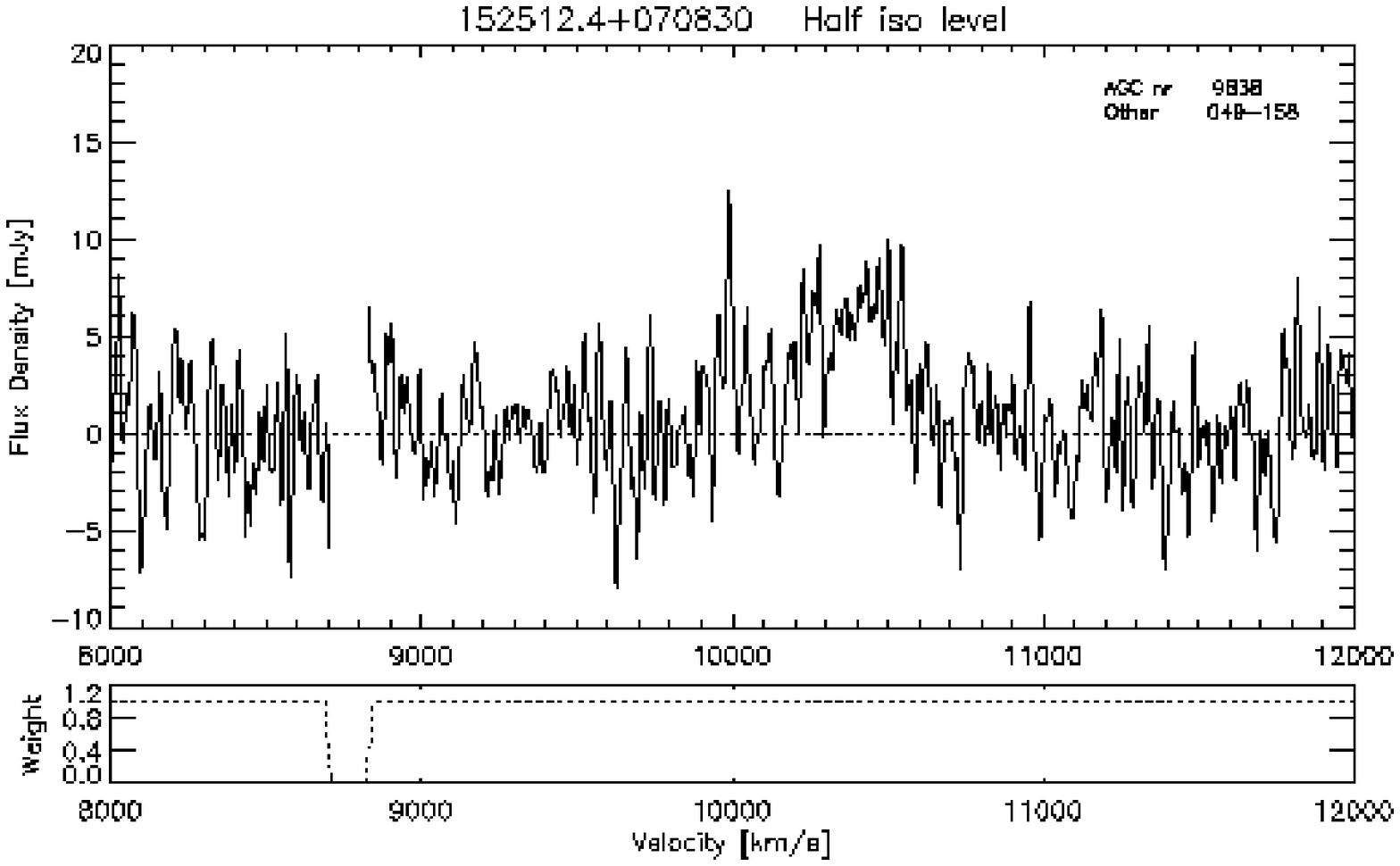}
\includegraphics[clip=,angle=0,width=7cm]{HI160448.0+140752_small.EPS}
\includegraphics[clip=,angle=0,width=7cm]{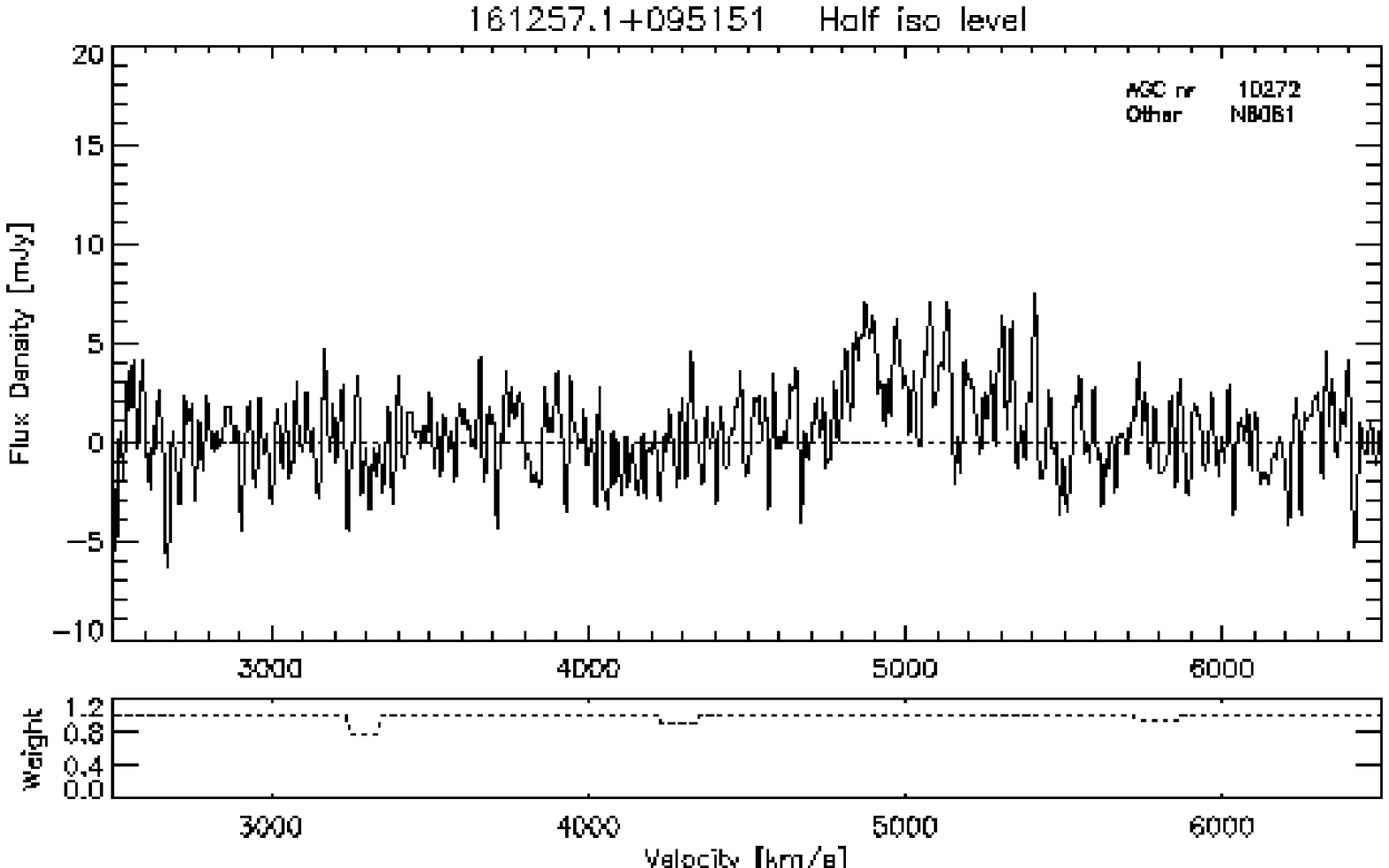}
\caption{Mosaic of HI profiles for possible high-mass galaxies (continued). The plots belong to the following ACG objects, with those objects retained as showing intrinsic wide HI profiles given in bold fonts: 009794 (top-left), 009838 (top-right), {\bf 260110} (bottom-left), and {\bf 010272} (bottom-right).}
\label{f:candidates3}
\end{figure}

\end{document}